\documentclass[a4paper,12pt]{article} 
\usepackage{pdflscape}
\usepackage{amssymb,amsmath,amsthm,mathrsfs,mathtools} 
\usepackage[all]{xy} 
\usepackage{caption} 
\usepackage{cite} 
\usepackage[colorlinks]{hyperref} 
\usepackage{eqparbox,graphicx} 
\usepackage[affil-it]{authblk} 
\usepackage{pgf,tikz,pgfplots,tkz-euclide,tikz-cd} 
\usetikzlibrary{arrows,positioning,calc,fadings,decorations.pathreplacing,decorations,shapes.geometric,fit} 
\tikzcdset{row sep/normal=7pt, column sep/normal=7pt} 
\tikzset{connect/.style={rounded corners=#1, 
        to path= 
($(\tikztostart)!-#1!(\tikztotarget)!-#1!-90:(\tikztotarget)$) -- 
($(\tikztotarget)!-#1!(\tikztostart)!-#1!90:(\tikztostart)$) -- 
        ($(\tikztotarget)!-#1!(\tikztostart)!#1!90:(\tikztostart)$) -- 
($(\tikztostart)!-#1!(\tikztotarget)!-#1!90:(\tikztotarget)$) -- cycle 
(\tikztotarget)}} 
\def\R{\mathbf{R}} 
\newcommand\kk{\mathbf{C}} 
\tikzstyle{block} = [rectangle, draw, fill=blue!80!black!20,  
    text width=6.3em, text centered, rounded corners, minimum height=3em] 
\newcommand\nodeone[1]{#1} 
\newcommand\nodetwo[2]{\begin{smallmatrix}#1\\\scriptstyle #2\end{smallmatrix}} 
\newcommand\nodethree[3]{\begin{smallmatrix} 
\scriptstyle #1\\\scriptstyle #2\\\scriptstyle #3 
\end{smallmatrix}} 
\newdimen\polrad 
\newcommand\Z{\mathbf{Z}} 
\newcommand\udim{\underline{\dim}} 
\renewcommand\mod[1]{{\operatorname{mod}\kk#1}} 
\newcommand\rep[1]{{\operatorname{rep} #1}} 
\newcommand\db[1]{\mathrm D^{\mathrm b}(\operatorname{rep} #1)} 
\newcommand\zz[1]{\Z{Q}} 
\newcommand\Hom[2][\empty]{\operatorname{Hom}_{#1}\left({#2}\right)} 
\newcommand\End[2][\empty]{\operatorname{End}_{#1}\left({#2}\right)} 
\newcommand\ind[1]{\operatorname{Ind}\left({#1}\right)} 
\def\H{\mathbf{H}} 
\def\A{\mathcal{A}} 
\def\B{\mathcal{B}} 
\def\D{\mathcal{D}} 
\def\K{\mathcal{K}} 
\def\P{\mathcal{P}} 
\def\q{\mathbf{q}} 
\def\F{\mathcal{F}} 
\def\T{\mathcal{T}} 
\def\C{\mathcal{C}} 
\newcommand\dg[1]{D^{\geq #1}} 
\newcommand\dl[1]{D^{\leq #1}} 
\newcommand\G[1]{\Gamma(#1)} 
\newcommand\ds{\ensuremath{\mathcal{M}_{\text{dS}}}} 
\newcommand\eq[1]{(\ref{#1})} 
\newcommand\diag{\operatorname{diag}} 
\def\res{\qopname\relax m{res}} 
\def\re{\qopname\relax m{Re}} 
\def\pa{\partial} 
\def\viz{{\it viz.\ }} 
\def\conf{{\mathcal{C}}} 
\def\ess{{\mathcal{S}}} 
\def\z{{\boldsymbol{z}}} 
\def\m{{\boldsymbol\mu}} 
\def\bxi{{\boldsymbol\xi}} 
\newcommand\inmu[3]{I^{{#2}}_{{#1}}({#3})} 
\def\inmuz{\inmu{N}{\m}{{\z}}} 
\newcommand\rt{\longrightarrow} 
\newcommand\lt{\longleftarrow} 
\def\M{\mathcal{M}} 
\def\I{\mathcal{I}} 
\def\M{\mathcal{M}} 
\def\Pid{\Pi_{\Delta}} 
\newcommand\norm[1]{\left\Vert #1\right\Vert} 
\def\Q{{\boldsymbol Q}} 
\def\Qd{Q^{\dagger}} 
\def\tr{\text{Tr\ }} 
\newcommand{\id}{\mathrm{id}}
\def\M{\mathcal{M}} 
\def\vt{\vartheta} 
\def\pp{{\mathbf{p}}} 
\def\xx{{\mathbf{x}}} 
\newcommand\op[1]{#1^{\text{op}}} 
\theoremstyle{definition} 
\newtheorem{example}{Example} 
 
\newcommand\dimvec[3]{\begin{smallmatrix}#1\\#2\\#3\end{smallmatrix}} 
\newcommand\singvec[3]{#1} 
\def\stackeqq#1#2{\underset{%
\displaystyle\overset{\displaystyle\shortparallel}{\scriptstyle 
#2}}{\scriptstyle #1}} 
\def\stackeq#1{\scriptstyle\overset{\rotatebox{90}{$=$}}{#1}} 
\newcommand\heart[1]{\langle #1\rangle} 
\newcommand\hsharp[1]{{#1}^{\sharp}} 
\newcommand\hflat[1]{{#1}^{\flat}} 
\newcommand\dleq[1]{{\D}^{\leqslant #1}} 
\newcommand\dgeq[1]{{\D}^{\geqslant #1}} 
\newcommand\cq[1]{{\mathcal{C}}_{{#1}}} 
\newcommand\dq[1]{{\mathcal{D}}_{{#1}}} 
\newcommand{\comment}[1]{{\sf\color{blue}[Comment] #1}} 

\newcommand\figref[1]{Fig.\ {\ref{#1}}} 
\def\lrar{\longleftrightarrow}
 
\title{Towards a categorification of scattering amplitudes} 
\author[1,2]{Severin Barmeier\thanks{s.barmeier@gmail.com}}  
\author[3]{Prafulla Oak\thanks{prafullaoakwork@gmail.com}} 
\author[3]{Aritra Pal\thanks{intap@iacs.res.in}} 
\author[3]{Koushik Ray\thanks{koushik@iacs.res.in}} 
\author[4]{Hipolito Treffinger\thanks{treffinger@imj-prg.fr}} 
\affil[1]{Hausdorff Research Institute for Mathematics, Poppelsdorfer Allee 45, 53115 Bonn, Germany}
\affil[2]{Universität zu Köln, Weyertal 86-90, 50931 Köln, Germany} 
\affil[3]{Indian Association for the Cultivation of Science, Calcutta 700 032, India}
\affil[4]{Institut de Mathématiques de Jussieu-Paris Rive Gauche, Université de Paris, 5 rue Thomas Mann, 75205 Paris, France}

\date{} 
\begin{document} 
\maketitle 
\begin{abstract} 
\noindent  
Categorification of scattering amplitudes for planar Feynman diagrams in
scalar field theories with a polynomial potential
is reported. Amplitudes for cubic theories are directly
written down in terms of projectives of hearts of intermediate $t$-structures 
restricted to the cluster category of quiver representations, without recourse to geometry. 
It is shown that for theories with $\phi^{m+2}$ potentials those 
corresponding to $m$-cluster categories are to be used.
The case of generic polynomial potentials is treated and our 
results suggest the existence of a generalization of higher cluster 
categories which we call pseudo-periodic categories. An algorithm
to obtain the projectives of hearts of intermediate $t$-structures 
for these types is presented. 
\end{abstract} 
\setcounter{page}{0} 
\thispagestyle{empty} 
\clearpage 
\section{Introduction} 
The ABHY scheme \cite{abhy} and the upsurge of activities it initiated 
unravelled novel connections between the quantum field 
theoretic computation of scattering amplitudes and a variety of mathematical 
notions and techniques. The ABHY programme provided a geometric way of 
writing scattering amplitudes of a biadjoint cubic scalar field theory in 
four dimensions as the volume form of the dual to the associahedron, a 
polyhedron associated to the combinatorics of grouping
the scattering particles according to the vertices of Feynman diagrams they
give rise to. This relationship  between the scattering amplitudes 
and the associahedron led to the connection of the former with the algebraic 
geometry of Grassmannians, tropical 
geometry \cite{dfgk}, cluster algebras \cite{bdmty,pppp}  
and representation theory of quivers and related subjects \cite{hbcgpt}.  
The ABHY scheme has been generalized in various directions. In one line of 
development scalar field theories with higher order interactions were 
studied \cite{raman,blr,abjjlm,kojima,ishan,ajk}.  
Another class of generalization is to obtain similar pictures for 
higher loop Feynman diagrams \cite{causal,jl3,ahs}.  
 
Various aspects of combinatorial computations are succinctly described 
within the framework of homological algebra using triangulated categories 
of quiver representations and related ideas. 
In an attempt to categorify the computation of scattering amplitudes
it has been shown that the scattering amplitude can be written 
down directly for a cubic scalar field theory from the cluster category of 
type $A$ quivers.  
In order to obtain the canonical form of $N$-particle scattering  
in a cubic scalar field theory one writes the cluster 
category of the $A_{N-3}$ quiver. The planar variables appearing  
in the different terms of the canonical form  
are obtained as the mass (modulus) of the 
central charges of the summands of cluster tilting objects, or equivalently central charges of projectives of intermediate $t$-structures in the corresponding derived category \cite{br}.
 
In the present article we complete the categorification of 
scattering amplitudes of tree-level planar Feynman diagrams
for arbitrary polynomial potentials.
The scattering amplitude of $N$ scalars  
in a $\phi^{m+2}$ theory for any integral $m$  
corresponding to such diagrams with $n$ vertices of valency $m$ 
is written in terms of the $m$-cluster tilting objects of an 
$A_{n-1}$ quiver, where $N = mn+2$, specifically,
in terms of  projectives of the hearts of the $m$-intermediate 
$t$-structures of the corresponding derived category.
Generalizing further, the amplitude for a theory with potential  
$\sum_{i}\lambda_i\phi^{m_i+2}$, where the sum is assumed to be over a finite 
number of terms, is written in terms of projectives of 
hearts in the intermediate $t$-structures restricted to what we call here a
pseudo-periodic
category, corresponding to the $N$-periodic triangulated 
thick subcategory of the derived category of an $A_{n-1}$ quiver, 
where $n=\sum_i n_i$ and $N=\sum_i m_i n_i+2$. This yields the amplitude
for planar Feynman diagrams with $n_i$ vertices of valency $m_i+2$, with a total number of $n$ vertices. One of the major motivations for the categorical 
approach is to facilitate effective bookkeeping, although this approach 
appears to indicate more. The pseudo-periodic category, which the
form of scattering amplitudes suggests is novel and has not
been studied earlier. 
All this is based on the relations between planar Feynman diagrams, 
polygons dual to them, their subdivisions and the connection of these to the 
representation theory of quivers, uncovered in previous studies.  
 
In section~\ref{sec:kin} we recall some of the known features which lead to the 
categorification using some examples. In section~\ref{sec:cat} we shall 
demonstrate how the categorical picture yields the amplitude directly from 
the field theoretic specification, after recalling the pertinent notions in
section~\ref{sec:dercat}.    
\section{Planar diagrams, angulations of polygons and meshes} 
\label{sec:kin} 
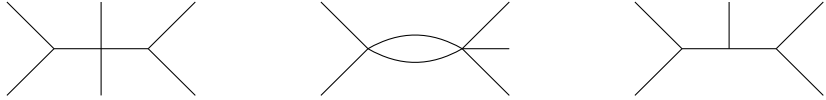
\begin{figure} 
\centering 
\begin{tikzpicture}[x=1.5em,y=1.5em]
\begin{scope} 
\draw (-1,0) to (1,0); 
\draw (-1,0) to (-2,1); 
\draw (-1,0) to (-2,-1); 
\draw (0,1) to (0,-1); 
\draw (1,0) to (2,1); 
\draw (1,0) to (2,-1); 
\end{scope} 
\begin{scope}[xshift=10em] 
\draw[bend right=30] (-1,0) to (1,0); 
\draw[bend left=30] (-1,0) to (1,0); 
\draw (-1,0) to (-2,1); 
\draw (-1,0) to (-2,-1); 
\draw (1,0) to (2,1); 
\draw (1,0) to (2,-1); 
\draw (1,0) to (2,0); 
\end{scope} 
\begin{scope}[xshift=20em] 
\draw (-1,0) to (1,0); 
\draw (-1,0) to (-2,1); 
\draw (-1,0) to (-2,-1); 
\draw (0,1) to (0,0); 
\draw (1,0) to (2,1); 
\draw (1,0) to (2,-1); 
\end{scope} 
\end{tikzpicture} 
\caption{Planar Feynman diagrams} 
\label{fig:34} 
\end{figure}

Let us begin with the planar Feynman diagrams of $N$-particle  
scattering in an interacting scalar field theory with a potential polynomial
in the field $\phi$,
\begin{equation} 
\label{vphi} 
V(\phi) = \sum_i\lambda_{m_i+2}\phi^{m_i+2},  
\end{equation}  
where the $m_i$'s are a finite set of positive integers and 
$\lambda_i$ is the  coupling at an interaction vertex of valency $m_i+2$.
The diagrams have $N$ external legs, satisfying   
\begin{equation} 
\label{NL} 
N = \sum_i m_i n_i +2-2L, 
\end{equation}  
where $n_i$ denotes the number of (internal) vertices in the 
diagram with valency 
$m_i+2$ and $L$ denotes the number of internal loops.  
This can be seen from the Euler characteristic of a planar graph with $V$ number of vertices, 
$E$ number of edges and $F$ faces, namely,
\begin{equation}
\label{EC}
V-E+F=2.
\end{equation}
We have $V=N+\sum_i n_i$ vertices in a planar Feynman diagram, $N$ being the number of 
monovalent vertices corresponding to the external legs. The total number of edges is 
given by  $2E=\sum_{i\neq 2} iV_i$, for $V_i$ number of vertices with valency 
$i$ in a graph. For the planar Feynman diagrams, this becomes
\begin{equation}
2E=N+\sum_i (m_i+2)n_i. 
\end{equation}
Also, counting the unbounded face in the plane as one, the 
total number of  faces is one more than the number of loops, $F=1+L$. Plugging these 
expressions  into \eq{EC} we obtain \eq{NL}.
Some examples are shown in Fig~\ref{fig:34}. The leftmost 
Feynman diagram has two 
vertices of valency $3$ and one of valency $4$. Thus, $m_1=1$, $n_1=2$, 
$m_2=2$, $n_2=1$ and no loop, accounting for the 
number of external particles $N=6$ by \eq{NL}.  
The middle one has one vertex of valency $4$, one of valency $5$ and 
one loop. Thus, $m_1=2$, $m_2=3$, $n_1=n_2=1$, and $L=1$, leading to $N=5$
external particles. 
In the special cases where the potential $V(\phi)$ has a single term,   
the Feynman diagrams have vertices of one type, with a fixed valency as 
indicated in the last diagram with $m=1,n=3,L=0,N=5$.  

Let us point out that in  
the special case of a cubic theory the 1-loop diagrams have $N=n$ by \eq{NL}. 
These may be related to the cluster category of $D_n$ quivers \cite{causal}. 
However, the $m$-cluster category of $D_n$ quivers are described in terms of 
$m$-angulations of a polygon with $(mn-m+1)$ sides \cite{BM1}. 
This number matches \eq{NL} with $L=1$ only if $m=1$. Thus, the 
analogy with  $D_n$ quivers may not hold beyond the cubic theory.  

\begin{figure}[h] 
\centering{%
\begin{tikzpicture}[x=12pt, y=12pt] 
\begin{scope}[yshift=3.3em] 
\node (a) at (0,.1) {\scriptsize $m=1, n=3$}; 
\end{scope} 
\begin{scope}[xshift=10em,yshift=3.3em] 
\node (a) at (0,.1) {\scriptsize $(m_1,m_2)=(1,2), (n_1,n_2)=(2,1)$}; 
\end{scope} 
\begin{scope} 
\draw[black!40] (0:\polrad) -- (72:\polrad) -- (144:\polrad) -- (216:\polrad)  
-- (288:\polrad) -- cycle;
\draw[black!40] (72:\polrad) -- (216:\polrad) node[above,pos=.4] {$\scriptstyle 1$}; 
\draw[black!40] (72:\polrad) -- (288:\polrad) node[right,pos=.3] 
{$\scriptstyle 2$}; 
%
\draw[thick](0:0) -- (0:.6\polrad);  
\draw[thick](0:.6\polrad) -- (36:1.2\polrad);  
\draw[thick](0:.6\polrad) -- (-36:1.2\polrad);  
\draw[thick ](0:0) -- (144:.6\polrad);  
\draw[thick] (144:.6\polrad) -- (108:1.2\polrad);  
\draw[thick](144:.6\polrad) -- (180:1.2\polrad);  
\draw[thick](0:0) -- (-108:1.2\polrad);  
\end{scope} 
\begin{scope}[xshift=10em] 
\draw[black!40] (0:\polrad) -- (60:\polrad) -- (120:\polrad) --  
(180:\polrad) --  (240:\polrad)  
-- (300:\polrad) -- (360:\polrad)  
-- cycle;
\draw[black!40] (240:\polrad) -- (360:\polrad) node[left,pos=.4] 
{$\scriptstyle 2$}; 
\draw[black!40] (120:\polrad) -- (360:\polrad) node[right,pos=.6] 
{$\scriptstyle 1$}; 
\draw[black!40,dashed] (180:\polrad) -- (360:\polrad) node[right,pos=.6] {}; 
\draw[thick](0:0) -- (67.5:.7*\polrad); 
\draw[thick](67.5:.7*\polrad) -- (40.5:1.2*\polrad); 
\draw[thick](67.5:.7*\polrad) -- (92.5:1.2*\polrad); 
\draw[thick](0:0) -- (292.5:.7*\polrad); 
\draw[thick](292.5:.7*\polrad) -- (325.5:1.2*\polrad); 
\draw[thick](292.5:.7*\polrad) -- (270.5:1.2*\polrad); 
\draw[thick](0:0) -- (202.5:1.2*\polrad); 
\draw[thick](0:0) -- (157.5:1.2*\polrad); 
\end{scope} 
%
%
\end{tikzpicture} 
}%
\caption{Angulation of polygons and Feynman diagrams} 
\label{tfp} 
\end{figure}
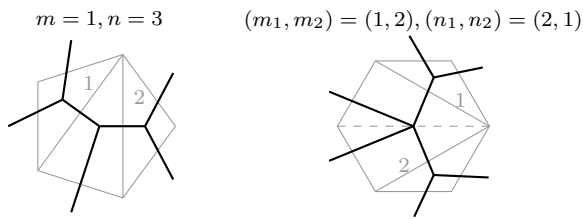

From now on we shall refer to the planar tree-level Feynman diagrams, 
that is the ones with no loops, $L=0$, as planar diagrams, unless 
otherwise qualified. For the planar diagrams \eq{NL} reads
\begin{equation}
\label{Nmn}
N=\sum_i m_in_i+2.
\end{equation}
Let us also define the total number of vertices,
\begin{equation}
\label{ni}
n=\sum_i n_i.
\end{equation}
A planar diagram for an $N$-particle scattering amplitude is dual to a 
decomposition of an $N$-gon by polygons with lesser number of sides. 
We shall refer to such a subdivision of a polygon as an
angulation. The internal lines of an angulation are referred to as diagonals.
If the angulation is with polygons all of which have the same number of sides, 
say $m$, it is called an $m$-angulation. 
The number of edges of an internal polygon in an angulation is equal to 
the valency of lines at a vertex of the Feynman diagram, the vertex 
being associated with the interior of the polygon. 
This is indicated in \figref{tfp}. 
The graphical connection between planar diagrams and angulation of polygons 
is algebraically realised by introducing the so-called planar variables. 
Indeed, the scattering amplitude is expressed  
in terms of these Lorentz invariant combinations of momenta of the external 
particles. We briefly recall the definition of kinematic variables 
for the scattering of a system of $N$ scalar particles \cite{abhy}. The 
 momenta of such particles are vectors in $\R^{1,3}$, 
denoted $\pp_i$ for $i = 1,\dotsc,N$ which satisfy the conservation equation 
\begin{equation} 
\label{cons} 
\sum_{i=1}^N \pp_i = 0. 
\end{equation} 
This equation can be solved by writing the momenta $\pp$ in terms of another set of $N$ 
four-vectors $\xx$ defined via
\begin{equation} 
\label{p2x} 
\pp_i = \xx_{i+1} - \xx_i, 
\end{equation} 
where the indices of $\xx$ are defined modulo $N$, in particular, 
$\xx_{N+1} = \xx_1$. 
Quadratic invariants in momenta for a pair of particles furnish the Mandelstam
variables
\begin{equation} 
\label{def:s} 
s_{ij}=(\pp_i+\pp_j)^2, 
\end{equation} 
where the norm of a four-vector $\pp = (p^0, p^1, p^2, p^3)$ is 
defined as 
$\pp^2=-(p^0)^2+(p^1)^2+(p^2)^2+(p^3)^2$. 
The planar variables are quadratic invariants defined similarly with 
$\mathbf x$'s, namely, 
\begin{equation} 
\label{x2X} 
X_{i,j}=(\xx_i-\xx_j)^2. 
\end{equation} 
The indices are symmetric by
definition, that is, $X_{j,i}=X_{i,j}$. 
Throughout this article we write the planar variables 
$X_{i,j}$ with subscripts ordered as $i<j$.
The planar variables are related to the Mandelstam variables by
\begin{equation}  
\label{X2s} 
s_{ij} = \pp_i^2 + \pp_j^2 + X_{i,j+1} + X_{i+1,j} - X_{i,j} - X_{i+1,j+1}, 
\end{equation} 
as can be derived from the definitions. 
If we further assume that the particles are massless, that is, their momentum 
vectors are null, $\pp_i^2=0$ for each 
$i$, then 
\begin{equation} 
\label{sX:null} 
s_{ii}=2\pp_i^2=0,\qquad X_{i,i+1}=\pp_i^2=0,
\end{equation} 
and the relation \eq{X2s} becomes 
\begin{equation} 
\label{sX} 
s_{ij}=X_{i,j+1}+X_{i+1,j}-X_{i,j}-X_{i+1,j+1}. 
\end{equation} 
The right-hand side is seen to coincide with the negative discrete Laplacian operating on the 
planar variables.  Some of the planar variables vanish by definition,  
\begin{equation} 
\label{planarvanish} 
\begin{aligned} 
X_{N,N+1} &= \eqmakebox[xN1]{$X_{N,1}$} = \eqmakebox[xN1]{$X_{1,N}$} = 0, \\ 
X_{2,N+1} &= \eqmakebox[xN1]{$X_{2,1}$} = \eqmakebox[xN1]{$X_{1,2}$} = 0. 
\end{aligned} 
\end{equation} 
Plotting $N$ points on the plane, each for an $x_i$,
joined by edges corresponding to 
the null variables \eq{planarvanish} yields a polygon. The non-vanishing $X_{i,j}$
are then identified with the diagonals of angulations of the polygon, as
shown for example, in Fig~\ref{fig:q3-5}. This is
the dual picture of planar diagrams described above. 
Quadratic invariants with more than a pair of particles are defined as
\begin{equation} 
\label{smult} 
s_{k_1k_2\cdots k_n} = \left(\sum_{i=1}^{n} \pp_{k_i}\right)^2,\quad 
\forall i, k_i\in\{1,\dotsc, N\}. 
\end{equation}  
Many of these  multi-indexed variables are 
related by momentum conservation \eq{cons}. 
For example, for $N=5$ we have $s_{123}=s_{45}$. 
These are related to the planar variables as
\begin{equation} 
\label{Xs} 
X_{i,j} = s_{i\ i+1\ \cdots\ j-1}. 
\end{equation}  
For a given $N$ and a potential \eq{vphi} 
a mesh diagram is obtained without alluding to the angulations.
The discrete Laplacian \eq{sX} is pictorially represented as  
\begin{equation} 
\label{mesh:X} 
s_{ij}=\begin{tikzpicture}[baseline=-2.6pt] 
\matrix (m) [matrix of math nodes, row sep={2.5em,between origins}, text 
height=1.4ex, font=\scriptsize, 
column sep={2.5em,between origins}, text depth=.15ex, 
ampersand replacement=\&]  
{ \& X_{i,j+1} \& \\ 
X_{i,j} \&\& X_{i+1,j+1}. \\ 
\& X_{i+1,j} \&\\ 
}; 
\draw[->] (m-2-1) -- (m-1-2);
\draw[->] (m-2-1) -- (m-3-2);
\draw[->] (m-1-2) -- (m-2-3);
\draw[->] (m-3-2) -- (m-2-3);
\end{tikzpicture}
\end{equation} 
For a given value of $N$, this unit can be woven into a mesh 
which has the form \cite{causal}  
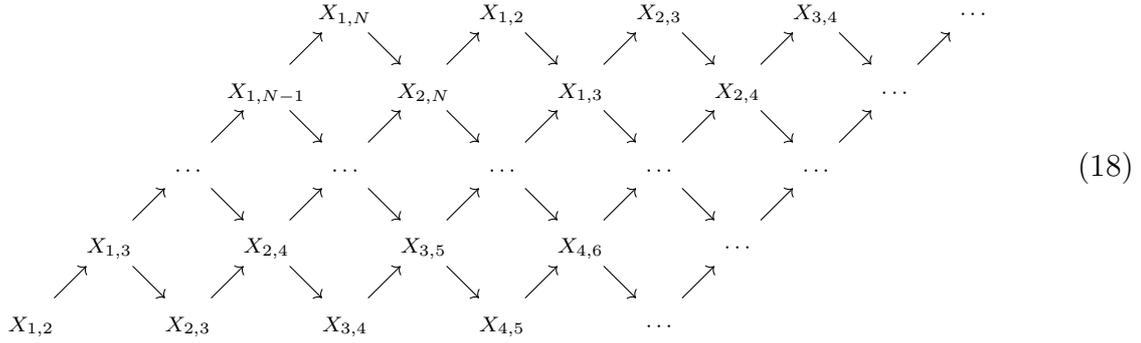
\begin{equation}
\label{mesh:big}
\begin{tikzpicture}[baseline=-4pt] 
\matrix (m) [matrix of math nodes, row sep={2.5em,between origins}, text 
height=1.6ex,  font=\scriptsize,
column sep={2.5em,between origins}, text depth=.15ex, 
ampersand replacement=\&]  
{ 
\&\&\&\& X_{1,N} \&\& X_{1,2} \&\& X_{2,3} \&\& X_{3,4} \&\& \cdots  \\ 
\&\&\& X_{1,N-1}\&\& X_{2,N} \&\& X_{1,3} \&\& X_{2,4} \&\& \cdots \& \\ 
\&\& \cdots \&\& \cdots \&\& \cdots \&\& \cdots \&\& \cdots \&\& \\ 
\& X_{1,3} \&\& X_{2,4} \&\& X_{3,5} \&\& X_{4,6} \&\& \cdots \&\&\& \\ 
X_{1,2}\&\& X_{2,3}\&\& X_{3,4}\&\&X_{4,5}\&\& \cdots\\ 
}; 
\foreach \i/\j in {1/2,3/4,5/6,7/8,9/10} 
\draw[->] (m-5-\i)--(m-4-\j); 
\foreach \i/\j in {2/3,4/5,6/7,8/9,10/11} 
\draw[->] (m-4-\i)--(m-3-\j); 
\foreach \i/\j in {3/4,5/6,7/8,9/10,11/12} 
\draw[->] (m-3-\i)--(m-2-\j); 
\foreach \i/\j in {4/5,6/7,8/9,10/11,12/13} 
\draw[->] (m-2-\i)--(m-1-\j); 
%
\foreach \i/\j in {5/6,7/8,9/10,11/12} 
\draw[->] (m-1-\i)--(m-2-\j); 
\foreach \i/\j in {4/5,6/7,8/9,10/11} 
\draw[->] (m-2-\i)--(m-3-\j); 
\foreach \i/\j in {3/4,5/6,7/8,9/10} 
\draw[->] (m-3-\i)--(m-4-\j); 
\foreach \i/\j in {2/3,4/5,6/7,8/9} 
\draw[->] (m-4-\i)--(m-5-\j); 
\end{tikzpicture} 
\end{equation} 
The top and the bottom rows are void, 
due to \eq{sX:null} and \eq{planarvanish}, 
the corresponding $X$'s 
being null. Discarding these rows we obtain a mesh of $X$'s with arrows.  
This is the mesh diagram for a $\phi^3$ theory with $m=1$.  However, 
including the rows of null planar variables in the 
bottom and top is convenient for weaving the mesh, especially for the general
cases.  Let us also point out that due to the periodicity of the indices 
such a mesh is always finite.
 
For $\phi^{m+2}$ theories, or, more generally for theories with a generic
polynomial potential 
\eq{vphi}, the knitting algorithm to obtain the mesh diagram of planar 
variables is to be modified. Instead of the discrete Laplacian, 
the basic unit is taken  to be \cite{jl3}
\begin{equation}
S_{ij}^{m,m'}:=\sum_{k'=0}^{m'-1}\sum_{k=0}^{m-1} s_{i+k,j+k'} =
X_{i,j+m'} + X_{i+m,j} -X_{i,j} - X_{i+m,j+m'}
\end{equation}
for a given pair $(m,m')$. This is represented pictorially as 
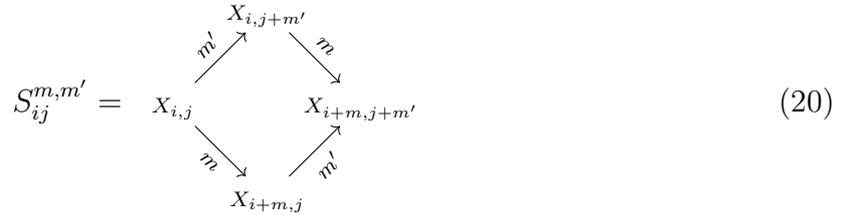
\begin{equation} 
\label{meshX:gen} 
S_{ij}^{m,m'}=\begin{tikzpicture}[baseline=-2.6pt] 
\matrix (m) [matrix of math nodes, row sep={3em,between origins}, text 
height=1.4ex, font=\scriptsize, 
column sep={3em,between origins}, text depth=.15ex, 
ampersand replacement=\&]  
{ \& X_{i,j+m'} \& \\ 
X_{i,j} \&\& X_{i+m,j+m'} \\ 
\& X_{i+m,j} \&\\ 
}; 
\draw[->] (m-2-1) -- (m-1-2) node[above,midway,sloped] {$\scriptstyle{m'}$} ; 
\draw[->] (m-2-1) -- (m-3-2) node[below,midway,sloped] {$\scriptstyle{m}$} ; 
\draw[->] (m-1-2) -- (m-2-3) node[above,midway,sloped] {$\scriptstyle{m}$} ; 
\draw[->] (m-3-2) -- (m-2-3) node[below,midway,sloped] {$\scriptstyle{m'}$} ; 
\end{tikzpicture} 
\end{equation} 
to be knit into a mesh. 

Weaving these units  into a mesh diagram is rather
cumbersome. Significant simplification is achieved by noting that the planar
variables appearing in the mesh are the diagonals of angulations of polygons.
While a planar diagram of the cubic theory is dual to a triangulation, 
which is a maximal subdivision of a polygon, the diagrams in theories with
higher indices of $\phi$ correspond to angulations with polygons having
number of sides more than three. These can be obtained by starting with the
triangulation and then systematically omitting diagonals, leaving a subset.
For example, the angulation of the polygon described in the right in Fig.~\ref{tfp} can be obtained by deleting the dashed diagonal from the
triangulation of the polygon. While enumeration of these angulations is
cumbersome, it translates to a combinatorial algorithm in terms of the mesh,
which we proceed to describe. 

Starting from the mesh \eq{mesh:big} corresponding to triangulations, 
entries are to be deleted to obtain the mesh for theories 
with general polynomial potentials. Given the number of vertices $n_i$ with 
valency $m_i$ in a planar diagram we first write a partition of 
the number $\sum_i m_in_i=N-2$  
such that $m_i$ appears $n_i$ number of times, namely,
\begin{equation} 
\label{mlist} 
\Lambda=(\overbrace{m_1, \dotsc, m_1}^{n_1 \text{ times}}, 
 \overbrace{m_2, \dotsc, m_2}^{n_2 \text{ times}}, 
      \dotsc, 
 \overbrace{m_i, \dotsc, m_i}^{n_i \text{ times}},\dotsc) .
\end{equation}
All permutations of the partition correspond one-to-one with 
the planar diagrams of a certain order in the couplings. 
Planar variables are then picked from the mesh \eq{mesh:big} using $\Lambda$
by going over the slices beginning with the first one
starting at $X_{1,2}$  according to the rules,
\begin{enumerate}
\item on a slice the entries obtained by incrementing the second index 
$j$ of $X_{i,j}$ are collected into a smaller slice, keeping their order
\item successive slices are chosen by adding $m_i$ from the partition to
both indices of the bottom row 
and shifting entries of $\Lambda$ periodically by one place
\item the above two steps are continued until the first slice is repeated.
\end{enumerate}
This is pictorially depicted as
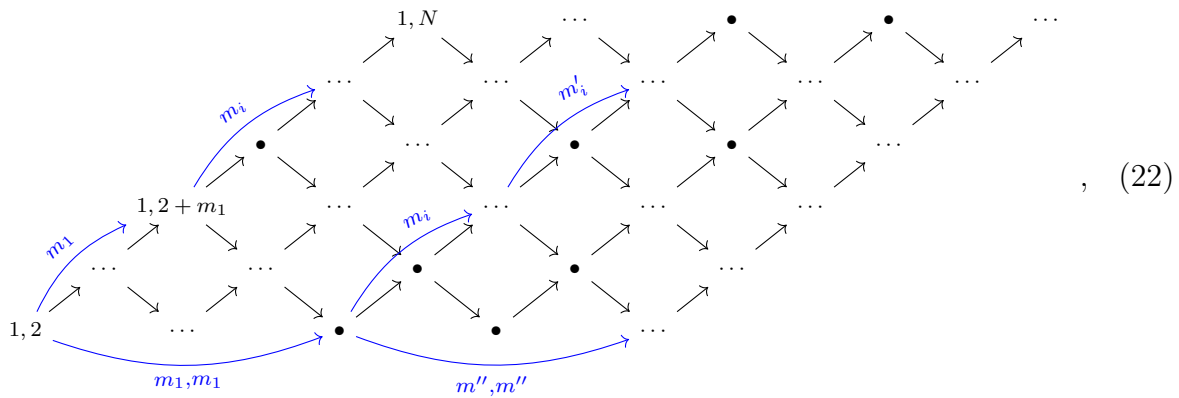
\begin{equation}
\label{mesh:hop}
\begin{tikzpicture}[baseline=-4pt] 
\matrix (m) [matrix of math nodes, row sep={2em,between origins}, text 
height=1ex,  font=\scriptsize,
column sep={2.5em,between origins}, text depth=.15ex, 
ampersand replacement=\&]  
{ 
\&\&\&\&\& {1,N} \&\&  \cdots \&\& \bullet \&\& \bullet \&\& \cdots  \\ 
\&\&\&\& \cdots \&\& \cdots \&\& \cdots \&\& \cdots \&\&  \cdots  \\ 
\&\&\& \bullet \&\& \cdots  \&\& \bullet \&\& \bullet \&\& \cdots \& \\ 
\&\& {1,2+m_1} \&\& \cdots \&\& \cdots \&\& \cdots 
\&\& \cdots \&\& \\ 
\& \cdots \&\& \cdots \&\& \bullet \&\& \bullet \&\& \cdots \&\&\& \\ 
{1,2}\&\& \cdots \&\& \bullet \&\& \bullet\&\& \cdots\\ 
}; 
\foreach \i/\j in {1/2,3/4,5/6,7/8,9/10} 
\draw[->] (m-6-\i)--(m-5-\j); 
\foreach \i/\j in {2/3,4/5,6/7,8/9,10/11} 
\draw[->] (m-5-\i)--(m-4-\j); 
\foreach \i/\j in {3/4,5/6,7/8,9/10,11/12} 
\draw[->] (m-4-\i)--(m-3-\j); 
\foreach \i/\j in {4/5,6/7,8/9,10/11,12/13} 
\draw[->] (m-3-\i)--(m-2-\j); 
\foreach \i/\j in {5/6,7/8,9/10,11/12,13/14} 
\draw[->] (m-2-\i)--(m-1-\j); 
%
\foreach \i/\j in {6/7,8/9,10/11,12/13} 
\draw[->] (m-1-\i)--(m-2-\j); 
\foreach \i/\j in {5/6,7/8,9/10,11/12} 
\draw[->] (m-2-\i)--(m-3-\j); 
\foreach \i/\j in {4/5,6/7,8/9,10/11} 
\draw[->] (m-3-\i)--(m-4-\j); 
\foreach \i/\j in {3/4,5/6,7/8,9/10} 
\draw[->] (m-4-\i)--(m-5-\j); 
\foreach \i/\j in {2/3,4/5,6/7,8/9} 
\draw[->] (m-5-\i)--(m-6-\j); 
\draw[blue,bend left=20,->] (m-6-1) to 
node[above,pos=.5,sloped]{$\scriptstyle m_1$}  (m-4-3) ;
\draw[blue,bend left=20,->] (m-4-3) to 
node[above,pos=.5,sloped]{$\scriptstyle m_i$} (m-2-5); 
\draw[blue,bend left=20,->] (m-6-5) to 
node[above,pos=.7,sloped]{$\scriptstyle m_i$} (m-4-7);
\draw[blue,bend left=20,->] (m-4-7) to 
node[above,pos=.7,sloped]{$\scriptstyle m_i'$} (m-2-9); 
\draw[blue,bend right=20,->] (m-6-1) to 
node[below,pos=.5,sloped]{$\scriptstyle m_1,m_1$} (m-6-5);
\draw[blue,bend right=20,->] (m-6-5) to 
node[below,pos=.5,sloped]{$\scriptstyle m'',m''$} (m-6-9);
\end{tikzpicture}, 
\end{equation} 
where only the indices $i,j$ of
$X_{i,j}$ are marked in the mesh diagrams, a convention followed in the
sequel. A concrete example of this procedure, which is more easier done 
than said, is discussed in Example~\ref{ex:34}.
Let us hasten to add that for a potential \eq{vphi} with a single term
$\phi^{m+2}$, the procedure is extremely simple. 
Starting with the first slice of
\eq{mesh:big}, slices are selected in steps of $m$ and 
entries of each of these are collected in steps of $m$  starting with the
null entry in the bottom row.

Enumeration of \emph{all} the angulations of $N$-gons and keeping track 
of these 
in  deriving the amplitude turns out to be as 
difficult a bookkeeping problem as drawing all the Feynman diagrams. In the categorical 
approach developed here it is possible to start from the combinatorics of the mesh diagram
obtained using the properties of the planar variables and the data \eq{Nmn}. 
Viewing the mesh as a periodic 
portion of the Auslander--Reiten (AR) quiver in the derived 
category of representation of type $A$ quivers, we obtain the 
scattering amplitudes directly.  This may be 
related to the geometric constructions through \cite{baur,BM2}.
Before proceeding to the categorical framework let us make the above 
constructions concrete through some illustrations studied previously in the
literature.
\begin{example}[Cubic theory] 
\label{ex:cubic} 
This case, with $m=1$, thus $N=n+2$, by \eq{NL},
has been dealt with earlier \cite{br}. Let us first consider the 
$N=5$ case, with $n=3$. There are five  planar diagrams obtained as duals of
five non-identical triangulations of a pentagon shown in 
Fig~\ref{fig:q3-5}. 
\begin{figure}[t] 
\centering 
\begin{tikzpicture}
\begin{scope}
  \foreach \a in {5}{ 
    \node [regular polygon, regular polygon sides=\a, minimum size=2cm, draw] 
at (0,0) (A) {}; 
    \foreach \i in {1,...,\a} 
    {%
      \pgfmathsetmacro\angle{360/\a} 
      \pgfmathsetmacro\b{\angle*\i+18}
      \node[font=\scriptsize] at (\b:1.25cm) {$\i$}; 
    } 
  } 
\draw  (A.corner 1) -- (A.corner 3) node[above=-.4ex, pos=.5,sloped] {$\scriptstyle X_{1,3}$}; 
\draw  (A.corner 1) -- (A.corner 4) node[above=-.4ex, pos=.5,sloped] {$\scriptstyle X_{1,4}$}; 
\end{scope} 
\begin{scope}[xshift=8em]
  \foreach \a in {5}{ 
    \node [regular polygon, regular polygon sides=\a, minimum size=2cm, draw] 
at (0,0) (A) {}; 
    \foreach \i in {1,...,\a} 
    {%
      \pgfmathsetmacro\angle{360/\a} 
      \pgfmathsetmacro\b{\angle*\i+18}
      \node[font=\scriptsize] at (\b:1.25cm) {$\i$}; 
    } 
  } 
\draw  (A.corner 1) -- (A.corner 4) node[above=-.4ex, pos=.5,sloped] {$\scriptstyle 
X_{1,4}$}; 
\draw  (A.corner 2) -- (A.corner 4) node[above=-.4ex, pos=.5,sloped] {$\scriptstyle 
X_{2,4}$}; 
\end{scope} 
\begin{scope}[xshift=16em]
  \foreach \a in {5}{ 
    \node [regular polygon, regular polygon sides=\a, minimum size=2cm, draw] 
at (0,0) (A) {}; 
    \foreach \i in {1,...,\a} 
    {%
      \pgfmathsetmacro\angle{360/\a} 
      \pgfmathsetmacro\b{\angle*\i+18}
      \node[font=\scriptsize] at (\b:1.25cm) {$\i$}; 
    } 
  } 
\draw  (A.corner 2) -- (A.corner 4) node[above=-.4ex, pos=.5,sloped] {$\scriptstyle 
X_{2,4}$}; 
\draw  (A.corner 2) -- (A.corner 5) node[above=-.4ex, pos=.5,sloped] {$\scriptstyle 
X_{2,5}$}; 
\end{scope} 
\begin{scope}[xshift=24em]
  \foreach \a in {5}{ 
    \node [regular polygon, regular polygon sides=\a, minimum size=2cm, draw] 
at (0,0) (A) {}; 
    \foreach \i in {1,...,\a} 
    {%
      \pgfmathsetmacro\angle{360/\a} 
      \pgfmathsetmacro\b{\angle*\i+18}
      \node[font=\scriptsize] at (\b:1.25cm) {$\i$}; 
    } 
  } 
\draw  (A.corner 3) -- (A.corner 5) node[above=-.4ex, pos=.5,sloped] {$\scriptstyle 
X_{3,5}$}; 
\draw  (A.corner 2) -- (A.corner 5) node[above=-.4ex, pos=.5,sloped] {$\scriptstyle 
X_{2,5}$}; 
\end{scope} 
\begin{scope}[xshift=32em]
  \foreach \a in {5}{ 
    \node [regular polygon, regular polygon sides=\a, minimum size=2cm, draw] 
at (0,0) (A) {}; 
    \foreach \i in {1,...,\a} 
    {%
      \pgfmathsetmacro\angle{360/\a} 
      \pgfmathsetmacro\b{\angle*\i+18}
      \node[font=\scriptsize] at (\b:1.25cm) {$\i$}; 
    } 
  } 
\draw  (A.corner 3) -- (A.corner 5) node[above=-.4ex, pos=.5,sloped] {$\scriptstyle 
X_{2,4}$}; 
\draw  (A.corner 1) -- (A.corner 3) node[above=-.4ex, pos=.5,sloped] {$\scriptstyle 
X_{2,5}$}; 
\end{scope} 
\end{tikzpicture}
\caption{Triangulations of a pentagon in $\phi^3$ theory $(m=1)$} 
\label{fig:q3-5} 
\end{figure}
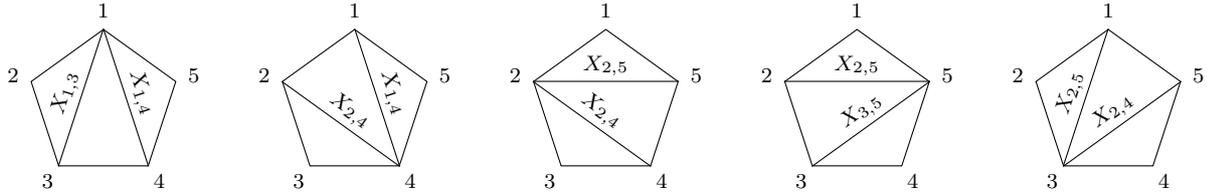
For example, the third triangulation  corresponds to the Feynman diagram  
\begin{equation}  
\label{FD:5pt} 
\begin{tikzpicture}[baseline=0,x=4.5ex, y=12pt] 
\draw(0,0) -- (0,-1) node[below] {4}; 
\draw(0,0) -- (1,0) node[pos=.7,below] {$\scriptstyle X_{2,5}$}; 
\draw(0,0) -- (-1,0) node[pos=.6,above] {$\scriptstyle X_{2,4}$}; 
\draw(-1,0) -- (-2,1) node[left] {2}; 
\draw(-1,0) -- (-2,-1) node[left] {3}; 
\draw(1,0) -- (2,-1) node[right] {5}; 
\draw(1,0) -- (2,1) node[right] {1}; 
\end{tikzpicture} 
\end{equation} 
where we used \eq{Xs} to write $s_{23}=X_{2,4}$ and $s_{15}=s_{234}=X_{2,5}$. 
The scattering amplitude is a sum of five terms, one for each 
Feynman diagram. In terms of the diagonals of the triangulation of the
pentagon, the amplitude reads
\begin{equation} 
\label{can:5} 
S_5 (\phi^3) =  
\frac{1}{X_{1,3}X_{1,4}}  
+\frac{1}{X_{1,4}X_{2,4}}  
+\frac{1}{X_{2,4}X_{2,5}}  
+\frac{1}{X_{2,5}X_{3,5}}  
+\frac{1}{X_{3,5}X_{1,3}}, 
\end{equation}  
the third term coming from \eq{FD:5pt}. 
The mesh diagram   
\begin{equation} 
\label{dbA2} 
\begin{tikzpicture}[baseline=-4pt] 
\matrix (m) [matrix of math nodes, minimum width=1.4em, row 
sep={2.6em,between origins}, text height=1.4ex, column sep={2.6em,between 
origins}, text depth=.15ex, font=\scriptsize
] 
{ 
& 1,4 &&  2,5 && 1,3 &&  \\ 
1,3 && 2,4 && 3,5&& 1,4 & \phantom{\hspace{2.6em}} \\ 
}; 
\foreach \i/\j in {1/2,3/4,5/6} 
\draw[->] (m-2-\i)--(m-1-\j); 
%
\foreach \i/\j in {2/3,4/5,6/7} 
\draw[->] (m-1-\i)--(m-2-\j); 
%
\end{tikzpicture} 
\end{equation}  
is obtained from \eq{mesh:big} after omitting the null entries. 
Let us stress that the mesh consists in 
all the diagonals that appear in all the triangulations 
corresponding to the planar diagrams. 
The terms in  \eq{can:5} are obtained directly by identifying the mesh 
as the cluster category 
of the AR quiver of an $A_{N-3}=A_2$  quiver \cite{br}. 
Each term contains the projectives of hearts of intermediate $t$-structures
of the derived category of representations of $A_2$ restricted to the cluster
category.
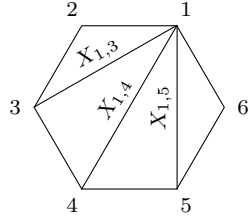
\begin{figure}
\centering
\begin{tikzpicture}
\begin{scope} 
  \foreach \a in {6}{ 
    \node [regular polygon, regular polygon sides=\a, minimum size=2.5cm, 
draw] at (0,0) (A)  {}; 
    \foreach \i in {1,...,\a} 
    {%
      \pgfmathsetmacro\angle{360/\a} 
      \pgfmathsetmacro\b{\angle*\i}
      \node[font=\scriptsize] at (\b:1.5cm) {$\i$}; 
    } 
  } 
\draw  (A.corner 1) -- (A.corner 3) node[above=-.4ex, pos=.5,sloped] {$\scriptstyle X_{1,3}$}; 
\draw  (A.corner 1) -- (A.corner 4) node[above=-.4ex, pos=.5,sloped] {$\scriptstyle X_{1,4}$}; 
\draw  (A.corner 5) -- (A.corner 1) node[above=-.4ex, pos=.5,sloped] {$\scriptstyle X_{1,5}$}; 
\end{scope} 
\end{tikzpicture} 
\caption{A triangulation of a hexagon in $\phi^3$ theory $(m=1)$} 
\label{fig:q3-6} 
\end{figure} 
Similarly, for  $N=6$,  
the three diagonals of all the triangulations of a hexagon, 
one of which is shown in Fig.~\ref{fig:q3-6}, appear in 
the scattering amplitude, given by 
\begin{multline}
\label{S3-6}
S_6(\phi^3) = \frac{1}{X_{1,3}X_{1,4}X_{1,5}} +
\frac{1}{X_{1,5}X_{2,4}X_{2,5}} +
\frac{1}{X_{1,5}X_{1,4}X_{2,4}} +
\frac{1}{X_{2,4}X_{2,5}X_{2,6}} +
\frac{1}{X_{2,5}X_{2,6}X_{3,5}} \\
+ \frac{1}{X_{2,6}X_{3,5}X_{3,6}} +
\frac{1}{X_{2,6}X_{3,6}X_{4,6}} +
\frac{1}{X_{1,5}X_{2,5}X_{3,5}} +
\frac{1}{X_{1,3}X_{1,4}X_{4,6}} +
\frac{1}{X_{1,3}X_{1,5}X_{3,5}} \\
+ \frac{1}{X_{1,4}X_{2,4}X_{4,6}} +
\frac{1}{X_{1,3}X_{3,6}X_{4,6}} +
\frac{1}{X_{2,4}X_{2,6}X_{4,6}} +
\frac{1}{X_{1,3}X_{3,5}X_{3,6}}. 
\end{multline}
The mesh diagram in this case is 
\begin{equation} 
\label{dbA3} 
\begin{tikzpicture}[baseline=-4pt] 
\matrix (m) [matrix of math nodes, row sep={2.5em,between origins}, text 
height=1.4ex, text centered, anchor=center,column sep={2.5em,between origins}, 
text depth=0.25ex, font=\scriptsize,ampersand replacement=\&, 
nodes in empty cells] 
{ 
\&\&  1,5 \&\& 2,6 \&\& 1,3 \&\&  \\ 
\&  1,4 \&\& 2,5 \&\& 3,6 \&\& 1,4 \& \\ 
1,3 \&\& 2,4 \&\& 3,5 
\&\& 4,6\&\& 1,5 \\ 
}; 
\foreach \i/\j in {1/2,3/4,5/6,7/8} 
\draw[thin,->] (m-3-\i)--(m-2-\j); 
\foreach \i/\j in {2/3,4/5,6/7} 
\draw[thin,->] (m-2-\i)--(m-1-\j); 
\foreach \i/\j in {3/4,5/6,7/8} 
\draw[thin,->] (m-1-\i)--(m-2-\j); 
\foreach \i/\j in {2/3,4/5,6/7,8/9} 
\draw[thin,->] (m-2-\i)--(m-3-\j); 
%
\end{tikzpicture} 
\end{equation}  
There are fourteen Feynman diagrams contributing to the 
scattering amplitude, which is again written directly in terms of the
projectives from the mesh \eq{dbA3} 
identified as the cluster category of an $A_{N-3}=A_3$ quiver
\cite{br}. 
\end{example} 
In this article we present a generalization of this approach to other, more
general, potentials. This brings in the paraphernalia of higher cluster
categories. In this section we obtain the mesh diagrams in a few more
general examples, postponing the categorical description 
to section~\ref{sec:cat}.
\begin{example}[Quartic theory] 
\label{ex:quartic} 
Planar diagrams of $N$-particle scattering in a quartic theory, 
with $V(\phi)=\lambda_4\phi^4$, that is, $m=2$, correspond to 
quadrangulations of $N$-gons, shown in Fig.~\ref{fig:q4} for $N=8,10$. 
\begin{figure}[h]
\centering
\begin{tikzpicture}
\begin{scope} 
  \foreach \a in {8}{ 
    \node [regular polygon, regular polygon sides=\a, minimum size=3cm, draw] 
at (0,0) (A) {}; 
    \foreach \i in {1,...,\a} 
    {%
      \pgfmathsetmacro\angle{360/\a} 
      \pgfmathsetmacro\b{\angle*\i+22.5}
      \node[font=\scriptsize] at (\b:1.75cm) {$\i$}; 
    } 
  } 
\draw  (A.corner 2) -- (A.corner 5) node[above=-.4ex, pos=.2,sloped] {$\scriptstyle 
X_{2,5}$}; 
\draw  (A.corner 5) -- (A.corner 8) node[above=-.4ex, pos=.8,sloped] {$\scriptstyle 
X_{5,8}$}; 
\end{scope}
\begin{scope}[xshift=12em]
  \foreach \a in {10}{ 
    \node [regular polygon, regular polygon sides=\a, minimum size=3cm, 
draw] at (0,0) (A)  {}; 
    \foreach \i in {1,...,\a} 
    {%
      \pgfmathsetmacro\angle{360/\a} 
      \pgfmathsetmacro\b{\angle*\i+36}
      \node[font=\scriptsize] at (\b:1.75cm) {$\i$}; 
    } 
  } 
\draw  (A.corner 2) -- (A.corner 5) node[above=-.4ex, pos=.5,sloped] {$\scriptstyle X_{2,5}$}; 
\draw  (A.corner 2) -- (A.corner 7) node[above=-.4ex, pos=.2,sloped] {$\scriptstyle X_{2,7}$}; 
\draw  (A.corner 7) -- (A.corner 10) node[above=-.4ex, pos=.5,sloped] {$\scriptstyle X_{7,10}$}; 
\end{scope}
\begin{scope}[xshift=24em]
  \foreach \a in {10}{ 
    \node [regular polygon, regular polygon sides=\a, minimum size=3cm, 
draw] at (0,0) (A)  {}; 
    \foreach \i in {1,...,\a} 
    {%
      \pgfmathsetmacro\angle{360/\a} 
      \pgfmathsetmacro\b{\angle*\i+36}
      \node[font=\scriptsize] at (\b:1.75cm) {$\i$}; 
    } 
  } 
\draw  (A.corner 2) -- (A.corner 5) node[above=-.4ex, pos=.5,sloped] {$\scriptstyle X_{2,5}$}; 
\draw  (A.corner 1) -- (A.corner 6) node[above=-.4ex, pos=.5,sloped] {$\scriptstyle X_{1,6}$}; 
\draw  (A.corner 7) -- (A.corner 10) node[above=-.4ex, pos=.5,sloped] {$\scriptstyle X_{7,10}$}; 
\end{scope} 
\end{tikzpicture}
\caption{Some quadrangulations of $(2n+2)$-gons in $\phi^4$ theory $(m=2)$} 
\label{fig:q4} 
\end{figure} 
The quadrangulations are classified into Stokes polytopes and the 
scattering amplitude 
can be obtained considering all such quadrangulations \cite{blr,ishan}. 
For $N=8$ and $n=3$, corresponding to three quartic vertices in a planar
diagram, the partition is $\Lambda=(2,2,2)$ yielding the mesh
\begin{equation}  
\label{mesh:42} 
\begin{tikzpicture}[baseline=-4pt] 
\matrix (m) [matrix of math nodes, 
minimum width=1.4em,
row sep={2.6em,between origins}, 
column sep={1.6em,between origins}, 
font=\scriptsize, 
anchor=center,
]  
{ 
& 1,6 && 3,8&& 2,5&& 4,7&& 1,6\\
1,4 && 3,6 && 5,8 && 2,7 && 1,4\\
};
\foreach \i/\j in {1/2,3/4,5/6,7/8,9/10}
{ \draw[->] (m-2-\i) to (m-1-\j); };
\foreach \i/\j in {2/3,4/5,6/7,8/9}
{
\draw[->] (m-1-\i) to (m-2-\j);
};
\end{tikzpicture} ,
\end{equation} 
while for $N=10$ the mesh is
\begin{equation}  
\label{mesh:43} 
\begin{tikzpicture}[baseline=-4pt] 
\matrix (m) [matrix of math nodes, 
minimum width=1.4em,
row sep={2.6em,between origins}, 
column sep={1.6em,between origins}, 
font=\scriptsize, 
anchor=center,
]  
{ 
&&1,8  &&3,10 && 2,5 && 4,7 && 6,9 && 1,8\\
&1,6 && 3,8 && 5,10 && 2,7 && 4,9&& 1,6 \\
1,4 && 3,6 && 5,8 && 7,10 && 2,9 && 1,4\\
};
\foreach \i/\j in {1/2,3/4,5/6,7/8,9/10,11/12}
{ \draw[->] (m-3-\i) to (m-2-\j); };
\foreach \i/\j in {2/3,4/5,6/7,8/9,10/11,12/13}
{ \draw[->] (m-2-\i) to (m-1-\j); };
\foreach \i/\j in {3/4,5/6,7/8,9/10,11/12}
{ \draw[->] (m-1-\i) to (m-2-\j); };
\foreach \i/\j in {2/3,4/5,6/7,8/9,10/11}
{ \draw[->] (m-2-\i) to (m-3-\j); };
\end{tikzpicture} ,
\end{equation}  
with $\Lambda=(2,2,2,2)$. 
As mentioned earlier, these are obtained from \eq{mesh:big}, starting
with the first slice, selecting entries in steps of $2$, starting from
$X_{1,2}$.
For example, with $N=10$, the entries selected from the first slice are
\[X_{1,2} - X_{1,4} - X_{1,6} - X_{1,8} -
X_{1,10}.\]
The slices are then chosen in steps of $2$, selecting the ones starting with 
the null entries $X_{3,4}$, $X_{5,6}$, $X_{7,8}$ and $X_{9,10}$. After this
the first slice is repeated. Omitting the null entries now yields the mesh
\eq{mesh:43}. In the categorical approach 
the amplitude is obtained by writing down the mesh directly. 
\end{example} 
Planar diagrams contributing to the scattering amplitude in a theory
with a generic polynomial potential \eq{vphi} are dual to   
mixed angulations of $N$-gons. Let us discuss an example. 
More general examples will be discussed
in section~\ref{sec:cat}.
\begin{example}[$(3,4)$ theory] 
\label{ex:34}
\begin{figure}[h] 
\centering 
\begin{tikzpicture} 
\begin{scope} 
  \foreach \a in {6}{ 
    \node [regular polygon, regular polygon sides=\a, minimum size=3cm, 
draw] at (0,0) (A)  {}; 
    \foreach \i in {1,...,\a} 
    {%
       \pgfmathsetmacro\angle{360/\a} 
      \pgfmathsetmacro\b{\angle*\i}
      \node[font=\scriptsize] at (\b:1.75cm) {$\i$}; 
    } 
  } 
\draw  (A.corner 1) -- (A.corner 3) node[above=-.4ex, pos=.5,sloped] {$\scriptstyle 
X_{1,3}$}; 
\draw  (A.corner 1) -- (A.corner 4) node[above=-.4ex, pos=.5,sloped] {$\scriptstyle 
X_{1,4}$}; 
\end{scope}
\begin{scope}[xshift=12em] 
  \foreach \a in {7}{ 
    \node [regular polygon, regular polygon sides=\a, minimum size=3cm, 
draw] at (0,0) (A)  {}; 
    \foreach \i in {1,...,\a} 
    {%
      \pgfmathsetmacro\angle{360/\a} 
      \pgfmathsetmacro\b{\angle*\i+38.6}
      \node[font=\scriptsize] at (\b:1.75cm) {$\i$}; 
    } 
  } 
\draw  (A.corner 1) -- (A.corner 3) node[below=-.4ex, pos=.5,sloped] {$\scriptstyle 
X_{1,3}$}; 
\draw  (A.corner 1) -- (A.corner 5) node[above=-.4ex, pos=.5,sloped] {$\scriptstyle 
X_{1,5}$}; 
\end{scope}
\begin{scope}[xshift=24em]
  \foreach \a in {7}{ 
    \node [regular polygon, regular polygon sides=\a, minimum size=3cm, draw] 
at (0,0) (A) {}; 
    \foreach \i in {1,...,\a} 
    {%
      \pgfmathsetmacro\angle{360/\a} 
      \pgfmathsetmacro\b{\angle*\i+38.6}
      \node[font=\scriptsize] at (\b:1.75cm) {$\i$}; 
    } 
  } 
\draw  (A.corner 2) -- (A.corner 4) node[above=-.4ex, pos=.5,sloped] {$\scriptstyle 
X_{2,4}$}; 
\draw  (A.corner 1) -- (A.corner 5) node[below=-.4ex, pos=.7,sloped] {$\scriptstyle 
X_{1,5}$}; 
\draw  (A.corner 1) -- (A.corner 6) node[below=-.4ex, pos=.7,sloped] {$\scriptstyle 
X_{1,6}$}; 
\end{scope} 
\end{tikzpicture} 
\caption{Mixed angulations for $N=6$ and $N=7$ in 
$V(\phi)=\lambda_3\phi^3+\lambda_4\phi^4$ theory} 
\label{fig:q34} 
\end{figure}
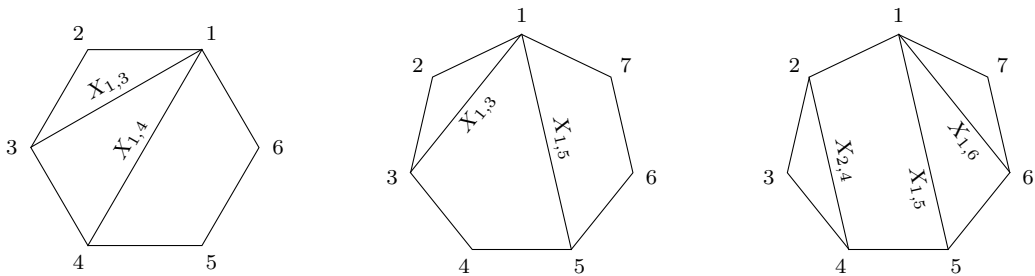 
Let us consider the interaction potential 
$V(\phi)=\lambda_3\phi^3+\lambda_4\phi^4$, so that $(m_1,m_2)=(1,2)$. The angulations involve triangles and quadrangles. 
For example, $(n_1,n_2)=(2,1)$ for $N=6$, so that the angulations 
have two triangles and one quadrangle \cite{ajk}, as in the first diagram of 
\figref{fig:q34}. 
For  $N=7$ two cases arise from the solutions of \eq{Nmn}, 
namely, $(n_1,n_2)=(1,2)$, corresponding to one 
triangle and two quadrangles and $(n_1,n_2)=(3,1)$, corresponding to three 
triangles and one quadrangle, as demonstrated in the second and third 
diagrams in \figref{fig:q34}, respectively. 
 
Let us describe the derivation of the mesh for $N=6$ with the partition
$\Lambda=(1,1,2)$  in detail. 
This is sufficiently general to cover all the aspects of the construction. 
The cases studied earlier are simpler special cases.
As described before, the first slice commences 
with $X_{1,2}$ and the second index is incremented successively 
by $\Lambda$, yielding
$X_{1,2} \rt X_{1,3} \rt X_{1,4}\rt X_{1,6}$, as shown in \eq{ex:mesh}. 
In the bottom row, again starting with $X_{1,2}$, both the  indices are
incremented by equal amounts by entries of $\Lambda$, 
as indicated by the blue arrows in \eq{ex:mesh}, proceeding till the first
slice is repeated. 
Let us remark that for odd values of $N$ the slice is 
repeated with same direction of arrows 
among the nodes, but  sense reversed, from top to bottom, 
instead of bottom to top, which is the order
in the first slice. 
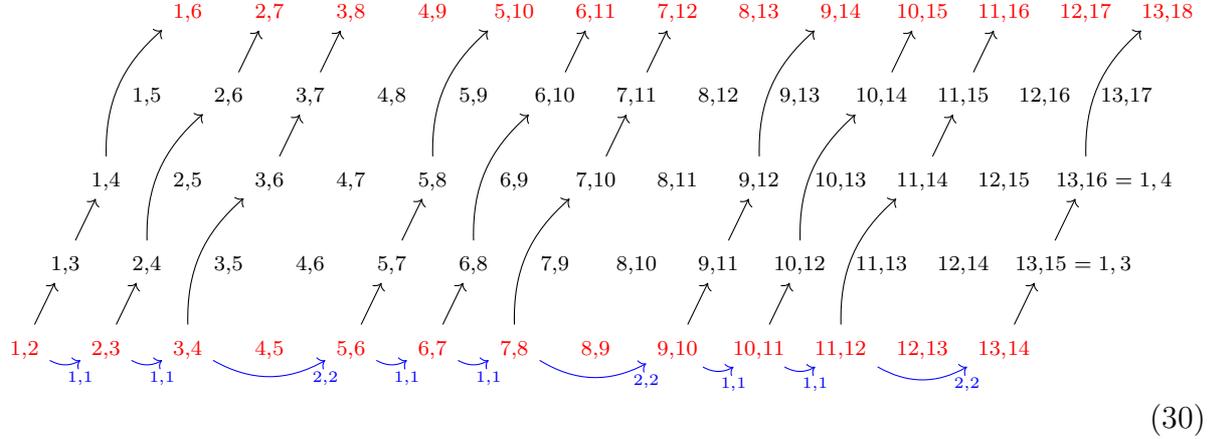
\begin{equation}  
\label{ex:mesh} 
\begin{tikzpicture}[baseline=-4pt] 
\matrix (m) [matrix of math nodes, row sep={2.7em,between origins}, text 
height=1.4ex,  
column sep={1.3em,between origins}, text depth=.1ex, font=\scriptsize, 
ampersand replacement=\&]  
{ 
\&\&\&\& \color{red} 1{,}6\&\&\color{red}  2{,}7 \&\& \color{red} 3{,}8 
\&\& \color{red} 4{,}9 \&\& \color{red} 5{,}10 \&\& \color{red} 6{,}11 \&\&
\color{red} 7{,}12\&\&\color{red}  8{,}13\&\&\color{red}  9{,}14\&\&
\color{red} 10{,}15 \&\&\color{red}  11{,}16\&\&\color{red}  12{,}17
\&\&\color{red} 13{,}18
\\ 
\&\&\&  1{,}5\&\& 2{,}6 \&\& 3{,}7 \&\& 4{,}8 \&\& 5{,}9  \&\& 6{,}10 \&\&
7{,}11\&\& 8{,}12\&\& 9{,}13\&\&10{,}14 \&\& 11{,}15\&\& 12{,}16\&\&13{,}17
\\ 
\&\&  1{,}4\&\& 2{,}5 \&\& 3{,}6 \&\& 4{,}7 \&\& 5{,}8  \&\& 6{,}9  \&\&
7{,}10\&\& 8{,}11\&\& 9{,}12\&\&10{,}13 \&\& 11{,}14\&\& 12{,}15\&\&13{,}16\
\mathrlap{=1,4}
\\ 
\&  1{,}3 \&\& 2{,}4 \&\& 3{,}5 \&\& 4{,}6 \&\& 5{,}7 \&\& 6{,}8 \&\& 
7{,}9 \&\& 8{,}10 \&\& 9{,}11 \&\&10{,}12\&\& 11{,}13 \&\& 12{,}14\&\& 13{,}15\
\mathrlap{=1,3} \\ 
\color{red} 1{,}2\&\&\color{red} 2{,}3\&\&\color{red} 3{,}4
\&\&\color{red} 4{,}5 \&\&\color{red}   5{,}6 
\&\&\color{red}  6{,}7\&\&\color{red}  7{,}8\&\&\color{red} 8{,}9\&\&\color{red}   9{,}10
\&\&\color{red}  10{,}11\&\&\color{red}  11{,}12\&\&\color{red}   12{,}13
\&\&\color{red}  13{,}14\\ 
}; 
\draw[->] (m-5-1)--(m-4-2);
\draw[->] (m-4-2)--(m-3-3);
\draw[bend left=25,->] (m-3-3) to (m-1-5);
\draw[->] (m-5-3)--(m-4-4);
\draw[bend left=25,->] (m-4-4) to (m-2-6);
\draw[->] (m-2-6)--(m-1-7);
\draw[bend left=25,->] (m-5-5) to (m-3-7);
\draw[->] (m-3-7)--(m-2-8);
\draw[->] (m-2-8)--(m-1-9);
\draw[->] (m-5-9)--(m-4-10);
\draw[->] (m-4-10)--(m-3-11);
\draw[bend left=25,->] (m-3-11) to (m-1-13);
\draw[->] (m-5-11)--(m-4-12);
\draw[bend left=25,->] (m-4-12) to (m-2-14);
\draw[->] (m-2-14)--(m-1-15);
\draw[bend left=25,->] (m-5-13) to (m-3-15);
\draw[->] (m-3-15)--(m-2-16);
\draw[->] (m-2-16)--(m-1-17);
\draw[->] (m-5-17)--(m-4-18);
\draw[->] (m-4-18)--(m-3-19);
\draw[bend left=25,->] (m-3-19) to (m-1-21);
\draw[->] (m-5-19)--(m-4-20);
\draw[bend left=25,->] (m-4-20) to (m-2-22);
\draw[->] (m-2-22)--(m-1-23);
\draw[bend left=25,->] (m-5-21) to (m-3-23);
\draw[->] (m-3-23)--(m-2-24);
\draw[->] (m-2-24)--(m-1-25);
\draw[->] (m-5-25)--(m-4-26);
\draw[->] (m-4-26)--(m-3-27);
\draw[bend left=25,->] (m-3-27) to (m-1-29);
%
\draw[blue,bend right=30,->] (m-5-1) to (m-5-3) 
node[below]  {$\scriptscriptstyle 1,1$}; 
\draw[blue,bend right=30,->] (m-5-3) to (m-5-5) 
node[below]  {$\scriptscriptstyle 1,1$}; 
\draw[blue,bend right=30,->] (m-5-5) to (m-5-9) 
node[below]  {$\scriptscriptstyle 2,2$}; 
\draw[blue,bend right=30,->] (m-5-9) to (m-5-11) 
node[below]  {$\scriptscriptstyle 1,1$}; 
\draw[blue,bend right=30,->] (m-5-11) to (m-5-13) 
node[below]  {$\scriptscriptstyle 1,1$}; 
\draw[blue,bend right=30,->] (m-5-13) to (m-5-17) 
node[below]  {$\scriptscriptstyle 2,2$}; 
\draw[blue,bend right=30,->] (m-5-17.320) to (m-5-19.220) 
node[below]  {$\scriptscriptstyle 1,1$}; 
\draw[blue,bend right=30,->] (m-5-19.320) to (m-5-21.220) 
node[below]  {$\scriptscriptstyle 1,1$}; 
\draw[blue,bend right=30,->] (m-5-21) to (m-5-25) 
node[below]  {$\scriptscriptstyle 2,2$}; 
\end{tikzpicture} 
\end{equation}
Discarding the top and bottom rows yields the mesh 
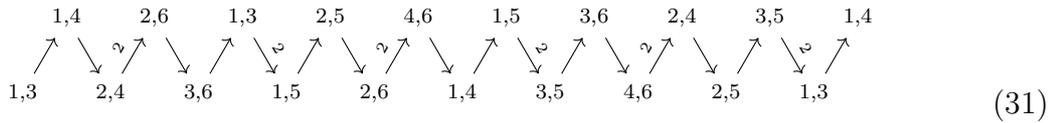
\begin{equation}  
\label{mesh:34-6} 
\begin{tikzpicture}[x=1.4em] 
\node at (0,1) (a1) {$\scriptstyle 1,4$}; 
\node at (2,1) (a2) {$\scriptstyle 2,6$}; 
\node at  (4,1) (a3) {$\scriptstyle 1,3$}; 
\node at  (6,1) (a4) {$\scriptstyle 2,5$}; 
\node at (8,1)  (a5) {$\scriptstyle 4,6$}; 
\node at (10,1)  (a6) {$\scriptstyle 1,5$}; 
\node at (12,1)  (a7) {$\scriptstyle 3,6$}; 
\node at (14,1)  (a8) {$\scriptstyle 2,4$}; 
\node at (16,1)  (a9) {$\scriptstyle 3,5$}; 
\node at (18,1)  (a10) {$\scriptstyle 1,4$}; 
\node at  (-1,0) (b1) {$\scriptstyle 1,3$}; 
\node at (1,0) (b2) {$\scriptstyle 2,4$}; 
\node at (3,0) (b3) {$\scriptstyle 3,6$}; 
\node at  (5,0)  (b4) {$\scriptstyle 1,5$}; 
\node at (7,0) (b5) {$\scriptstyle 2,6$}; 
\node at  (9,0) (b6) {$\scriptstyle 1,4$}; 
\node at (11,0) (b7) {$\scriptstyle 3,5$}; 
\node at (13,0) (b8) {$\scriptstyle 4,6$}; 
\node at  (15,0)  (b9) {$\scriptstyle 2,5$}; 
\node at (17,0) (b10) {$\scriptstyle 1,3$}; 
\draw[->] (b1) -- (a1); 
\draw[->] (b2) -- (a2) node[above,midway,sloped] {$\scriptscriptstyle 2$}; 
\draw[->] (b3) -- (a3); 
\draw[->] (b4) -- (a4); 
\draw[->] (b5) -- (a5) node[above,midway,sloped] {$\scriptscriptstyle 2$}; 
\draw[->] (b6) -- (a6); 
\draw[->] (b7) -- (a7); 
\draw[->] (b8) -- (a8) node[above,midway,sloped] {$\scriptscriptstyle 2$}; 
\draw[->] (b9) -- (a9); 
\draw[->] (b10) -- (a10); 
\draw[->] (a1) -- (b2); 
\draw[->] (a2) -- (b3); 
\draw[->] (a3) -- (b4) node[above,midway,sloped] {$\scriptscriptstyle 2$}; 
\draw[->] (a4) -- (b5); 
\draw[->] (a5) -- (b6); 
\draw[->] (a6) -- (b7) node[above,midway,sloped] {$\scriptscriptstyle 2$}; 
\draw[->] (a7) -- (b8); 
\draw[->] (a8) -- (b9); 
\draw[->] (a9) -- (b10) node[above,midway,sloped] {$\scriptscriptstyle 2$}; 
\end{tikzpicture} 
\end{equation} 
Let us emphasize that the blue arrows, discarded along with the bottom row,
are not part of the mesh diagram.
We have thus obtained the mesh diagram in terms of the planar variables.
While these may be associated  to angulations and thence to the dual planar 
diagrams, the present construction is algebraic, without alluding to 
the angulations. The mesh diagrams are identical to certain orbit 
categories within the bounded derived category of representations 
of quivers of $A$-type up to further identifications. 
Let us now discuss various notions leading to this identification. 
\end{example}
\section{Derived categories and quiver representations} 
\label{sec:dercat}
We briefly recollect various notions pertinent to the categorical
approach. In order to keep the article more-or-less self-contained we include
some pedagogical material as well
\cite{schiffler,keller,bmrrt,air,thomas,subir}. 
\subsection{Quiver representations}
A {\it quiver} $Q$ is a directed graph consisting of a set of directed edges between a set of vertices. In the context of this article, 
we work exclusively with quivers of type $A$, which are directed graphs 
whose underlying undirected graphs are Dynkin diagrams of type $A$. 
Moreover, we choose the following linear orientation for a quiver of 
type $A_n$
\begin{equation}
\label{an}
\stackrel{1}{\bullet} \, \longrightarrow \,\stackrel{2}{\bullet}\, \longrightarrow \dotsb \longrightarrow \,\stackrel{n}{\bullet}
\end{equation}
where we label the vertices by $1, \dotsc, n$.

A {\it representation} of a quiver $Q$ is given by assigning a finite-dimensional $\kk$-vector space $V_i$ to each vertex $\stackrel{i}{\bullet}$ and a linear map $f_{ij} \colon V_i \longrightarrow V_j$ to each edge $\stackrel{i}{\bullet}\, \longrightarrow \,\stackrel{j}{\bullet}$. Representations of a quiver $Q$ form an Abelian category denoted $\rep{Q}$ with morphisms $\phi \colon V \to W$ between two representations $V = ((V_i)_i, (f_{ij})_{i,j})$ and $W = ((W_i)_i, (g_{ij})_{i,j})$ given by commutative diagrams
\[
\begin{tikzpicture}[baseline=-4pt] 
\matrix (m) [matrix of math nodes, minimum width=1.4em, row 
sep={3.6em,between origins}, text height=1.4ex, column sep={4.6em,between 
origins}, text depth=.15ex, ampersand replacement=\&] 
{ 
V_i \& V_j \\
W_i \& W_j \\
};
\path[->,line width=.5pt,font=\scriptsize]
(m-1-1) edge node[above=-.2ex] {$f_{ij}$} (m-1-2)
(m-2-1) edge node[above=-.2ex] {$g_{ij}$} (m-2-2)
(m-1-1) edge (m-2-1)
(m-1-2) edge (m-2-2)
;
\end{tikzpicture}
\]
This category turns out to be Krull--Schmidt, i.e.\ every representation can be written uniquely (up to ordering of the direct summands) as the direct sum of indecomposable representations. In general, an explicit description of $\rep{Q}$ can be quite complicated. However, for quivers of type $A$ there are only finitely many indecomposable representations up to isomorphism which for the linearly oriented quiver \eqref{an} are all of the form
\[
\begin{smallmatrix} i \\[-.3em] \vdots \\[.1em] j \end{smallmatrix} := 0 \longrightarrow \dotsb \longrightarrow 0 \longrightarrow \overset{\raise.3em\hbox{\scriptsize $i$}}{\kk} \stackrel{\id}{\longrightarrow} \dotsb \stackrel{\id}{\longrightarrow} \overset{\raise.3em\hbox{\scriptsize $j$}}{\kk} \longrightarrow 0 \longrightarrow \dotsb \longrightarrow 0.
\]
Fig.~\ref{a2a3} lists all the  indecomposable representations for quivers 
of type $A_2$ and type $A_3$.
\begin{figure}
\[
\begin{tikzpicture}[x=1em,y=1em,baseline=-4pt]
\node at (1.2,5) {\it type $A_2$};
\node at (-5.5,2.6) {\it quiver};
\node at (-7.4,.5) {\it indecomposable};
\node at (-7.4,-.7) {\it representations};
\matrix (m) [matrix of math nodes, minimum width=1.4em, row 
sep=.25em, text height=1.4ex, column sep={0em,between 
origins}, text depth=.15ex, ampersand replacement=\&]
{ 
\&[1em] \&[1.3em] \stackrel{1}{\bullet} \&[3em] \stackrel{2}{\bullet} \\[.5em]
\nodeone{1}    \& := \& \kk \&  0  \\
\nodeone{2}    \& := \&  0  \& \kk \\
\nodetwo{1}{2} \& := \& \kk \& \kk \\
};
\path[->,line width=.5pt,font=\scriptsize]
(m-1-3) edge (m-1-4)
(m-2-3) edge (m-2-4)
(m-3-3) edge (m-3-4)
(m-4-3) edge node[above=-.2ex] {$\id$} (m-4-4)
;
\begin{scope}[xshift=10.5em]
\node at (1.1,5) {\it type $A_3$};
\matrix (m) [matrix of math nodes, minimum width=1.4em, row 
sep=.25em, text height=1.4ex, column sep={0em,between 
origins}, text depth=.15ex, ampersand replacement=\&]
{ 
\&[1em] \&[1.3em] \stackrel{1}{\bullet} \&[3em] \stackrel{2}{\bullet} \&[3em] \stackrel{3}{\bullet} \\[.5em]
\nodeone{1}    \& := \& \kk \&  0  \&  0  \\
\nodeone{2}    \& := \&  0  \& \kk \&  0  \\
\nodeone{3}    \& := \&  0  \&  0  \& \kk \\
};
\path[->,line width=.5pt,font=\scriptsize]
(m-1-3) edge (m-1-4)
(m-1-4) edge (m-1-5)
(m-2-3) edge (m-2-4)
(m-2-4) edge (m-2-5)
(m-3-3) edge (m-3-4)
(m-3-4) edge (m-3-5)
(m-4-3) edge (m-4-4)
(m-4-4) edge (m-4-5)
;
\begin{scope}[xshift=11em]
\matrix (m) [matrix of math nodes, minimum width=1.4em, row 
sep=.25em, text height=1.4ex, column sep={0em,between 
origins}, text depth=.15ex, ampersand replacement=\&]
{ 
\&[1em] \&[1.3em] \&[3em] \&[3em] \phantom{\stackrel{3}{\bullet}} \\[.5em]
\nodetwo{1}{2}      \& := \& \kk \& \kk \&  0  \\
\nodetwo{2}{3}      \& := \&  0  \& \kk \& \kk \\
\nodethree{1}{2}{3} \& := \& \kk \& \kk \& \kk \\
};
\path[->,line width=.5pt,font=\scriptsize]
(m-2-3) edge node[above=-.2ex] {$\id$} (m-2-4)
(m-2-4) edge (m-2-5)
(m-3-3) edge (m-3-4)
(m-3-4) edge node[above=-.2ex] {$\id$} (m-3-5)
(m-4-3) edge node[above=-.2ex] {$\id$} (m-4-4)
(m-4-4) edge node[above=-.2ex] {$\id$} (m-4-5)
;
\end{scope}
\end{scope}
\end{tikzpicture}
\]
\caption{The indecomposable representations of the quivers of type $A_2$ and type $A_3$}
\label{a2a3}
\end{figure}

Since $\rep{Q}$ is a {\it hereditary} category --- i.e.\ $\operatorname{Ext}_{\rep{Q}}^i (V, W) = 0$ for all $i \geq 2$ and all representations $V, W$ in $\rep{Q}$ --- the classification of indecomposable representations together with the description of the short exact sequences between them completely characterizes the Abelian category $\rep{Q}$.

The indecomposable representations for $\rep{Q}$ can be woven into a mesh indicating the so-called {\it almost-split exact} sequences, which are building blocks of the short exact sequences in $\rep{Q}$. This mesh is called the Auslander--Reiten quiver of the Abelian category $\rep{Q}$. In case of the quiver \eqref{an} of type $A_n$, the Auslander--Reiten quiver of $\rep{Q}$ looks as follows
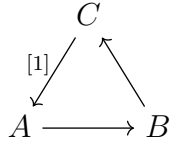
\begin{equation}
\label{mesh:an}
\begin{tikzpicture}[baseline=-4pt] 
\matrix (m) [matrix of math nodes, row sep={2em,between origins}, text height=1.4ex,  
column sep={2em,between origins}, text depth=.15ex, 
ampersand replacement=\&]  
{ 
\&\&\&\& \begin{smallmatrix} 1 \\[-.3em] \vdots \\[.1em] n \end{smallmatrix} \&\&\&\& \\ 
\&\&\& \phantom{\cdots} \&\& \phantom{\cdots} \&\&\& \\ 
\&\& \phantom{\cdots} \&\& \phantom{\cdots} \&\& \nodethree{1}{2}{3} \&\& \\ 
\& \phantom{\cdots} \&\& \phantom{\cdots} \&\& \nodetwo{2}{3} \&\& \nodetwo{1}{2} \& \\ 
n \&\& \cdots \&\& 3 \&\& 2 \&\& 1 \\ 
}; 
\path[dash pattern=on 0pt off 3pt, line width=.6pt, line cap=round] (m-4-2) ++(-.5em,-.5em) edge (m-2-4.45);
\path[dash pattern=on 0pt off 3pt, line width=.6pt, line cap=round] (m-2-6.135) ++ (.2em,-.2em) edge (m-2-6.315);
\path[dash pattern=on 0pt off 3pt, line width=.6pt, line cap=round] (m-3-5.135) ++ (.2em,-.2em) edge (m-3-5.315);
\path[dash pattern=on 0pt off 3pt, line width=.6pt, line cap=round] (m-4-4.135) ++ (.2em,-.2em) edge (m-4-4.315);
\foreach \i/\j in {1/2,5/6,7/8} 
\draw[->] (m-5-\i)--(m-4-\j); 
\foreach \i/\j in {6/7} 
\draw[->] (m-4-\i)--(m-3-\j); 
\foreach \i/\j in {4/5} 
\draw[->] (m-2-\i)--(m-1-\j); 
%
\foreach \i/\j in {5/6} 
\draw[->] (m-1-\i)--(m-2-\j); 
\foreach \i/\j in {6/7} 
\draw[->] (m-2-\i)--(m-3-\j); 
\foreach \i/\j in {5/6,7/8} 
\draw[->] (m-3-\i)--(m-4-\j); 
\foreach \i/\j in {4/5,6/7,8/9} 
\draw[->] (m-4-\i)--(m-5-\j); 
%
\end{tikzpicture} 
\end{equation} 
In order to relate this mesh to the one obtained in \eqref{mesh:big} from the planar variables, we have to enlarge the mesh in \eqref{mesh:an}, which can be achieved by considering the Auslander--Reiten quiver of the {\it derived category} $D^b (\rep{Q})$.

\subsection{Derived categories}
Let us recall \cite{thomas,subir} that the {\it derived category} 
$D(\A)$ of an Abelian category $\A$ consists of cochain complexes 
$A^{\bullet}$ of objects of $\A$, \viz 
\begin{equation}
\label{complex}
A^{\bullet} = \;\cdots \stackrel{d^{i-2}}{\longrightarrow} A^{i-1} \stackrel{d^{i-1}}{\longrightarrow} A^i \stackrel{d^i}{\longrightarrow} A^{i+1} \stackrel{d^{i+1}}{\longrightarrow} \cdots
\end{equation}
such that $d^i \circ d^{i-1} = 0$, where each $A^i$ in degree $i$ is an object of $\A$, localized in the class of quasi-isomorphisms. A \emph{quasi-isomorphism} is a morphism of complexes inducing an isomorphism on cohomology. A 
derived category is \emph{bounded} if the complexes only have cohomology in a finite 
number of degrees. The bounded derived category of $\A$ is denoted $D^b (\A)$.

The $k$-th {\it shift} of a complex $A^\bullet$ as in \eqref{complex} is denoted by $A^\bullet [k]$ with $A^i [k] := A^{i+k}$, i.e.\ the entries of $A^\bullet$ are shifted $k$ positions to the left. This shift defines an automorphism $[k] \colon D^b (\A) \to D^b (\A)$.

Triangles in the derived category, on the other hand, are counterparts 
of the exact sequences in an Abelian category. 

\begin{example}
Let us consider a short exact sequence $0 \to A \to B \to C \to 0$ in the Abelian category $\A$. Let us form a 
complex $C' = \cdots \to 0 \to A \to B \to 0 \to \cdots$, 
where $B$ is the degree $0$ term and view $C$ as the complex $\cdots \to 0 \to C \to 0 \to \cdots$, concentrated in degree $0$, as an element in the derived category. Then $C'$ and $C$ are quasi-isomorphic as complexes and thus isomorphic in $D (\A)$. We get a natural map $C \to A [1] = \cdots \to 0 \to A \to 0 \to \cdots$, where $A$ is placed in degree $-1$. This map is given by
\[
\begin{tikzpicture}[baseline=-4pt] 
\matrix (m) [matrix of math nodes, minimum width=1.4em, row 
sep={3.6em,between origins}, text height=1.4ex, column sep={3.6em,between 
origins}, text depth=.15ex, ampersand replacement=\&] 
{ 
\cdots \& 0 \& A \& B \& 0 \& \cdots \\
\cdots \& 0 \& A \& 0 \& 0 \& \cdots \\
};
\path[->,line width=.5pt,font=\scriptsize]
(m-1-1) edge (m-1-2)
(m-1-2) edge (m-1-3)
(m-1-3) edge (m-1-4)
(m-1-4) edge (m-1-5)
(m-1-5) edge (m-1-6)
(m-2-1) edge (m-2-2)
(m-2-2) edge (m-2-3)
(m-2-3) edge (m-2-4)
(m-2-4) edge (m-2-5)
(m-2-5) edge (m-2-6)
(m-1-2) edge (m-2-2)
(m-1-3) edge node[left=-.2ex] {$\id$} (m-2-3)
(m-1-4) edge node[left=-.2ex] {$0$} (m-2-4)
(m-1-5) edge (m-2-5)
;
\end{tikzpicture}
\]
where we replaced $C$ by $C'$ since they are isomorphic. In this way a short exact sequence in $\A$ gives rise to a \emph{triangle} $A \to B \to C \to A [1]$, also written in a more 
picturesque form as 
\[
\begin{tikzpicture}[baseline=-4pt] 
\matrix (m) [matrix of math nodes, minimum width=1.4em, row 
sep={3.6em,between origins}, text height=1.4ex, column sep={2.2em,between 
origins}, text depth=.15ex, ampersand replacement=\&] 
{ 
\& C \& \\
A \&\& B \\
};
\path[->,line width=.5pt,font=\scriptsize]
(m-1-2) edge node[pos=.4,left=-.2ex] {$[1]$} (m-2-1)
(m-2-1) edge (m-2-3)
(m-2-3) edge (m-1-2)
;
\end{tikzpicture}
\]
\end{example} 

More generally, the triangulated structure of $D (\A)$ allows one to look at the cone of a morphism $A^\bullet \to B^\bullet$ between two arbitrary complexes, not necessarily concentrated in degree $0$, which may be completed to a so-called distinguished triangle $A^\bullet \to B^\bullet \to C^\bullet \to A^\bullet [1]$ (see e.g.\ \cite{thomas} for more details).

\subsubsection{Derived categories of quiver representations}

Since $\rep{Q}$ is hereditary, any complex in the derived category $D (\rep{Q})$ is quasi-isomorphic to a complex with all differentials $d^i$ being the zero map. The indecomposable objects of $D (\rep{Q})$ are thus of the form $V [k]$ where $V$ is an indecomposable representation in $\rep{Q}$ viewed as a complex concentrated in degree $-k$.

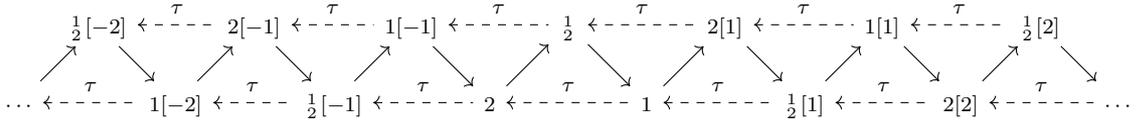
\begin{figure}
\begin{center}
\begin{tikzpicture}[baseline=-4pt] 
\matrix (m) [matrix of math nodes, row sep={2.5em,between origins}, text
height=1.4ex,  font=\scriptsize,
column sep={2.5em,between origins}, text depth=.15ex, 
ampersand replacement=\&]  
{ 
\& \begin{smallmatrix} 1 \\ 2 \end{smallmatrix} [-2] \&\& 2 [-1] \&\& 1 [-1] \&\& \begin{smallmatrix} 1 \\ 2 \end{smallmatrix} \&\& 2 [1] \&\& 1 [1] \&\& \begin{smallmatrix} 1 \\ 2 \end{smallmatrix} [2] \& \\
\cdots \&\& 1 [-2] \&\& \begin{smallmatrix} 1 \\ 2 \end{smallmatrix} [-1] \&\& 2 \&\& 1 \&\& \begin{smallmatrix} 1 \\ 2 \end{smallmatrix} [1] \&\& 2 [2] \&\& \cdots \\
};
\draw[->] (m-2-1)--(m-1-2);
\draw[->] (m-1-2)--(m-2-3);
\draw[->] (m-2-3)--(m-1-4);
\draw[->] (m-1-4)--(m-2-5);
\draw[->] (m-2-5)--(m-1-6);
\draw[->] (m-1-6)--(m-2-7);
\draw[->] (m-2-7)--(m-1-8);
\draw[->] (m-1-8)--(m-2-9);
\draw[->] (m-2-9)--(m-1-10);
\draw[->] (m-1-10)--(m-2-11);
\draw[->] (m-2-11)--(m-1-12);
\draw[->] (m-1-12)--(m-2-13);
\draw[->] (m-2-13)--(m-1-14);
\draw[->] (m-1-14)--(m-2-15);
\draw[<-,dashed] (m-1-2)--(m-1-4) node[above,midway,font=\scriptsize] {$\tau$};
\draw[<-,dashed] (m-1-4)--(m-1-6) node[above,midway,font=\scriptsize] {$\tau$};
\draw[<-,dashed] (m-1-6)--(m-1-8) node[above,midway,font=\scriptsize] {$\tau$};
\draw[<-,dashed] (m-1-8)--(m-1-10) node[above,midway,font=\scriptsize] {$\tau$};
\draw[<-,dashed] (m-1-10)--(m-1-12) node[above,midway,font=\scriptsize] {$\tau$};
\draw[<-,dashed] (m-1-12)--(m-1-14) node[above,midway,font=\scriptsize] {$\tau$};
\draw[<-,dashed] (m-2-1)--(m-2-3) node[above,midway,font=\scriptsize] {$\tau$};
\draw[<-,dashed] (m-2-3)--(m-2-5) node[above,midway,font=\scriptsize] {$\tau$};
\draw[<-,dashed] (m-2-5)--(m-2-7) node[above,midway,font=\scriptsize] {$\tau$};
\draw[<-,dashed] (m-2-7)--(m-2-9) node[above,midway,font=\scriptsize] {$\tau$};
\draw[<-,dashed] (m-2-9)--(m-2-11) node[above,midway,font=\scriptsize] {$\tau$};
\draw[<-,dashed] (m-2-11)--(m-2-13) node[above,midway,font=\scriptsize] {$\tau$};
\draw[<-,dashed] (m-2-13)--(m-2-15) node[above,midway,font=\scriptsize] {$\tau$};
\end{tikzpicture}
\caption{The Auslander--Reiten quiver of the derived category of the $A_2$ quiver}
\label{ar2}
\end{center}
\end{figure}

As in the case of the Abelian category $\rep{Q}$, the derived category is completely characterized by its Auslander--Reiten quiver. The derived category has another natural automorphism $\tau \colon D (\rep{Q}) \to D (\rep{Q})$, called the {\it Auslander--Reiten translation} which is characterized by 
$\Hom{V, \tau (V) [1]} \neq 0$, leading to  
the existence of a triangle 
$\tau (V) \to W \to V \to \tau (V) [1]$. Figs.~\ref{ar2} and \ref{ar3} 
exhibit the Auslander--Reiten quivers of the derived categories of the 
$A_2$ and $A_3$ quivers, respectively. Let us point out that 
the Auslander--Reiten quiver of the Abelian categories $\rep{A_2}$ 
and $\rep{A_3}$ \eqref{mesh:an} appears in the middle of the diagram, 
as complexes concentrated in degree $0$.
\begin{figure}
\begin{center}
\begin{tikzpicture}[baseline=-4pt] 
\matrix (m) [matrix of math nodes, row sep={2.5em,between origins}, text
height=1.4ex, font=\scriptsize, 
column sep={2.5em,between origins}, text depth=.15ex, 
ampersand replacement=\&]  
{ 
\& 3 [-1] \&\& 2 [-1] \&\& 1 [-1] \&\& \begin{smallmatrix} 1 \\ 2 \\ 3 \end{smallmatrix} \&\& 3 [1] \&\& 2 [1] \&\& 1 [1] \& \\
\cdots \&\& \begin{smallmatrix} 2 \\ 3 \end{smallmatrix} [-1] \&\& \begin{smallmatrix} 1 \\ 2 \end{smallmatrix} [-1] \&\& \begin{smallmatrix} 2 \\ 3 \end{smallmatrix} \&\& \begin{smallmatrix} 1 \\ 2 \end{smallmatrix} \&\& \begin{smallmatrix} 2 \\ 3 \end{smallmatrix} [1] \&\& \begin{smallmatrix} 1 \\ 2 \end{smallmatrix} [1] \&\& \cdots \\
\& 1 [-2] \&\& \begin{smallmatrix} 1 \\ 2 \\ 3 \end{smallmatrix} [-1] \&\& 3 \&\& 2 \&\& 1 \&\& \begin{smallmatrix} 1 \\ 2 \\ 3 \end{smallmatrix} [1] \&\& 3 [2] \& \\
};
\draw[->] (m-2-1)--(m-1-2);
\draw[->] (m-1-2)--(m-2-3);
\draw[->] (m-2-3)--(m-1-4);
\draw[->] (m-1-4)--(m-2-5);
\draw[->] (m-2-5)--(m-1-6);
\draw[->] (m-1-6)--(m-2-7);
\draw[->] (m-2-7)--(m-1-8);
\draw[->] (m-1-8)--(m-2-9);
\draw[->] (m-2-9)--(m-1-10);
\draw[->] (m-1-10)--(m-2-11);
\draw[->] (m-2-11)--(m-1-12);
\draw[->] (m-1-12)--(m-2-13);
\draw[->] (m-2-13)--(m-1-14);
\draw[->] (m-1-14)--(m-2-15);
\draw[->] (m-2-1)--(m-3-2);
\draw[->] (m-3-2)--(m-2-3);
\draw[->] (m-2-3)--(m-3-4);
\draw[->] (m-3-4)--(m-2-5);
\draw[->] (m-2-5)--(m-3-6);
\draw[->] (m-3-6)--(m-2-7);
\draw[->] (m-2-7)--(m-3-8);
\draw[->] (m-3-8)--(m-2-9);
\draw[->] (m-2-9)--(m-3-10);
\draw[->] (m-3-10)--(m-2-11);
\draw[->] (m-2-11)--(m-3-12);
\draw[->] (m-3-12)--(m-2-13);
\draw[->] (m-2-13)--(m-3-14);
\draw[->] (m-3-14)--(m-2-15);
\draw[<-,dashed] (m-1-2)--(m-1-4) node[above,midway,font=\scriptsize] {$\tau$};
\draw[<-,dashed] (m-1-4)--(m-1-6) node[above,midway,font=\scriptsize] {$\tau$};
\draw[<-,dashed] (m-1-6)--(m-1-8) node[above,midway,font=\scriptsize] {$\tau$};
\draw[<-,dashed] (m-1-8)--(m-1-10) node[above,midway,font=\scriptsize] {$\tau$};
\draw[<-,dashed] (m-1-10)--(m-1-12) node[above,midway,font=\scriptsize] {$\tau$};
\draw[<-,dashed] (m-1-12)--(m-1-14) node[above,midway,font=\scriptsize] {$\tau$};
\draw[<-,dashed] (m-2-1)--(m-2-3) node[above,midway,font=\scriptsize] {$\tau$};
\draw[<-,dashed] (m-2-3)--(m-2-5) node[above,midway,font=\scriptsize] {$\tau$};
\draw[<-,dashed] (m-2-5)--(m-2-7) node[above,midway,font=\scriptsize] {$\tau$};
\draw[<-,dashed] (m-2-7)--(m-2-9) node[above,midway,font=\scriptsize] {$\tau$};
\draw[<-,dashed] (m-2-9)--(m-2-11) node[above,midway,font=\scriptsize] {$\tau$};
\draw[<-,dashed] (m-2-11)--(m-2-13) node[above,midway,font=\scriptsize] {$\tau$};
\draw[<-,dashed] (m-2-13)--(m-2-15) node[above,midway,font=\scriptsize] {$\tau$};
\draw[<-,dashed] (m-3-2)--(m-3-4) node[above,midway,font=\scriptsize] {$\tau$};
\draw[<-,dashed] (m-3-4)--(m-3-6) node[above,midway,font=\scriptsize] {$\tau$};
\draw[<-,dashed] (m-3-6)--(m-3-8) node[above,midway,font=\scriptsize] {$\tau$};
\draw[<-,dashed] (m-3-8)--(m-3-10) node[above,midway,font=\scriptsize] {$\tau$};
\draw[<-,dashed] (m-3-10)--(m-3-12) node[above,midway,font=\scriptsize] {$\tau$};
\draw[<-,dashed] (m-3-12)--(m-3-14) node[above,midway,font=\scriptsize] {$\tau$};
\end{tikzpicture}
\caption{The Auslander--Reiten quiver of the derived category of the $A_3$ quiver}
\label{ar3}
\end{center}
\end{figure}
\subsection{$t$-structures} 
The formalism of $t$-structures is a means to vivisect a triangulated 
category.  A $t$-structure on a triangulated 
category $\mathcal D$ is a pair $(\mathcal D^{\preceq 0}, \mathcal D^{\succeq 0})$ of strictly full subcategories, 
satisfying the following conditions. 
\begin{enumerate} 
\item $\mathcal D^{\preceq 0} [1] \subset \mathcal D^{\preceq 0}$ and $\mathcal D^{\succeq 0} \subset \mathcal D^{\succeq 0} [1]$
\item $\operatorname{Hom}_{\mathcal D} (\mathcal D^{\preceq 0}, \mathcal D^{\succeq 0} [-1]) = 0$
\item For each object $K$ of $\mathcal D$, there exists a 
distinguished triangle $X \to K \to Y [-1] \to X [1]$ with $X \in \mathcal D^{\preceq 0}$ and $Y \in \mathcal D^{\succeq 0}$.
\end{enumerate}

Further, a $t$-structure is called \emph{bounded} if 
each $K$ in $\mathcal D$ is contained in $\mathcal D^{\preceq 0} [m] \cap \mathcal D^{\succeq 0} [n]$ 
for some integers $m$ and $n$ and any bounded $t$-structure is determined 
by its \emph{heart} $\mathcal D^{\preceq 0} \cap \mathcal D^{\succeq 0}$,
which is always an Abelian category. 
All $t$-structures considered in this article are bounded.

Finally, an object $P \in \mathcal D$ is called a \emph{projective} of the heart if for all $M \in \mathcal H$ and all $k
\neq 0$ one has $\operatorname{Hom}_{\mathcal D} (P, M [k]) = 0$. For type
$A$ quivers the dimension of the Hom space between indecomposable objects can
be read off the AR quiver \cite[\S 3.1.4]{schiffler}.

\subsubsection{Standard $t$-structure} 

Let $\mathcal D = D (\A)$ be the derived category of an Abelian category $\A$ with its usual triangulated structure. Let $\mathcal D^{\geq n}$ denote the full subcategory of $\mathcal D$ formed by complexes 
$A^{\bullet}$ with cohomology only beyond $n$, that is, 
$\mathrm H^i(A^{\bullet})=0$ for $i<n$. 
Similarly, let $\mathcal D^{\leq n}$ denote the full subcategory of $D(\A)$ of 
complexes $A^{\bullet}$ with cohomology only below $n$, that is 
$\mathrm H^i(A^{\bullet})=0$ for $i>n$. 

Then $(\mathcal D^{\leq 0}, \mathcal D^{\geq 0})$ is called the \emph{standard $t$-structure} and its heart $\mathcal D^{\leq 0} \cap \mathcal D^{\geq 0} \simeq \A$ recovers the original Abelian category viewed as complexes concentrated in degree $0$.

\subsubsection{Intermediate $t$-structures} 
\label{int}

Let $(\mathcal D^{\leq 0}, \mathcal D^{\geq 0})$ denote the standard $t$-structure on $\mathcal D = D (\A)$ with heart $\mathcal D^{\leq 0} \cap \mathcal D^{\geq 0} \simeq \A$. For any positive integer $m$, an \emph{$m$-intermediate} $t$-structure is any $t$-structure $(\mathcal D^{\preceq 0}, \mathcal D^{\succeq 0})$ satisfying $\mathcal D^{\leq 0} \subset \mathcal D^{\preceq 0} \subset \mathcal D^{\leq 0} [m]$ \cite{by}, and a $1$-intermediate $t$-structure is usually simply called \emph{intermediate}. The heart of this new $t$-structure is again an Abelian category which may or may not be equivalent to $\A$.

\begin{figure}
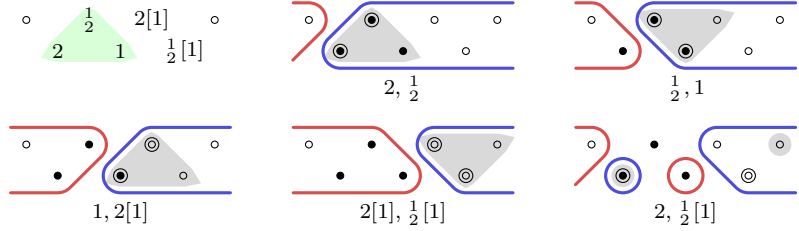
 
\begin{center} 
 
\end{center} 
\caption{The five intermediate $t$-structures of $\dq{A_2}$, their hearts (shaded) and projective 
objects (circled vertices). The standard heart is marked in Green.}
\label{hearts1} 
\end{figure}

Figs.~\ref{hearts1} and \ref{hearts2} give an illustration of the $1$- and $2$-intermediate $t$-structures on $\dq{A_2} = \db{A_2}$. Each diagram is a picture of the Auslander--Reiten quiver of the derived category $\dq{A_2}$ described in Fig.~\ref{ar2}. The vertices correspond to indecomposable objects, with the filled vertices corresponding to the complexes concentrated in degree $0$, i.e.\ the indecomposable objects in the heart of the standard $t$-structure. Each diagram illustrates some $t$-structure $(\dq{A_2}^{\preceq 0}, \dq{A_2}^{\succeq 0})$, the blue part corresponds to $\dq{A_2}^{\succeq 0}$ and the red part to $\dq{A_2}^{\preceq 0} [-1]$ of the $t$-structure. The 
vertices in the shaded part correspond to the indecomposable objects in the heart of the $t$-structure, which is given by $\dq{A_2}^{\preceq 0} \cap 
\dq{A_2}^{\succeq 0}$. Diagrammatically, the heart is obtained as the intersection of the blue part and the shift (given by glide reflection to the right) of the red part.

As we show in this article, each $m$-intermediate $t$-structure  corresponds 
to a term in the scattering amplitude in a $\phi^{m+2}$ theory.

\begin{figure} 
\begin{center} 
%
 
\end{center} 
\caption{The twelve 2-intermediate $t$-structures of $\dq{A_2}$, their hearts 
(shaded), projective objects (circled vertices). The standard heart is marked 
in Green.} 
\label{hearts2} 
\end{figure} 
\begin{figure}[h]
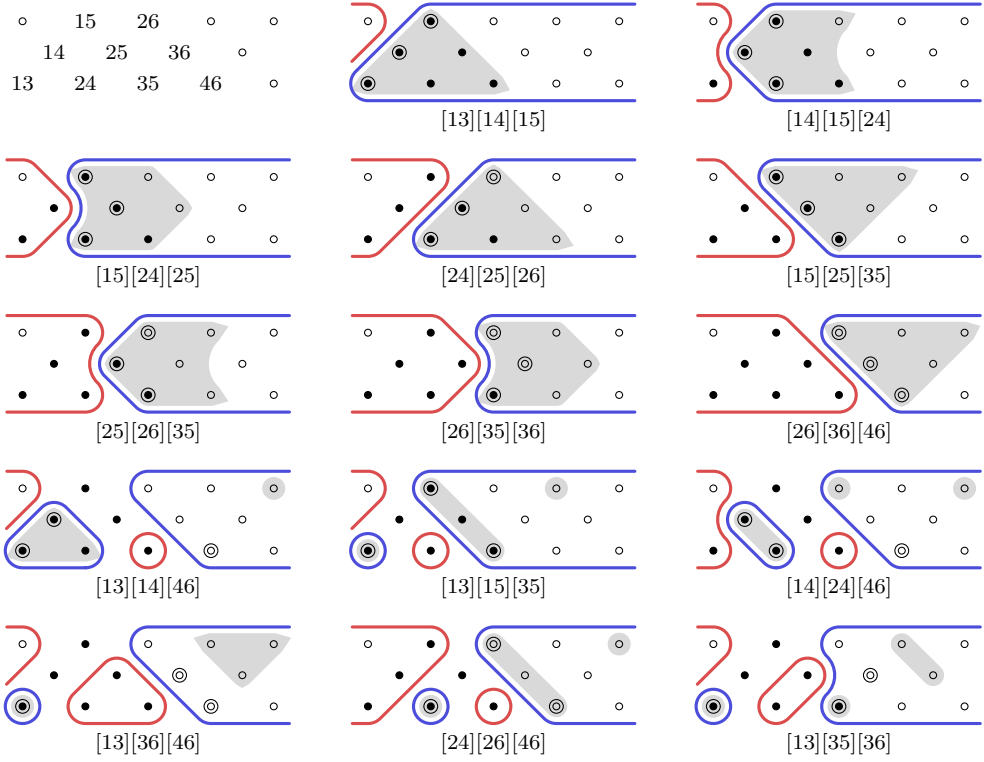

\begin{center}
%
\end{center}
\caption{The 14 intermediate $t$-structures of $\dq{A_3}$ obtained by tilting from the standard heart (filled vertices), their hearts (shaded), projective objects (circled vertices) and their contribution to the scattering amplitude}
\label{hearts3}
\end{figure}
 
\begin{figure} 
\begin{center} 
%
 
\end{center} 
\caption{The 22 3-intermediate $t$-structures of $\dq{A_2}$, their hearts 
(shaded), projective objects (circled vertices). The standard heart is marked 
in green.} 
\label{hearts23} 
\end{figure} 
\subsection{Orbit categories} 
\label{dercat:orb}
Given an additive category $\mathcal D$ --- in particular the derived category 
of quiver representations ---
together with an automorphism 
$g \colon \mathcal D \to \mathcal D$, an \emph{orbit category} 
$\mathcal D / g$ (also denoted $\mathcal D/g^{\Z}$) is defined by taking the same objects as 
$\mathcal D$ and taking morphisms from $X$ to $Y$ to be in bijection with
\[
\bigoplus_{n \in \Z} \operatorname{Hom}_{\mathcal D} (X, g^n (Y)),
\]
where $\Z$ denotes the set of integers.
Two important examples of such automorphisms are
\begin{enumerate}
\item the $N$-th power $[1]^N = [N] \colon \mathcal D \to \mathcal D$ of the shift functor
\item the $m$-cluster automorphism $F^m = \tau^{-1} \circ [m] \colon \mathcal D \to \mathcal D$, where $\tau$ is the Auslander--Reiten translation.
\end{enumerate}
The corresponding orbit categories are
\begin{enumerate}
\item the \emph{$N$-periodic derived category} $\mathcal P^N = \mathcal D / [N]$
\item the \emph{$m$-cluster category} $\mathcal C^m = \mathcal D / (\tau^{-1} \circ [m])$
\end{enumerate}
which both naturally inherit a triangulated structure from $\mathcal D$ \cite{keller5,saito}. 

Thus, in an $m$-cluster category
the AR translation and the shift functors are related by
\begin{equation}
\label{tm1}
\tau^{-1}\circ [m] \simeq \id.
\end{equation}
The AR quiver of these categories appear as the mesh diagrams for
potentials of the form $V(\phi)=\phi^{m+2}$. 
In an $N$-periodic category, on the other hand,
\begin{equation}
\label{N1}
[1]^N \simeq \id.
\end{equation}
The quotient that appears for a generic polynomial is a variant of
the periodic category. In addition to the identification \eq{N1}, the
corresponding mesh diagram entails identification of intervening
entries. 
We refer to it suggestively as a \emph{pseudo-periodic category}, 
although we have not studied its categorical structure here. 
As the scattering amplitude for polynomial potentials can be categorified via
certain partial cluster tilting objects in a cluster category, the unknown
structure of the pseudo-periodic category, however, 
does not undermine the rigour of the categorification we show.

Let us point out some features of this, arising from the combinatorics of
planar diagrams.   
First, we have, for the AR quiver of an $A_{n-1}$ quiver, a relation between
the shift functor and the AR translation,
\begin{equation}
\label{arp}
[1]^2 \simeq \tau^{-n},
\end{equation}
by the very definition of the functors. 
Using \eq{Nmn} and \eq{ni} arising in the combinatorics of Feynman diagrams
in \eq{N1} and \eq{arp}  we obtain
\begin{equation}
\label{tmn}
\tau^{-\sum_in_i}\big[{\textstyle\sum_i m_in_i}\big] \simeq \id. 
\end{equation}
Since the AR translation $\tau$ and the shift $[1]$ commute, 
this can be written as 
\begin{equation}
\label{Feq}
\prod_i(F^{m_i} )^{n_i}=1,
\end{equation}
where the product  means successive composition of the
functors $(F^{m_i})^{n_i}$. However, the pseudo-periodic category has further identifications besides the periodic structure, which are not realized by the identifications given by the cyclic group generated by a single automorphism.
In the case of a single term in the potential, 
$V(\phi)=\phi^{m+2}$, \eq{Feq} reduces to 
\begin{equation}
\big(F^m\big)^n \simeq \id,
\end{equation}
which is solved by \eq{tm1}.

Let us now list some examples of cluster categories of $A_{n-1}$ quivers 
to be used to derive the scattering amplitudes.
We denote the $m$-cluster category of an $A_n$ quiver by $\C^m_{A_n} = \D_{A_n} / (\tau^{-1} \circ [m])$. 
For $m=1$, it is called a cluster category and the superscript is suppressed. 
The $N$-periodic category of an $A_n$ quiver is denoted $\P^N_{A_n} = \D_{A_n} / [N]$.
\subsubsection{Cluster categories} 
The cluster category $\C_{A_2}$ is obtained by setting 
$F^1 = \tau^{-1} \circ [1] \simeq \id$, namely,
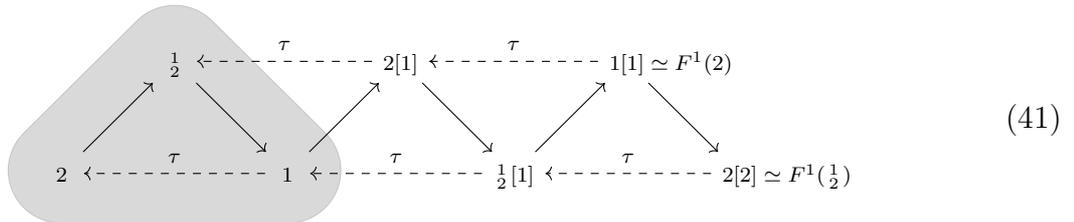
\begin{equation}
\label{A2:C1} 
\begin{tikzpicture}[baseline=-4pt] 
\matrix (m) [matrix of math nodes, minimum width=1.4em, row 
sep={3.6em,between origins}, text height=1.4ex, font=\scriptsize,
column sep={3.6em,between 
origins}, text depth=.15ex, ampersand replacement=\&] 
{ 
\& \nodetwo{1}{2} \&\& \nodeone{2}[1] \&\& \nodeone{1}[1] 
\mathrlap{{}\simeq F^1 (\nodeone{2})} \&\&  \\ 
2 \&\& 1 \&\& \nodetwo{1}{2}[1] \&\& \nodeone{2}[2] \mathrlap{{} \simeq F^1
(\nodetwo{1}{2})} \& \phantom{\hspace{2.6em}} \\ 
}; 
\foreach \i/\j in {1/2,3/4,5/6} 
\draw[->] (m-2-\i)--(m-1-\j); 
%
\foreach \i/\j in {2/3,4/5,6/7} 
\draw[->] (m-1-\i)--(m-2-\j); 
%
\draw[->,dashed] (m-1-6) -- (m-1-4) node[above,midway,font=\scriptsize] 
{$\tau$}; 
\draw[->,dashed] (m-1-4) -- (m-1-2) node[above,midway,font=\scriptsize] 
{$\tau$}; 
\draw[->,dashed] (m-2-7) -- (m-2-5) node[above,midway,font=\scriptsize] 
{$\tau$}; 
\draw[->,dashed] (m-2-5) -- (m-2-3) node[above,midway,font=\scriptsize] 
{$\tau$}; 
\draw[->,dashed] (m-2-3) -- (m-2-1) node[above,midway,font=\scriptsize] 
{$\tau$}; 
\draw[opacity=.15,fill,line width=0] ([yshift=-1.6em]m-2-1.center) -- 
++(7.2em,0) arc[start angle=-90, end angle=45, radius=1.7em] -- 
++(-3.6em,3.6em) arc[start angle=45, end angle=135, radius=1.7em] -- 
++(-3.6em,-3.6em) arc[start angle=135, end angle=270, radius=1.7em] -- cycle; 
\end{tikzpicture} 
\end{equation}
where $1[1]$ is identified with $2$ and $2[2]$ with $\nodetwo{1}{2}$. 
This diagram can be written geometrically by associating to each 
indecomposable object a diagonal in a triangulation of a pentagon, i.e.
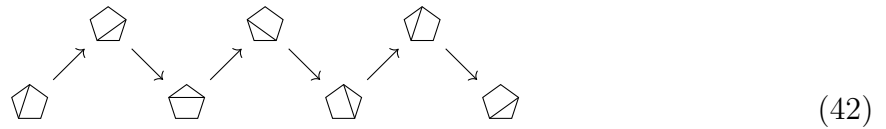
\begin{equation}
\label{A2:C1g}
\begin{tikzpicture}[x=2.5em,y=2.5em,line width=.4pt]
\begin{scope}
\draw (18:.6em) -- (90:.6em) -- (162:.6em) -- (234:.6em) -- (306:.6em) -- cycle;
\draw (234:.6em) -- (90:.6em);
\end{scope}
\begin{scope}[shift={(1,1)}]
\draw (18:.6em) -- (90:.6em) -- (162:.6em) -- (234:.6em) -- (306:.6em) -- cycle;
\draw (234:.6em) -- (18:.6em);
\end{scope}
\begin{scope}[shift={(2,0)}]
\draw (18:.6em) -- (90:.6em) -- (162:.6em) -- (234:.6em) -- (306:.6em) -- cycle;
\draw (18:.6em) -- (162:.6em);
\end{scope}
\begin{scope}[shift={(3,1)}]
\draw (18:.6em) -- (90:.6em) -- (162:.6em) -- (234:.6em) -- (306:.6em) -- cycle;
\draw (162:.6em) -- (306:.6em);
\end{scope}
\begin{scope}[shift={(4,0)}]
\draw (18:.6em) -- (90:.6em) -- (162:.6em) -- (234:.6em) -- (306:.6em) -- cycle;
\draw (306:.6em) -- (90:.6em);
\end{scope}
\begin{scope}[shift={(5,1)}]
\draw (18:.6em) -- (90:.6em) -- (162:.6em) -- (234:.6em) -- (306:.6em) -- cycle;
\draw (234:.6em) -- (90:.6em);
\end{scope}
\begin{scope}[shift={(6,0)}]
\draw (18:.6em) -- (90:.6em) -- (162:.6em) -- (234:.6em) -- (306:.6em) -- cycle;
\draw (234:.6em) -- (18:.6em);
\end{scope}
\draw[->] (.3,.3) -- (.7,.7);
\draw[->] (1.3,.7) -- (1.7,.3);
\draw[->] (2.3,.3) -- (2.7,.7);
\draw[->] (3.3,.7) -- (3.7,.3);
\draw[->] (4.3,.3) -- (4.7,.7);
\draw[->] (5.3,.7) -- (5.7,.3);
\end{tikzpicture}
\end{equation}
Labelling each vertex with a positive natural number and each diagonal by the two vertices 
it connects, e.g.
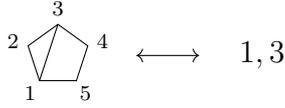
\begin{equation*}
\begin{tikzpicture}[x=1em,y=1em,line width=.4pt]
\begin{scope}[shift={(-3.5,0)}]
\draw (18:1em) -- (90:1em) -- (162:1em) -- (234:1em) -- (306:1em) -- cycle;
\draw (234:1em) -- (90:1em);
\node[font=\scriptsize] at (234:1.5em) {$1$};
\node[font=\scriptsize] at (162:1.5em) {$2$};
\node[font=\scriptsize] at (90:1.5em)  {$3$};
\node[font=\scriptsize] at (18:1.5em)  {$4$};
\node[font=\scriptsize] at (306:1.5em) {$5$};
\end{scope}
\draw[<->] (-1,0) -- (1,0);
\node at (3,0) {$1,3$};
\end{tikzpicture}
\end{equation*}
the mesh diagram \eqref{dbA2} is recovered.

The cluster category $\C_{A_3}$ is similarly given by 
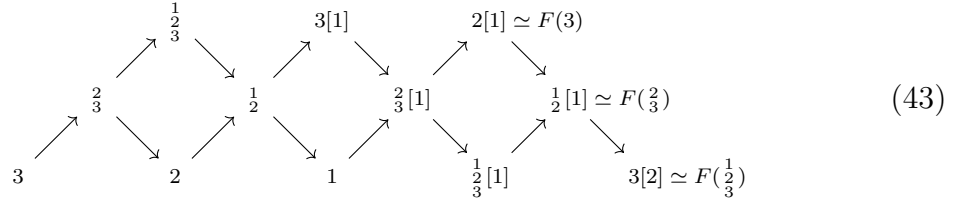
\begin{equation} 
\label{A3:C1} 
\begin{tikzpicture}[baseline=-4pt] 
\matrix (m) [matrix of math nodes, row sep={2.5em,between origins}, text 
height=1.4ex, font=\scriptsize,
text centered, anchor=center,column sep={2.5em,between origins}, 
text depth=0.25ex, ampersand replacement=\&, nodes in empty cells] 
{ 
\&\& \nodethree{1}{2}{3} \&\& 
\nodeone{3}[1] \&\& \nodeone{2}[1]\mathrlap{{} \simeq F(\nodeone{3})} \&\&  \\ 
\& \nodetwo{2}{3} \&\& \nodetwo{1}{2} \&\& 
\nodetwo{2}{3}[1] \&\& \nodetwo{1}{2}[1]\mathrlap{{} \simeq F(\nodetwo{2}{3})} \& \\ 
\nodeone{3} \&\& \nodeone{2} \&\& \nodeone{1} 
\&\& \nodethree{1}{2}{3}[1] \&\& 
\nodeone{3}[2]\mathrlap{{} \simeq F(\nodethree{1}{2}{3})} \\ 
}; 
\foreach \i/\j in {1/2,3/4,5/6,7/8} 
\draw[thin,->] (m-3-\i)--(m-2-\j); 
\foreach \i/\j in {2/3,4/5,6/7} 
\draw[thin,->] (m-2-\i)--(m-1-\j); 
\foreach \i/\j in {3/4,5/6,7/8} 
\draw[thin,->] (m-1-\i)--(m-2-\j); 
\foreach \i/\j in {2/3,4/5,6/7,8/9} 
\draw[thin,->] (m-2-\i)--(m-3-\j); 
\end{tikzpicture} 
\end{equation} 
and can likewise be written geometrically as
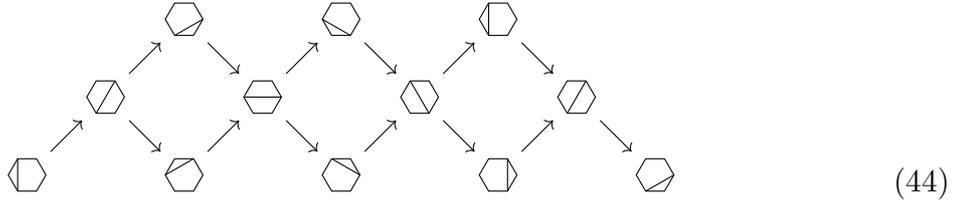
\begin{equation}
\label{A3:C1g}
\begin{tikzpicture}[x=2.5em,y=2.5em,line width=.4pt]
\begin{scope}
\draw (0:.6em) -- (60:.6em) -- (120:.6em) -- (180:.6em) -- (240:.6em) -- (300:.6em) -- cycle;
\draw (240:.6em) -- (120:.6em);
\end{scope}
\begin{scope}[shift={(1,1)}]
\draw (0:.6em) -- (60:.6em) -- (120:.6em) -- (180:.6em) -- (240:.6em) -- (300:.6em) -- cycle;
\draw (240:.6em) -- (60:.6em);
\end{scope}
\begin{scope}[shift={(2,2)}]
\draw (0:.6em) -- (60:.6em) -- (120:.6em) -- (180:.6em) -- (240:.6em) -- (300:.6em) -- cycle;
\draw (240:.6em) -- (0:.6em);
\end{scope}
\begin{scope}[shift={(2,0)}]
\draw (0:.6em) -- (60:.6em) -- (120:.6em) -- (180:.6em) -- (240:.6em) -- (300:.6em) -- cycle;
\draw (180:.6em) -- (60:.6em);
\end{scope}
\begin{scope}[shift={(3,1)}]
\draw (0:.6em) -- (60:.6em) -- (120:.6em) -- (180:.6em) -- (240:.6em) -- (300:.6em) -- cycle;
\draw (180:.6em) -- (0:.6em);
\end{scope}
\begin{scope}[shift={(4,2)}]
\draw (0:.6em) -- (60:.6em) -- (120:.6em) -- (180:.6em) -- (240:.6em) -- (300:.6em) -- cycle;
\draw (180:.6em) -- (300:.6em);
\end{scope}
\begin{scope}[shift={(4,0)}]
\draw (0:.6em) -- (60:.6em) -- (120:.6em) -- (180:.6em) -- (240:.6em) -- (300:.6em) -- cycle;
\draw (0:.6em) -- (120:.6em);
\end{scope}
\begin{scope}[shift={(5,1)}]
\draw (0:.6em) -- (60:.6em) -- (120:.6em) -- (180:.6em) -- (240:.6em) -- (300:.6em) -- cycle;
\draw (120:.6em) -- (300:.6em);
\end{scope}
\begin{scope}[shift={(6,2)}]
\draw (0:.6em) -- (60:.6em) -- (120:.6em) -- (180:.6em) -- (240:.6em) -- (300:.6em) -- cycle;
\draw (120:.6em) -- (240:.6em);
\end{scope}
\begin{scope}[shift={(6,0)}]
\draw (0:.6em) -- (60:.6em) -- (120:.6em) -- (180:.6em) -- (240:.6em) -- (300:.6em) -- cycle;
\draw (60:.6em) -- (300:.6em);
\end{scope}
\begin{scope}[shift={(7,1)}]
\draw (0:.6em) -- (60:.6em) -- (120:.6em) -- (180:.6em) -- (240:.6em) -- (300:.6em) -- cycle;
\draw (60:.6em) -- (240:.6em);
\end{scope}
\begin{scope}[shift={(8,0)}]
\draw (0:.6em) -- (60:.6em) -- (120:.6em) -- (180:.6em) -- (240:.6em) -- (300:.6em) -- cycle;
\draw (0:.6em) -- (240:.6em);
\end{scope}
\draw[->] (.3,.3) -- (.7,.7);
\draw[->] (1.3,.7) -- (1.7,.3);
\draw[->] (2.3,.3) -- (2.7,.7);
\draw[->] (3.3,.7) -- (3.7,.3);
\draw[->] (4.3,.3) -- (4.7,.7);
\draw[->] (5.3,.7) -- (5.7,.3);
\draw[->] (6.3,.3) -- (6.7,.7);
\draw[->] (7.3,.7) -- (7.7,.3);
\draw[->] (1.3,1.3) -- (1.7,1.7);
\draw[->] (2.3,1.7) -- (2.7,1.3);
\draw[->] (3.3,1.3) -- (3.7,1.7);
\draw[->] (4.3,1.7) -- (4.7,1.3);
\draw[->] (5.3,1.3) -- (5.7,1.7);
\draw[->] (6.3,1.7) -- (6.7,1.3);
\end{tikzpicture}
\end{equation}
leading to the mesh diagram \eq{dbA3}.
\subsubsection{Higher cluster categories}
The 2-cluster category $\C^2_{A_2}$ is obtained by
identifying $F^2 = \tau^{-1} \circ [2] \simeq \id$, leading to
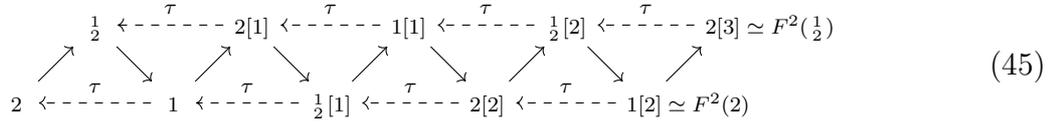
\begin{equation}
\label{A2:C2} 
\begin{tikzpicture}[baseline=-4pt] 
\matrix (m) [matrix of math nodes, minimum width=1.4em, row 
sep={2.5em,between origins}, text height=1.4ex, font=\scriptsize,
column sep={2.5em,between 
origins}, text depth=.15ex, ampersand replacement=\&] 
{ 
\& \nodetwo{1}{2} \&\& \nodeone{2}[1] \&\& \nodeone{1}[1] 
\&\&  \nodeone{\nodetwo{1}{2}[2]}\&\& 
\nodeone{2}[3]\mathrlap{{} \simeq F^2(\nodetwo{1}{2})} \&\&\\ 
2 \&\& 1 \&\& \nodetwo{1}{2}[1] \&\& \nodeone{2}[2]  
\&\& \nodeone{1}[2] 
\mathrlap{{} \simeq F^{2}(\nodeone{2})} \&\&\\ 
}; 
\foreach \i/\j in {1/2,3/4,5/6,7/8,9/10} 
\draw[->] (m-2-\i)--(m-1-\j); 
%
\foreach \i/\j in {2/3,4/5,6/7,8/9} 
\draw[->] (m-1-\i)--(m-2-\j); 
%
\draw[->,dashed] (m-1-10) -- (m-1-8) node[above,midway,font=\scriptsize] 
{$\tau$}; 
\draw[->,dashed] (m-1-8) -- (m-1-6) node[above,midway,font=\scriptsize] 
{$\tau$}; 
\draw[->,dashed] (m-1-6) -- (m-1-4) node[above,midway,font=\scriptsize] 
{$\tau$}; 
\draw[->,dashed] (m-1-4) -- (m-1-2) node[above,midway,font=\scriptsize] 
{$\tau$}; 
\draw[->,dashed] (m-2-9) -- (m-2-7) node[above,midway,font=\scriptsize] 
{$\tau$}; 
\draw[->,dashed] (m-2-7) -- (m-2-5) node[above,midway,font=\scriptsize] 
{$\tau$}; 
\draw[->,dashed] (m-2-5) -- (m-2-3) node[above,midway,font=\scriptsize] 
{$\tau$}; 
\draw[->,dashed] (m-2-3) -- (m-2-1) node[above,midway,font=\scriptsize] 
{$\tau$}; 
\end{tikzpicture} 
\end{equation}  
The 3-cluster category $\C^3_{A_2}$ with $F^3=\tau^{-1} \circ [3] \simeq \id$ is given by  
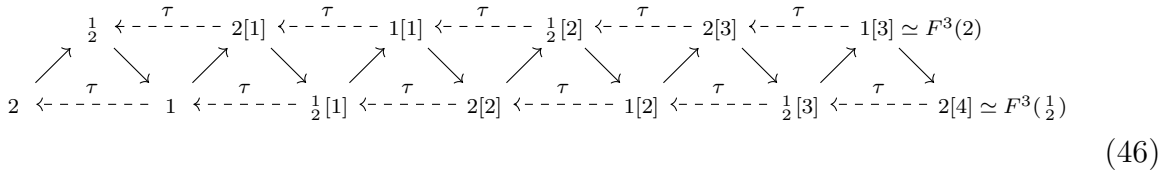
\begin{equation}
\label{A2:C3} 
\begin{tikzpicture}[baseline=-4pt] 
\matrix (m) [matrix of math nodes, minimum width=1.4em, row 
sep={2.5em,between origins}, text height=1.4ex, font=\scriptsize,
column sep={2.5em,between 
origins}, text depth=.15ex, ampersand replacement=\&] 
{ 
\& \nodetwo{1}{2} \&\& \nodeone{2}[1] \&\& \nodeone{1}[1] 
\&\&  \nodeone{\nodetwo{1}{2}[2]}\&\& 
\nodeone{2}[3]\&\&\nodeone{1}[3]\mathrlap{{} \simeq F^3(\nodeone{2})} \&\&\\ 
2 \&\& 1 \&\& \nodetwo{1}{2}[1] \&\& \nodeone{2}[2]  
\&\& \nodeone{1}[2]\&\& 
\nodetwo{1}{2}[3]\&\&\nodeone{2}[4]\mathrlap{{} \simeq F^{3}(\nodetwo{1}{2})} \&\&\\ 
}; 
\foreach \i/\j in {1/2,3/4,5/6,7/8,9/10,11/12} 
\draw[->] (m-2-\i)--(m-1-\j); 
%
\foreach \i/\j in {2/3,4/5,6/7,8/9,10/11,12/13} 
\draw[->] (m-1-\i)--(m-2-\j); 
%
\draw[->,dashed] (m-1-12) -- (m-1-10) node[above,midway,font=\scriptsize] 
{$\tau$}; 
\draw[->,dashed] (m-1-10) -- (m-1-8) node[above,midway,font=\scriptsize] 
{$\tau$}; 
\draw[->,dashed] (m-1-8) -- (m-1-6) node[above,midway,font=\scriptsize] 
{$\tau$}; 
\draw[->,dashed] (m-1-6) -- (m-1-4) node[above,midway,font=\scriptsize] 
{$\tau$}; 
\draw[->,dashed] (m-1-4) -- (m-1-2) node[above,midway,font=\scriptsize] 
{$\tau$}; 
\draw[->,dashed] (m-2-13) -- (m-2-11) node[above,midway,font=\scriptsize] 
{$\tau$}; 
\draw[->,dashed] (m-2-11) -- (m-2-9) node[above,midway,font=\scriptsize] 
{$\tau$}; 
\draw[->,dashed] (m-2-9) -- (m-2-7) node[above,midway,font=\scriptsize] 
{$\tau$}; 
\draw[->,dashed] (m-2-7) -- (m-2-5) node[above,midway,font=\scriptsize] 
{$\tau$}; 
\draw[->,dashed] (m-2-5) -- (m-2-3) node[above,midway,font=\scriptsize] 
{$\tau$}; 
\draw[->,dashed] (m-2-3) -- (m-2-1) node[above,midway,font=\scriptsize] 
{$\tau$}; 
\end{tikzpicture} 
\end{equation}
The geometric analogue of the higher cluster categories are diagonals of an 
$N$-gon that form part of an $(m+2)$-angulation, where $N = nm + 2$. 
Here $n$ corresponds to the number of $(m+2)$-gons in 
an $(m+2)$-angulation \cite{baur}. For example, considering a quadrangulation 
($m = 2$) of an octagon ($N = 8$), we can write \eqref{A2:C2} as
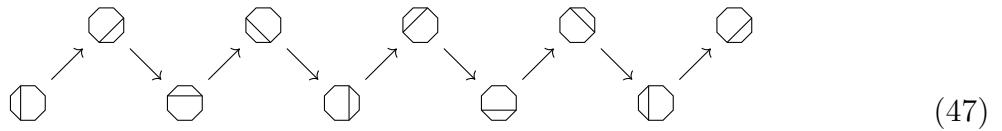
\begin{equation}
\label{A2:C2g}
\begin{tikzpicture}[x=2.5em,y=2.5em,line width=.4pt]
\begin{scope}
\draw (22.5:.6em) -- (67.5:.6em) -- (112.5:.6em) -- (157.5:.6em) -- (202.5:.6em) -- (247.5:.6em) -- (292.5:.6em) -- (337.5:.6em) -- cycle;
\draw (247.5:.6em) -- (112.5:.6em);
\end{scope}
\begin{scope}[shift={(1,1)}]
\draw (22.5:.6em) -- (67.5:.6em) -- (112.5:.6em) -- (157.5:.6em) -- (202.5:.6em) -- (247.5:.6em) -- (292.5:.6em) -- (337.5:.6em) -- cycle;
\draw (247.5:.6em) -- (22.5:.6em);
\end{scope}
\begin{scope}[shift={(2,0)}]
\draw (22.5:.6em) -- (67.5:.6em) -- (112.5:.6em) -- (157.5:.6em) -- (202.5:.6em) -- (247.5:.6em) -- (292.5:.6em) -- (337.5:.6em) -- cycle;
\draw (22.5:.6em) -- (157.5:.6em);
\end{scope}
\begin{scope}[shift={(3,1)}]
\draw (22.5:.6em) -- (67.5:.6em) -- (112.5:.6em) -- (157.5:.6em) -- (202.5:.6em) -- (247.5:.6em) -- (292.5:.6em) -- (337.5:.6em) -- cycle;
\draw (157.5:.6em) -- (292.5:.6em);
\end{scope}
\begin{scope}[shift={(4,0)}]
\draw (22.5:.6em) -- (67.5:.6em) -- (112.5:.6em) -- (157.5:.6em) -- (202.5:.6em) -- (247.5:.6em) -- (292.5:.6em) -- (337.5:.6em) -- cycle;
\draw (292.5:.6em) -- (67.5:.6em);
\end{scope}
\begin{scope}[shift={(5,1)}]
\draw (22.5:.6em) -- (67.5:.6em) -- (112.5:.6em) -- (157.5:.6em) -- (202.5:.6em) -- (247.5:.6em) -- (292.5:.6em) -- (337.5:.6em) -- cycle;
\draw (67.5:.6em) -- (202.5:.6em);
\end{scope}
\begin{scope}[shift={(6,0)}]
\draw (22.5:.6em) -- (67.5:.6em) -- (112.5:.6em) -- (157.5:.6em) -- (202.5:.6em) -- (247.5:.6em) -- (292.5:.6em) -- (337.5:.6em) -- cycle;
\draw (202.5:.6em) -- (337.5:.6em);
\end{scope}
\begin{scope}[shift={(7,1)}]
\draw (22.5:.6em) -- (67.5:.6em) -- (112.5:.6em) -- (157.5:.6em) -- (202.5:.6em) -- (247.5:.6em) -- (292.5:.6em) -- (337.5:.6em) -- cycle;
\draw (337.5:.6em) -- (112.5:.6em);
\end{scope}
\begin{scope}[shift={(8,0)}]
\draw (22.5:.6em) -- (67.5:.6em) -- (112.5:.6em) -- (157.5:.6em) -- (202.5:.6em) -- (247.5:.6em) -- (292.5:.6em) -- (337.5:.6em) -- cycle;
\draw (112.5:.6em) -- (247.5:.6em);
\end{scope}
\begin{scope}[shift={(9,1)}]
\draw (22.5:.6em) -- (67.5:.6em) -- (112.5:.6em) -- (157.5:.6em) -- (202.5:.6em) -- (247.5:.6em) -- (292.5:.6em) -- (337.5:.6em) -- cycle;
\draw (247.5:.6em) -- (22.5:.6em);
\end{scope}
\draw[->] (.3,.3) -- (.7,.7);
\draw[->] (1.3,.7) -- (1.7,.3);
\draw[->] (2.3,.3) -- (2.7,.7);
\draw[->] (3.3,.7) -- (3.7,.3);
\draw[->] (4.3,.3) -- (4.7,.7);
\draw[->] (5.3,.7) -- (5.7,.3);
\draw[->] (6.3,.3) -- (6.7,.7);
\draw[->] (7.3,.7) -- (7.7,.3);
\draw[->] (8.3,.3) -- (8.7,.7);
\end{tikzpicture}
\end{equation}
where the identical diagrams on the left and right are identified.

As discussed in section~\ref{sec:kin}, the diagonals of an $(m+2)$-angulation
are but a proper subset of those of triangulations of the same $N$-gon,
obtained by omitting some of the diagonals of the latter. Concretely, the
diagonals in \eq{A2:C2g} appear in the cluster category $\mathcal C_{A_5}$ 
of an $A_5$ quiver, describing the triangulations of an octagon.
The 2-cluster category $\mathcal C^2_{A_2}$ can thus be viewed 
as a full extension-closed subcategory of $\mathcal C_{A_5}$ 
given by the objects on the second and fourth rows of the following diagram
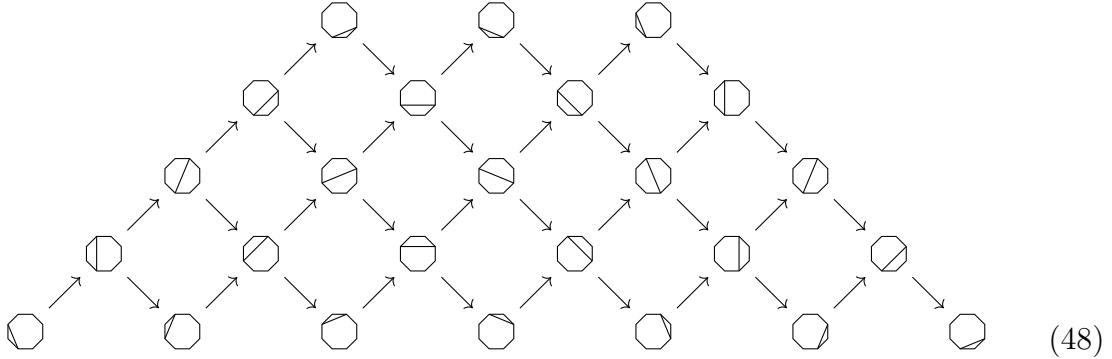
\begin{equation}
\label{A5:C1g}
\begin{tikzpicture}[x=2.5em,y=2.5em,line width=.4pt]
\begin{scope}
\draw (22.5:.6em) -- (67.5:.6em) -- (112.5:.6em) -- (157.5:.6em) -- (202.5:.6em) -- (247.5:.6em) -- (292.5:.6em) -- (337.5:.6em) -- cycle;
\draw (247.5:.6em) -- (157.5:.6em);
\end{scope}
\begin{scope}[shift={(1,1)}]
\draw (22.5:.6em) -- (67.5:.6em) -- (112.5:.6em) -- (157.5:.6em) -- (202.5:.6em) -- (247.5:.6em) -- (292.5:.6em) -- (337.5:.6em) -- cycle;
\draw (247.5:.6em) -- (112.5:.6em);
\end{scope}
\begin{scope}[shift={(2,2)}]
\draw (22.5:.6em) -- (67.5:.6em) -- (112.5:.6em) -- (157.5:.6em) -- (202.5:.6em) -- (247.5:.6em) -- (292.5:.6em) -- (337.5:.6em) -- cycle;
\draw (247.5:.6em) -- (67.5:.6em);
\end{scope}
\begin{scope}[shift={(3,3)}]
\draw (22.5:.6em) -- (67.5:.6em) -- (112.5:.6em) -- (157.5:.6em) -- (202.5:.6em) -- (247.5:.6em) -- (292.5:.6em) -- (337.5:.6em) -- cycle;
\draw (247.5:.6em) -- (22.5:.6em);
\end{scope}
\begin{scope}[shift={(4,4)}]
\draw (22.5:.6em) -- (67.5:.6em) -- (112.5:.6em) -- (157.5:.6em) -- (202.5:.6em) -- (247.5:.6em) -- (292.5:.6em) -- (337.5:.6em) -- cycle;
\draw (247.5:.6em) -- (337.5:.6em);
\end{scope}
\begin{scope}[shift={(2,0)}]
\draw (22.5:.6em) -- (67.5:.6em) -- (112.5:.6em) -- (157.5:.6em) -- (202.5:.6em) -- (247.5:.6em) -- (292.5:.6em) -- (337.5:.6em) -- cycle;
\draw (202.5:.6em) -- (112.5:.6em);
\end{scope}
\begin{scope}[shift={(3,1)}]
\draw (22.5:.6em) -- (67.5:.6em) -- (112.5:.6em) -- (157.5:.6em) -- (202.5:.6em) -- (247.5:.6em) -- (292.5:.6em) -- (337.5:.6em) -- cycle;
\draw (202.5:.6em) -- (67.5:.6em);
\end{scope}
\begin{scope}[shift={(4,2)}]
\draw (22.5:.6em) -- (67.5:.6em) -- (112.5:.6em) -- (157.5:.6em) -- (202.5:.6em) -- (247.5:.6em) -- (292.5:.6em) -- (337.5:.6em) -- cycle;
\draw (202.5:.6em) -- (22.5:.6em);
\end{scope}
\begin{scope}[shift={(5,3)}]
\draw (22.5:.6em) -- (67.5:.6em) -- (112.5:.6em) -- (157.5:.6em) -- (202.5:.6em) -- (247.5:.6em) -- (292.5:.6em) -- (337.5:.6em) -- cycle;
\draw (202.5:.6em) -- (337.5:.6em);
\end{scope}
\begin{scope}[shift={(6,4)}]
\draw (22.5:.6em) -- (67.5:.6em) -- (112.5:.6em) -- (157.5:.6em) -- (202.5:.6em) -- (247.5:.6em) -- (292.5:.6em) -- (337.5:.6em) -- cycle;
\draw (202.5:.6em) -- (292.5:.6em);
\end{scope}
\begin{scope}[shift={(4,0)}]
\draw (22.5:.6em) -- (67.5:.6em) -- (112.5:.6em) -- (157.5:.6em) -- (202.5:.6em) -- (247.5:.6em) -- (292.5:.6em) -- (337.5:.6em) -- cycle;
\draw (157.5:.6em) -- (67.5:.6em);
\end{scope}
\begin{scope}[shift={(5,1)}]
\draw (22.5:.6em) -- (67.5:.6em) -- (112.5:.6em) -- (157.5:.6em) -- (202.5:.6em) -- (247.5:.6em) -- (292.5:.6em) -- (337.5:.6em) -- cycle;
\draw (157.5:.6em) -- (22.5:.6em);
\end{scope}
\begin{scope}[shift={(6,2)}]
\draw (22.5:.6em) -- (67.5:.6em) -- (112.5:.6em) -- (157.5:.6em) -- (202.5:.6em) -- (247.5:.6em) -- (292.5:.6em) -- (337.5:.6em) -- cycle;
\draw (157.5:.6em) -- (337.5:.6em);
\end{scope}
\begin{scope}[shift={(7,3)}]
\draw (22.5:.6em) -- (67.5:.6em) -- (112.5:.6em) -- (157.5:.6em) -- (202.5:.6em) -- (247.5:.6em) -- (292.5:.6em) -- (337.5:.6em) -- cycle;
\draw (157.5:.6em) -- (292.5:.6em);
\end{scope}
\begin{scope}[shift={(8,4)}]
\draw (22.5:.6em) -- (67.5:.6em) -- (112.5:.6em) -- (157.5:.6em) -- (202.5:.6em) -- (247.5:.6em) -- (292.5:.6em) -- (337.5:.6em) -- cycle;
\draw (157.5:.6em) -- (247.5:.6em);
\end{scope}
\begin{scope}[shift={(6,0)}]
\draw (22.5:.6em) -- (67.5:.6em) -- (112.5:.6em) -- (157.5:.6em) -- (202.5:.6em) -- (247.5:.6em) -- (292.5:.6em) -- (337.5:.6em) -- cycle;
\draw (112.5:.6em) -- (22.5:.6em);
\end{scope}
\begin{scope}[shift={(7,1)}]
\draw (22.5:.6em) -- (67.5:.6em) -- (112.5:.6em) -- (157.5:.6em) -- (202.5:.6em) -- (247.5:.6em) -- (292.5:.6em) -- (337.5:.6em) -- cycle;
\draw (112.5:.6em) -- (337.5:.6em);
\end{scope}
\begin{scope}[shift={(8,2)}]
\draw (22.5:.6em) -- (67.5:.6em) -- (112.5:.6em) -- (157.5:.6em) -- (202.5:.6em) -- (247.5:.6em) -- (292.5:.6em) -- (337.5:.6em) -- cycle;
\draw (112.5:.6em) -- (292.5:.6em);
\end{scope}
\begin{scope}[shift={(9,3)}]
\draw (22.5:.6em) -- (67.5:.6em) -- (112.5:.6em) -- (157.5:.6em) -- (202.5:.6em) -- (247.5:.6em) -- (292.5:.6em) -- (337.5:.6em) -- cycle;
\draw (112.5:.6em) -- (247.5:.6em);
\end{scope}
\begin{scope}[shift={(8,0)}]
\draw (22.5:.6em) -- (67.5:.6em) -- (112.5:.6em) -- (157.5:.6em) -- (202.5:.6em) -- (247.5:.6em) -- (292.5:.6em) -- (337.5:.6em) -- cycle;
\draw (67.5:.6em) -- (337.5:.6em);
\end{scope}
\begin{scope}[shift={(9,1)}]
\draw (22.5:.6em) -- (67.5:.6em) -- (112.5:.6em) -- (157.5:.6em) -- (202.5:.6em) -- (247.5:.6em) -- (292.5:.6em) -- (337.5:.6em) -- cycle;
\draw (67.5:.6em) -- (292.5:.6em);
\end{scope}
\begin{scope}[shift={(10,2)}]
\draw (22.5:.6em) -- (67.5:.6em) -- (112.5:.6em) -- (157.5:.6em) -- (202.5:.6em) -- (247.5:.6em) -- (292.5:.6em) -- (337.5:.6em) -- cycle;
\draw (67.5:.6em) -- (247.5:.6em);
\end{scope}
\begin{scope}[shift={(10,0)}]
\draw (22.5:.6em) -- (67.5:.6em) -- (112.5:.6em) -- (157.5:.6em) -- (202.5:.6em) -- (247.5:.6em) -- (292.5:.6em) -- (337.5:.6em) -- cycle;
\draw (22.5:.6em) -- (292.5:.6em);
\end{scope}
\begin{scope}[shift={(11,1)}]
\draw (22.5:.6em) -- (67.5:.6em) -- (112.5:.6em) -- (157.5:.6em) -- (202.5:.6em) -- (247.5:.6em) -- (292.5:.6em) -- (337.5:.6em) -- cycle;
\draw (22.5:.6em) -- (247.5:.6em);
\end{scope}
\begin{scope}[shift={(12,0)}]
\draw (22.5:.6em) -- (67.5:.6em) -- (112.5:.6em) -- (157.5:.6em) -- (202.5:.6em) -- (247.5:.6em) -- (292.5:.6em) -- (337.5:.6em) -- cycle;
\draw (337.5:.6em) -- (247.5:.6em);
\end{scope}
\draw[->] (.3,.3) -- (.7,.7);
\draw[->] (1.3,.7) -- (1.7,.3);
\draw[->] (2.3,.3) -- (2.7,.7);
\draw[->] (3.3,.7) -- (3.7,.3);
\draw[->] (4.3,.3) -- (4.7,.7);
\draw[->] (5.3,.7) -- (5.7,.3);
\draw[->] (6.3,.3) -- (6.7,.7);
\draw[->] (7.3,.7) -- (7.7,.3);
\draw[->] (8.3,.3) -- (8.7,.7);
\draw[->] (9.3,.7) -- (9.7,.3);
\draw[->] (10.3,.3) -- (10.7,.7);
\draw[->] (11.3,.7) -- (11.7,.3);
\draw[->] (1.3,1.3) -- (1.7,1.7);
\draw[->] (2.3,1.7) -- (2.7,1.3);
\draw[->] (3.3,1.3) -- (3.7,1.7);
\draw[->] (4.3,1.7) -- (4.7,1.3);
\draw[->] (5.3,1.3) -- (5.7,1.7);
\draw[->] (6.3,1.7) -- (6.7,1.3);
\draw[->] (7.3,1.3) -- (7.7,1.7);
\draw[->] (8.3,1.7) -- (8.7,1.3);
\draw[->] (9.3,1.3) -- (9.7,1.7);
\draw[->] (10.3,1.7) -- (10.7,1.3);
\draw[->] (2.3,2.3) -- (2.7,2.7);
\draw[->] (3.3,2.7) -- (3.7,2.3);
\draw[->] (4.3,2.3) -- (4.7,2.7);
\draw[->] (5.3,2.7) -- (5.7,2.3);
\draw[->] (6.3,2.3) -- (6.7,2.7);
\draw[->] (7.3,2.7) -- (7.7,2.3);
\draw[->] (8.3,2.3) -- (8.7,2.7);
\draw[->] (9.3,2.7) -- (9.7,2.3);
\draw[->] (3.3,3.3) -- (3.7,3.7);
\draw[->] (4.3,3.7) -- (4.7,3.3);
\draw[->] (5.3,3.3) -- (5.7,3.7);
\draw[->] (6.3,3.7) -- (6.7,3.3);
\draw[->] (7.3,3.3) -- (7.7,3.7);
\draw[->] (8.3,3.7) -- (8.7,3.3);
\end{tikzpicture}
\end{equation}
depicting the cluster category ${\mathcal C}_{A_5}$.
Let us note that on the level of the representations, 
the objects of $\mathcal C^2_{A_2}$ viewed as a subcategory of 
$\mathcal C_{A_5}$ correspond precisely to the even-dimensional 
representations of the $A_5$ quiver.

The $2$-cluster category of $A_3$ is given by 
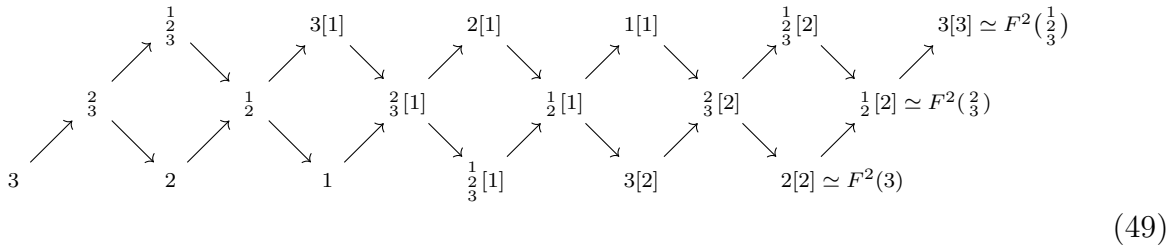
\begin{equation} 
\label{A3:C2} 
\begin{tikzpicture}[baseline=2.5em] 
\matrix (m) [matrix of math nodes, row sep={2.5em,between origins}, text 
height=1.4ex, text centered, font=\scriptsize, 
anchor=center,column sep={2.5em,between origins}, 
text depth=0.25ex, ampersand replacement=\&, 
nodes in empty cells] 
{ 
\&\& \nodethree{1}{2}{3} \&\& 
\nodeone{3}[1] \&\& \nodeone{2}[1]
\&\& 1[1] \&\& \nodethree{1}{2}{3}[2] 
\&\& 3[3]\mathrlap{{} \simeq F^2 \bigl( \nodethree{1}{2}{3} \bigr)}
\&\&  \\ 
\& \nodetwo{2}{3} \&\& \nodetwo{1}{2} \&\& 
\nodetwo{2}{3}[1] \&\& \nodetwo{1}{2}[1]
\&\&\nodetwo{2}{3}[2]\&\&\nodetwo{1}{2}[2]
\mathrlap{{} \simeq F^2(\nodetwo{2}{3})} 
\& \\ 
\nodeone{3} \&\& \nodeone{2} \&\& \nodeone{1} 
\&\& \nodethree{1}{2}{3}[1] \&\& 
\nodeone{3}[2]
\&\& 2[2]\mathrlap{{} \simeq F^2(3)} \& \\ 
}; 
\foreach \i/\j in {1/2,3/4,5/6,7/8,9/10,11/12} 
\draw[thin,->] (m-3-\i)--(m-2-\j); 
\foreach \i/\j in {2/3,4/5,6/7,8/9,10/11,12/13} 
\draw[thin,->] (m-2-\i)--(m-1-\j); 
\foreach \i/\j in {3/4,5/6,7/8,9/10,11/12} 
\draw[thin,->] (m-1-\i)--(m-2-\j); 
\foreach \i/\j in {2/3,4/5,6/7,8/9,10/11} 
\draw[thin,->] (m-2-\i)--(m-3-\j); 
\end{tikzpicture} 
\end{equation}
which can be written as diagonals forming part of a quadrangulation of a decagon and $\mathcal C^2_{A_3}$ can be viewed as a full extension-closed subcategory of $\mathcal C_{A_{7}}$.

\subsubsection{Cluster tilting objects}

For $N = mn + 2$, the $m$-intermediate $t$-structures discussed in
\S\ref{int} are in one-to-one correspondence to $m$-cluster tilting objects
in $\mathcal C^m_{A_{n-1}}$ and the diagonals of $m$-angulations of an
$N$-gon as in \eqref{A2:C2g} \cite{by,brt,BM2}. Concretely, the direct sum of the projectives of the heart of an $m$-intermediate $t$-structure is an $m$-cluster tilting object for $\mathcal C^m_{A_{n-1}}$ and every $m$-cluster tilting object arises in this way.

Viewing the $m$-cluster category as an extension-closed subcategory of the cluster category $\mathcal C_{A_{N-3}}$, the $m$-cluster tilting objects of $\mathcal C^m_{A_{n-1}}$ correspond to partial cluster tilting objects of $\mathcal C_{A_{N-3}}$. This point of view is also useful in categorifying the scattering amplitudes for general polynomial potentials.

\subsubsection{Periodic triangulated categories} 
The pseudo-periodic categories that appear in the computation of scattering
amplitudes are modelled after the periodic categories. 
Let us list some periodic categories to be used in the examples later on.
The $6$-periodic category $\P^6_{A_2}$ of the $A_2$ quiver is 
\begin{equation}
\label{A2:P6}
\begin{tikzpicture}[baseline=-4pt]
\matrix (m) [matrix of math nodes, minimum width=1.4em, row
sep={3.6em,between origins}, text height=1.4ex, column sep={1.3em,between
origins}, text depth=.15ex, ampersand replacement=\&,font=\scriptsize]
{
\& \nodetwo{1}{2} \&\&   \nodeone{2}[1] \&\& \nodeone{1}[1]
\&\&\nodetwo{1}{2}[2] \&\&  \nodeone{2}[3]  \&\&  \nodeone{1}[3]
\&\& \nodetwo{1}{2}[4] \&\&   \nodeone{2}[5] \&\& \nodeone{1}[5]
\&\&\nodetwo{1}{2}[6]\mathrlap{{}=\nodetwo{1}{2}}
\\
\nodeone{2} \&\& \nodeone{1} \&\& \nodetwo{1}{2}[1]
\&\& \nodeone{2}[2] \&\& \nodeone{1}[2]  \&\& \nodetwo{1}{2}[3]
\&\& \nodeone{2}[4] \&\& \nodeone{1}[4]  \&\& \nodetwo{1}{2}[5]
\&\& \nodeone{2}[6]\mathrlap{{}=\nodeone{2}}
\\
};
\foreach \i/\j in {1/2,3/4,5/6,7/8,9/10,11/12,13/14,15/16,17/18,19/20} 
\draw[thin,->] (m-2-\i)--(m-1-\j); 
\foreach \i/\j in {2/3,4/5,6/7,8/9,10/11,12/13,14/15,16/17,18/19} 
\draw[thin,->] (m-1-\i)--(m-2-\j); 
\draw[->,dashed] (m-2-3)--(m-2-1) node[above,midway,font=\scriptsize] {$\tau$};
\end{tikzpicture}
\end{equation}
The $7$-periodic category $\P^7_{A_2}$ of the $A_2$ quiver is 
\begin{equation}
\label{A2:P7}
\begin{tikzpicture}[baseline=-4pt]
\matrix (m) [matrix of math nodes, minimum width=1.4em, row
sep={3.6em,between origins}, text height=1.4ex, column sep={1.5em,between
origins}, text depth=.15ex, ampersand replacement=\&,font=\scriptsize]
{
\& \nodetwo{1}{2} 
\&\& \nodeone{2}[1] 
\&\& \nodeone{1}[1]
\&\& \nodetwo{1}{2}[2] 
\&\& \nodeone{2}[3]  
\&\& \nodeone{1}[3]
\&\& \nodetwo{1}{2}[4] 
\&\& \nodeone{2}[5] 
\&\& \nodeone{1}[5]
\&\& \nodetwo{1}{2}[6] 
\&\& \nodeone{2}[7] \mathrlap{{}=\nodeone{2}}
\&\&
\\
\nodeone{2} 
\&\& \nodeone{1} 
\&\& \nodetwo{1}{2}[1]
\&\& \nodeone{2}[2] 
\&\& \nodeone{1}[2]  
\&\& \nodetwo{1}{2}[3]
\&\& \nodeone{2}[4] 
\&\& \nodeone{1}[4]  
\&\& \nodetwo{1}{2}[5]
\&\& \nodeone{2}[6] 
\&\& \nodeone{1}[6]  
\&\& \nodetwo{1}{2}[7]\mathrlap{{}=\nodetwo{1}{2}}
\&\&
\\
};
\foreach \i/\j in {1/2,3/4,5/6,7/8,9/10,11/12,13/14,15/16,17/18,19/20,21/22} 
\draw[thin,->] (m-2-\i)--(m-1-\j); 
\foreach \i/\j in {2/3,4/5,6/7,8/9,10/11,12/13,14/15,16/17,18/19,20/21,22/23} 
\draw[thin,->] (m-1-\i)--(m-2-\j); 
\draw[->,dashed] (m-2-3)--(m-2-1) node[above,midway,font=\scriptsize] {$\tau$};
\end{tikzpicture}
\end{equation}
The $8$-periodic category $\P^8_{A_2}$ of the $A_2$ quiver is 
\begin{equation}  
\label{A2:P8} 
\begin{tikzpicture}[baseline=-4pt] 
\matrix (m) [matrix of math nodes, minimum width=1.4em, row 
sep={3.6em,between origins}, text height=1.4ex, column sep={1.3em,between 
origins}, text depth=.15ex, ampersand replacement=\&,font=\scriptsize] 
{ 
\& \nodetwo{1}{2} \&\&   \nodeone{2}[1] \&\& \nodeone{1}[1] 
\&\&\nodetwo{1}{2}[2] \&\&  \nodeone{2}[3]  \&\&  \nodeone{1}[3] 
\&\& \nodetwo{1}{2}[4] \&\&   \nodeone{2}[5] \&\& \nodeone{1}[5] 
\&\&\nodetwo{1}{2}[6] \&\&  \nodeone{2}[7]  \&\&  \nodeone{1}[7] 
\&\&   \nodetwo{1}{2}[8]\mathrlap{{}=\nodetwo{1}{2}} \&\&
\\ 
\nodeone{2} \&\& \nodeone{1} \&\& \nodetwo{1}{2}[1] 
\&\& \nodeone{2}[2] \&\& \nodeone{1}[2]  \&\& \nodetwo{1}{2}[3] 
\&\& \nodeone{2}[4] \&\& \nodeone{1}[4]  \&\& \nodetwo{1}{2}[5] 
\&\& \nodeone{2}[6] \&\& \nodeone{1}[6]  \&\& \nodetwo{1}{2}[7] 
\&\&  \nodeone{2}[8]\mathrlap{{}=\nodeone{2}} \&\&
\\ 
}; 
\draw[->](m-2-1)--(m-1-2);
\draw[->](m-1-4)--(m-2-5);
\draw[->](m-2-7)--(m-1-8);
\draw[->](m-1-10)--(m-2-11);
\draw[->](m-2-13)--(m-1-14);
\draw[->](m-1-16)--(m-2-17);
\draw[->] (m-2-19)--(m-1-20);
\draw[->] (m-1-22)--(m-2-23);
\draw[->] (m-2-25)--(m-1-26);
\draw[->] (m-2-3)--(m-1-4);
\draw[->] (m-1-6)--(m-2-7);
\draw[->] (m-2-9)--(m-1-10);
\draw[->] (m-1-12)--(m-2-13);
\draw[->] (m-2-15)--(m-1-16);
\draw[->] (m-1-18)--(m-2-19);
\draw[->] (m-2-21)--(m-1-22);
\draw[->] (m-1-24)--(m-2-25);
\draw[->] (m-1-2)--(m-2-3);
\draw[->] (m-2-5)--(m-1-6);
\draw[->] (m-1-8)--(m-2-9);
\draw[->] (m-2-11)--(m-1-12);
\draw[->] (m-1-14)--(m-2-15);
\draw[->] (m-2-17)--(m-1-18);
\draw[->] (m-1-20)--(m-2-21);
\draw[->] (m-2-23)--(m-1-24);
\draw[->,dashed] (m-2-3)--(m-2-1) node[above,midway,font=\scriptsize] {$\tau$}; 
\end{tikzpicture} 
\end{equation} 
The category $\P^{7}_{A_3}$ is  
\begin{equation}  
\label{A3:P7}
\begin{tikzpicture}[baseline=-4pt] 
\matrix (m) [matrix of math nodes, minimum width=1.4em, row 
sep={3.6em,between origins}, text height=1.4ex, column sep={1.1em,between 
origins}, text depth=.15ex, ampersand replacement=\&,font=\scriptsize] 
{ 
\&\&   \nodethree{1}{2}{3} \&\&   3[1] \&\&   2[1] 
\&\&   1[1] \&\&   \nodethree{1}{2}{3}[2] \&\&  3[3] 
\&\&   2[3] \&\&   1[3] \&\&   \nodethree{1}{2}{3}[4]
\&\&   3[5] \&\&   2[5] \&\&   1[5] 
\&\&   \nodethree{1}{2}{3}[6] \&\&  
3[7]\mathrlap{{}=\nodeone{3}} \&\&\\
%
\& \nodetwo{2}{3} \&\&  \nodetwo{1}{2} 
\&\&  \nodetwo{2}{3}[1] \&\&  \nodetwo{1}{2}[1] 
\&\&  \nodetwo{2}{3}[2] \&\&  \nodetwo{1}{2}[2] 
\&\&  \nodetwo{2}{3}[3] \&\&  \nodetwo{1}{2}[3] 
\&\&  \nodetwo{2}{3}[4] \&\&  \nodetwo{1}{2}[4] 
\&\&  \nodetwo{2}{3}[5] \&\&  \nodetwo{1}{2}[5] 
\&\&  \nodetwo{2}{3}[6] \&\&  \nodetwo{1}{2}[6] 
\&\&  \nodetwo{2}{3}[7] \mathrlap{{}=\nodetwo{2}{3}} \&\& \\
%
\nodeone{3} \&\& \nodeone{2} \&\& \nodeone{1} 
\&\&  \nodethree{1}{2}{3}[1] \&\&  \nodeone{3}[2] \&\&  2[2]
\&\&  1[2] \&\& \nodethree{1}{2}{3}[3] \&\&  3[4] 
\&\&  2[4] \&\&  1[4] \&\&  \nodethree{1}{2}{3}[5] 
\&\&  3[6] \&\&  2[6] \&\&  1[6] 
\&\&  \nodethree{1}{2}{3}[7]\mathrlap{{}=\nodethree{1}{2}{3}} \\
}; 
\draw[->] (m-3-1)--(m-2-2); 
\draw[->] (m-3-3)--(m-2-4); 
\draw[->] (m-3-5)--(m-2-6); 
\draw[->] (m-3-7)--(m-2-8); 
\draw[->] (m-3-9)--(m-2-10); 
\draw[->] (m-3-11)--(m-2-12); 
\draw[->] (m-3-13)--(m-2-14); 
\draw[->] (m-3-15)--(m-2-16); 
\draw[->] (m-3-17)--(m-2-18); 
\draw[->] (m-3-19)--(m-2-20); 
\draw[->] (m-3-21)--(m-2-22); 
\draw[->] (m-3-23)--(m-2-24); 
\draw[->] (m-3-25)--(m-2-26); 
\draw[->] (m-3-27)--(m-2-28); 
\draw[->] (m-3-29)--(m-2-30); 
\draw[->] (m-2-2)--(m-1-3); 
\draw[->] (m-2-4)--(m-1-5); 
\draw[->] (m-2-6)--(m-1-7); 
\draw[->] (m-2-8)--(m-1-9); 
\draw[->] (m-2-10)--(m-1-11); 
\draw[->] (m-2-12)--(m-1-13); 
\draw[->] (m-2-14)--(m-1-15); 
\draw[->] (m-2-16)--(m-1-17); 
\draw[->] (m-2-18)--(m-1-19); 
\draw[->] (m-2-20)--(m-1-21); 
\draw[->] (m-2-22)--(m-1-23); 
\draw[->] (m-2-24)--(m-1-25); 
\draw[->] (m-2-26)--(m-1-27); 
\draw[->] (m-2-28)--(m-1-29); 
%
%
\draw[->] (m-2-2)--(m-3-3); 
\draw[->] (m-2-4)--(m-3-5); 
\draw[->] (m-2-6)--(m-3-7); 
\draw[->] (m-2-8)--(m-3-9); 
\draw[->] (m-2-10)--(m-3-11); 
\draw[->] (m-2-12)--(m-3-13); 
\draw[->] (m-2-14)--(m-3-15); 
\draw[->] (m-2-16)--(m-3-17); 
\draw[->] (m-2-18)--(m-3-19); 
\draw[->] (m-2-20)--(m-3-21); 
\draw[->] (m-2-22)--(m-3-23); 
\draw[->] (m-2-24)--(m-3-25); 
\draw[->] (m-2-26)--(m-3-27); 
\draw[->] (m-2-28)--(m-3-29); 
\draw[->] (m-2-30)--(m-3-31); 
%
%
\draw[->] (m-1-3)--(m-2-4); 
\draw[->] (m-1-5)--(m-2-6); 
\draw[->] (m-1-7)--(m-2-8); 
\draw[->] (m-1-9)--(m-2-10); 
\draw[->] (m-1-11)--(m-2-12); 
\draw[->] (m-1-13)--(m-2-14); 
\draw[->] (m-1-15)--(m-2-16); 
\draw[->] (m-1-17)--(m-2-18); 
\draw[->] (m-1-19)--(m-2-20); 
\draw[->] (m-1-21)--(m-2-22); 
\draw[->] (m-1-23)--(m-2-24); 
\draw[->] (m-1-25)--(m-2-26); 
\draw[->] (m-1-27)--(m-2-28); 
\draw[->] (m-1-29)--(m-2-30); 
\end{tikzpicture} 
\end{equation} 
The category $\P^{12}_{A_3}$ to be used later 
is given in \eq{A3:P12} in the Appendix.
\section{Categorical description of scattering amplitudes} 
\label{sec:cat} 
In this section we point out the identification of the mesh diagrams obtained 
from \eq{mesh:big} with the cluster and pseudo-periodic categories of  
$A$-type quivers. We then use the intermediate $t$-structure for the former 
and a variant for the latter case to write down the scattering amplitude in 
terms of projectives of hearts of intermediate $t$-structures.  
\begin{example}[Cubic theory \cite{br}]
Amplitudes of $N$-particle scattering in the cubic theory arise from the 
cluster category of $A_{n-1}$ quivers. For $N=5$ the mesh diagram \eq{dbA2}
can be identified as the cluster category $\C_{A_2}$ in \eq{A2:C1}. 
Let us point out that the objects mapped under the automorphism $F$ are not
indicated in the mesh diagram. The planar variables in \eq{dbA2} are in
one-to-one correspondence with the representations in \eq{A2:C1} and
\figref{hearts1} according to the dictionary
\begin{align}
2 &\lrar X_{1,3} \\ 
\nodetwo{1}{2} &\lrar X_{1,4}\\
1 &\lrar X_{2,4} \\
2[1] &\lrar X_{2,5}\\
\nodetwo{1}{2}[1] &\lrar X_{3,5},
\end{align}
relating representations and diagonals as in \eq{A2:C1g}.
Restricting the intermediate
$t$-structures \figref{hearts1} to the cluster category and using the
correspondence of $X$'s and the objects of $\C_{A_2}$ we find that the
amplitude \eq{can:5} is given term by term in terms of the projectives 
of the hearts of intermediate $t$-structures.

For the $N=6$ case, similarly, the mesh diagram \eq{dbA3} is identified with
the  cluster category $\C_{A_3}$ in \eq{A3:C1}. The amplitude \eq{S3-6} is
then obtained term by term through the correspondence between $X$'s and the
representations in the cluster category from the projectives of hearts of
the intermediate $t$-structures \figref{hearts3} restricted to $\C_{A_3}$.
\end{example}
\begin{example}[Quartic theory]
The amplitudes of $N$-particle scattering 
for the quartic theory with $m=2$ are obtained from the
$2$-cluster category $\C^2_{A_{n-1}}$.
For example, for $N=8$, $n=3$, $\Lambda=(2,2,2)$. 
The mesh \eq{mesh:42} is identified with
$\C^2_{A_2}$ in \eq{A2:C2}. The $2$-intermediate $t$-structures are shown in
\figref{hearts2} and \figref{hearts2X} in terms of the representations and
the planar variables, respectively. 
The scattering amplitude is obtained as twelve terms corresponding to the
projectives of the hearts in the twelve $2$-intermediate $t$-structures
restricted to $\C^2_{A_2}$ as
\begin{multline} 
\label{S84}
S_8(\phi^4)=
\frac{1}{X_{1,4}X_{1,6}}+ \frac{1}{X_{1,6}X_{3,6}}
+ \frac{1}{X_{3,6}X_{3,8}}+ \frac{1}{X_{3,8}X_{5,8}} \\
+ \frac{1}{X_{5,8}X_{2,5}} +\frac{1}{X_{2,5}X_{2,7}} 
+\frac{1}{X_{2,7}X_{4,7}} +\frac{1}{X_{4,7}X_{1,4}}\\
+ \frac{1}{X_{1,4}X_{5,8}} +\frac{1}{X_{1,6}X_{2,5}} 
+ \frac{1}{X_{3,6}X_{2,7}} + \frac{1}{X_{3,8}X_{4,7}}.
\end{multline}  
 
Similarly, for $N=10$, $\Lambda=(2,2,2,2)$ and 
the mesh \eq{mesh:43} matches the $2$-cluster category
$\C^2_{A_3}$ given in \eq{A3:C2}. Using the $2$-intermediate $t$-structures
we obtain the scattering amplitude
\begin{multline}
\label{S104}
S_{10}(\phi^4)=
\frac{1}{X_{1,4}X_{1,6}X_{1,8}}+\frac{1}{X_{3,6}X_{3,8}X_{3,10}}
+\frac{1}{X_{5,8}X_{5,10}X_{2,5}}+\frac{1}{X_{7,10}X_{2,7}X_{4,7}}
+\frac{1}{X_{2,9}X_{4,9}X_{6,9}}\\
+\frac{1}{X_{1,8}X_{3,8}X_{5,8}}+\frac{1}{X_{3,10}X_{5,10}X_{7,10}}
+\frac{1}{X_{2,5}X_{2,7}X_{2,9}}+\frac{1}{X_{4,7}X_{4,9}X_{1,4}}
+\frac{1}{X_{6,9}X_{1,6}X_{3,6}}\\
+\frac{1}{X_{1,8}X_{3,8}X_{3,6}}+\frac{1}{X_{3,10}X_{5,10}X_{5,8}}
+\frac{1}{X_{2,5}X_{2,7}X_{7,10}}+\frac{1}{X_{4,7}X_{4,9}X_{2,9}}
+\frac{1}{X_{6,9}X_{1,6}X_{1,4}}\\
+\frac{1}{X_{1,8}X_{1,6}X_{3,6}}+\frac{1}{X_{3,10}X_{3,8}X_{5,8}}
+\frac{1}{X_{2,5}X_{5,10}X_{7,10}}+\frac{1}{X_{4,7}X_{2,7}X_{2,9}}
+\frac{1}{X_{6,9}X_{4,9}X_{1,4}}\\
+\frac{1}{X_{1,4}X_{1,6}X_{7,10}}+\frac{1}{X_{3,6}X_{3,8}X_{2,9}}
+\frac{1}{X_{5,8}X_{5,10}X_{1,4}}+\frac{1}{X_{7,10}X_{2,7}X_{3,6}}
+\frac{1}{X_{2,9}X_{4,9}X_{5,8}}\\
+\frac{1}{X_{1,4}X_{5,10}X_{7,10}}+\frac{1}{X_{3,6}X_{2,7}X_{2,9}}
+\frac{1}{X_{5,8}X_{4,9}X_{1,4}}+\frac{1}{X_{7,10}X_{1,6}X_{3,6}}
+\frac{1}{X_{2,9}X_{3,8}X_{5,8}}\\
+\frac{1}{X_{1,8}X_{3,8}X_{4,7}}+\frac{1}{X_{3,10}X_{5,10}X_{6,9}}
+\frac{1}{X_{2,5}X_{2,7}X_{1,8}}+\frac{1}{X_{4,7}X_{4,9}X_{3,10}}
+\frac{1}{X_{6,9}X_{1,6}X_{2,5}}\\
+\frac{1}{X_{1,8}X_{2,7}X_{4,7}}+\frac{1}{X_{3,10}X_{4,9}X_{6,9}}
+\frac{1}{X_{2,5}X_{1,6}X_{1,8}}+\frac{1}{X_{4,7}X_{3,8}X_{3,10}}
+\frac{1}{X_{6,9}X_{5,10}X_{2,5}}\\
+\frac{1}{X_{1,4}X_{5,10}X_{6,9}}+\frac{1}{X_{3,6}X_{2,7}X_{1,8}}
+\frac{1}{X_{5,8}X_{4,9}X_{3,10}}+\frac{1}{X_{7,10}X_{1,6}X_{2,5}}
+\frac{1}{X_{2,9}X_{3,8}X_{4,7}}\\
+\frac{1}{X_{1,4}X_{1,8}X_{5,8}}+\frac{1}{X_{3,6}X_{3,10}X_{7,10}}
+\frac{1}{X_{5,8}X_{2,5}X_{2,9}}+\frac{1}{X_{7,10}X_{4,7}X_{1,4}}
+\frac{1}{X_{2,9}X_{6,9}X_{3,6}}\\
+\frac{1}{X_{1,8}X_{5,8}X_{2,5}}+\frac{1}{X_{3,10}X_{7,10}X_{4,7}}
+\frac{1}{X_{2,5}X_{2,9}X_{6,9}}+\frac{1}{X_{4,7}X_{1,4}X_{1,8}}
+\frac{1}{X_{6,9}X_{3,6}X_{3,10}}.
\end{multline}
\end{example}
\begin{figure} 
\begin{center} 
\begin{tikzpicture}[x=1em,y=1em] 
\begin{scope}[shift={(12em,4em)}] 
\draw[line width=.4pt] (-2,1) circle(.25ex); 
\node[font=\scriptsize] at (-1,0) {$14$}; 
\node[font=\scriptsize] at  (0,1) {$16$}; 
\node[font=\scriptsize] at  (1,0) {$36$}; 
\node[font=\scriptsize] at  (2,1) {$38$}; 
\node[font=\scriptsize] at  (3,0) {$58$}; 
\node[font=\scriptsize] at  (4,1) {$25$}; 
\node[font=\scriptsize] at  (5,0) {$27$}; 
\node[font=\scriptsize] at  (6,1) {$47$}; 
\draw[line width=.4pt]  (7,0) circle(.25ex); 
\end{scope} 
\begin{scope} 
\draw[line width=.7em,fill=black!15,black!15,rounded corners=.001em] (1,0) -- (-1,0) -- (0,1) -- cycle; 
\draw[line width=.4pt] (-2,1) circle(.25ex); 
\draw[line width=.4pt,fill=black] (-1,0) circle(.25ex); 
\draw[line width=.4pt] (-1,0) circle(.5ex); 
\draw[line width=.4pt,fill=black]  (0,1) circle(.25ex); 
\draw[line width=.4pt]  (0,1) circle(.5ex); 
\draw[line width=.4pt,fill=black]  (1,0) circle(.25ex); 
\draw[line width=.4pt]  (2,1) circle(.25ex); 
\draw[line width=.4pt]  (3,0) circle(.25ex); 
\draw[line width=.4pt]  (4,1) circle(.25ex); 
\draw[line width=.4pt]  (5,0) circle(.25ex); 
\draw[line width=.4pt]  (6,1) circle(.25ex); 
\draw[line width=.4pt]  (7,0) circle(.25ex); 
\draw[line width=.1em,red!80!black!70, line cap=round] (-2.5,1.55) -- (-2,1.55) arc[start angle=90, end angle=-45, radius=.55em] -- ++(-.9,-.9); 
\draw[line width=.1em,blue!80!black!70,line cap=round] (7.5,1.55) -- (0,1.55) arc[start angle=90, end angle=135, radius=.55em] -- ++(-1,-1) arc[start angle=135, end angle=270, radius=.55em] -- ++(8.5,0); 
\node[font=\scriptsize] at (3,-1.2) {$[14] [16]$}; 
\end{scope} 
\begin{scope}[yshift=-4em] 
\draw[line width=.7em,fill=black!15,black!15,rounded corners=.001em] (1,0) -- (2,1) -- (0,1) -- cycle; 
\draw[line width=.4pt] (-2,1) circle(.25ex); 
\draw[line width=.4pt,fill=black] (-1,0) circle(.25ex); 
\draw[line width=.4pt,fill=black]  (0,1) circle(.25ex); 
\draw[line width=.4pt]  (0,1) circle(.5ex); 
\draw[line width=.4pt,fill=black]  (1,0) circle(.25ex); 
\draw[line width=.4pt]  (1,0) circle(.5ex); 
\draw[line width=.4pt]  (2,1) circle(.25ex); 
\draw[line width=.4pt]  (3,0) circle(.25ex); 
\draw[line width=.4pt]  (4,1) circle(.25ex); 
\draw[line width=.4pt]  (5,0) circle(.25ex); 
\draw[line width=.4pt]  (6,1) circle(.25ex); 
\draw[line width=.4pt]  (7,0) circle(.25ex); 
\draw[line width=.1em,red!80!black!70, line cap=round] (-2.5,1.55) -- (-2,1.55) arc[start angle=90, end angle=45, radius=.55em] -- ++(1,-1) arc[start angle=45, end angle=-90, radius=.55em] -- ++(-1.5,0); 
\draw[line width=.1em,blue!80!black!70,line cap=round] (7.5,1.55) -- (0,1.55) arc[start angle=90, end angle=225, radius=.55em] -- ++(1,-1) arc[start angle=225, end angle=270, radius=.55em] -- ++(6.5,0); 
\node[font=\scriptsize] at (3,-1.2) {$[16] [36]$}; 
\end{scope} 
\begin{scope}[yshift=-8em] 
\draw[line width=.7em,fill=black!15,black!15,rounded corners=.001em] (1,0) -- (2,1) -- (3,0) -- cycle; 
\draw[line width=.4pt] (-2,1) circle(.25ex); 
\draw[line width=.4pt,fill=black] (-1,0) circle(.25ex); 
\draw[line width=.4pt,fill=black]  (0,1) circle(.25ex); 
\draw[line width=.4pt,fill=black]  (1,0) circle(.25ex); 
\draw[line width=.4pt]  (1,0) circle(.5ex); 
\draw[line width=.4pt]  (2,1) circle(.25ex); 
\draw[line width=.4pt]  (2,1) circle(.5ex); 
\draw[line width=.4pt]  (3,0) circle(.25ex); 
\draw[line width=.4pt]  (4,1) circle(.25ex); 
\draw[line width=.4pt]  (5,0) circle(.25ex); 
\draw[line width=.4pt]  (6,1) circle(.25ex); 
\draw[line width=.4pt]  (7,0) circle(.25ex); 
\draw[line width=.1em,red!80!black!70, line cap=round] (-2.5,1.55) -- (0,1.55) arc[start angle=90, end angle=-45, radius=.55em] -- ++(-1,-1) arc[start angle=-45, end angle=-90, radius=.55em] -- ++(-1.5,0); 
\draw[line width=.1em,blue!80!black!70,line cap=round] (7.5,1.55) -- (2,1.55) arc[start angle=90, end angle=135, radius=.55em] -- ++(-1,-1) arc[start angle=135, end angle=270, radius=.55em] -- ++(6.5,0); 
\node[font=\scriptsize] at (3,-1.2) {$[36] [38]$}; 
\end{scope} 
\begin{scope}[yshift=-12em] 
\draw[line width=.7em,fill=black!15,black!15,rounded corners=.001em] (3,0) -- (2,1) -- (4,1) -- cycle; 
\draw[line width=.4pt] (-2,1) circle(.25ex); 
\draw[line width=.4pt,fill=black] (-1,0) circle(.25ex); 
\draw[line width=.4pt,fill=black]  (0,1) circle(.25ex); 
\draw[line width=.4pt,fill=black]  (1,0) circle(.25ex); 
\draw[line width=.4pt]  (2,1) circle(.25ex); 
\draw[line width=.4pt]  (2,1) circle(.5ex); 
\draw[line width=.4pt]  (3,0) circle(.25ex); 
\draw[line width=.4pt]  (3,0) circle(.5ex); 
\draw[line width=.4pt]  (4,1) circle(.25ex); 
\draw[line width=.4pt]  (5,0) circle(.25ex); 
\draw[line width=.4pt]  (6,1) circle(.25ex); 
\draw[line width=.4pt]  (7,0) circle(.25ex); 
\draw[line width=.1em,red!80!black!70, line cap=round] (-2.5,1.55) -- (-0,1.55) arc[start angle=90, end angle=45, radius=.55em] -- ++(1,-1) arc[start angle=45, end angle=-90, radius=.55em] -- ++(-3.5,0); 
\draw[line width=.1em,blue!80!black!70,line cap=round] (7.5,1.55) -- (2,1.55) arc[start angle=90, end angle=225, radius=.55em] -- ++(1,-1) arc[start angle=225, end angle=270, radius=.55em] -- ++(4.5,0); 
\node[font=\scriptsize] at (3,-1.2) {$[38] [58]$}; 
\end{scope} 
\begin{scope}[xshift=12em] 
\draw[line width=.7em,fill=black!15,black!15,rounded corners=.001em] (5,0) -- (3,0) -- (4,1) -- cycle; 
\draw[line width=.4pt] (-2,1) circle(.25ex); 
\draw[line width=.4pt,fill=black] (-1,0) circle(.25ex); 
\draw[line width=.4pt,fill=black]  (0,1) circle(.25ex); 
\draw[line width=.4pt,fill=black]  (1,0) circle(.25ex); 
\draw[line width=.4pt]  (2,1) circle(.25ex); 
\draw[line width=.4pt]  (3,0) circle(.25ex); 
\draw[line width=.4pt]  (3,0) circle(.5ex); 
\draw[line width=.4pt]  (4,1) circle(.25ex); 
\draw[line width=.4pt]  (4,1) circle(.5ex); 
\draw[line width=.4pt]  (5,0) circle(.25ex); 
\draw[line width=.4pt]  (6,1) circle(.25ex); 
\draw[line width=.4pt]  (7,0) circle(.25ex); 
\draw[line width=.1em,red!80!black!70, line cap=round] (-2.5,1.55) -- (2,1.55) arc[start angle=90, end angle=-45, radius=.55em] -- ++(-1,-1) arc[start angle=-45, end angle=-90, radius=.55em] -- ++(-3.5,0); 
\draw[line width=.1em,blue!80!black!70,line cap=round] (7.5,1.55) -- (4,1.55) arc[start angle=90, end angle=135, radius=.55em] -- ++(-1,-1) arc[start angle=135, end angle=270, radius=.55em] -- ++(4.5,0); 
\node[font=\scriptsize] at (3,-1.2) {$[58] [25]$}; 
\end{scope} 
\begin{scope}[xshift=12em,yshift=-4em] 
\draw[line width=.7em,fill=black!15,black!15,rounded corners=.001em] (5,0) -- (6,1) -- (4,1) -- cycle; 
\draw[line width=.4pt] (-2,1) circle(.25ex); 
\draw[line width=.4pt,fill=black] (-1,0) circle(.25ex); 
\draw[line width=.4pt,fill=black]  (0,1) circle(.25ex); 
\draw[line width=.4pt,fill=black]  (1,0) circle(.25ex); 
\draw[line width=.4pt]  (2,1) circle(.25ex); 
\draw[line width=.4pt]  (3,0) circle(.25ex); 
\draw[line width=.4pt]  (4,1) circle(.25ex); 
\draw[line width=.4pt]  (4,1) circle(.5ex); 
\draw[line width=.4pt]  (5,0) circle(.25ex); 
\draw[line width=.4pt]  (5,0) circle(.5ex); 
\draw[line width=.4pt]  (6,1) circle(.25ex); 
\draw[line width=.4pt]  (7,0) circle(.25ex); 
\draw[line width=.1em,red!80!black!70, line cap=round] (-2.5,1.55) -- (2,1.55) arc[start angle=90, end angle=45, radius=.55em] -- ++(1,-1) arc[start angle=45, end angle=-90, radius=.55em] -- ++(-5.5,0); 
\draw[line width=.1em,blue!80!black!70,line cap=round] (7.5,1.55) -- (4,1.55) arc[start angle=90, end angle=225, radius=.55em] -- ++(1,-1) arc[start angle=225, end angle=270, radius=.55em] -- ++(2.5,0); 
\node[font=\scriptsize] at (3,-1.2) {$[25] [27]$}; 
\end{scope} 
\begin{scope}[xshift=12em,yshift=-8em] 
\draw[line width=.7em,fill=black!15,black!15,rounded corners=.001em] (5,0) -- (6,1) -- (7,0) -- cycle; 
\draw[line width=.4pt] (-2,1) circle(.25ex); 
\draw[line width=.4pt,fill=black] (-1,0) circle(.25ex); 
\draw[line width=.4pt,fill=black]  (0,1) circle(.25ex); 
\draw[line width=.4pt,fill=black]  (1,0) circle(.25ex); 
\draw[line width=.4pt]  (2,1) circle(.25ex); 
\draw[line width=.4pt]  (3,0) circle(.25ex); 
\draw[line width=.4pt]  (4,1) circle(.25ex); 
\draw[line width=.4pt]  (5,0) circle(.25ex); 
\draw[line width=.4pt]  (5,0) circle(.5ex); 
\draw[line width=.4pt]  (6,1) circle(.25ex); 
\draw[line width=.4pt]  (6,1) circle(.5ex); 
\draw[line width=.4pt]  (7,0) circle(.25ex); 
\draw[line width=.1em,red!80!black!70, line cap=round] (-2.5,1.55) -- (4,1.55) arc[start angle=90, end angle=-45, radius=.55em] -- ++(-1,-1) arc[start angle=-45, end angle=-90, radius=.55em] -- ++(-5.5,0); 
\draw[line width=.1em,blue!80!black!70,line cap=round] (7.5,1.55) -- (6,1.55) arc[start angle=90, end angle=135, radius=.55em] -- ++(-1,-1) arc[start angle=135, end angle=270, radius=.55em] -- ++(2.5,0); 
\node[font=\scriptsize] at (3,-1.2) {$[27] [47]$}; 
\end{scope} 
\begin{scope}[xshift=12em,yshift=-12em] 
\draw[line width=0,fill=black!15,black!15] (-1,0) circle(.35em); 
\draw[line width=0,fill=black!15,black!15] (7,0) circle(.35em); 
\draw[line width=.4pt] (-2,1) circle(.25ex); 
\draw[line width=.4pt,fill=black] (-1,0) circle(.25ex); 
\draw[line width=.4pt] (-1,0) circle(.5ex); 
\draw[line width=.4pt,fill=black]  (0,1) circle(.25ex); 
\draw[line width=.4pt,fill=black]  (1,0) circle(.25ex); 
\draw[line width=.4pt]  (2,1) circle(.25ex); 
\draw[line width=.4pt]  (3,0) circle(.25ex); 
\draw[line width=.4pt]  (4,1) circle(.25ex); 
\draw[line width=.4pt]  (5,0) circle(.25ex); 
\draw[line width=.4pt]  (6,1) circle(.25ex); 
\draw[line width=.4pt]  (6,1) circle(.5ex); 
\draw[line width=.4pt]  (7,0) circle(.25ex); 
\draw[line width=.1em,red!80!black!70, line cap=round] (-2.5,1.55) -- (-2,1.55) arc[start angle=90, end angle=-45, radius=.55em] -- ++(-.9,-.9); 
\draw[line width=.1em,blue!80!black!70,line cap=round] (7.5,1.55) -- (6,1.55) arc[start angle=90, end angle=135, radius=.55em] -- ++(-1,-1) arc[start angle=135, end angle=270, radius=.55em] -- ++(2.5,0); 
\draw[line width=.1em,blue!80!black!70,line cap=round] (-1,0) circle(.55em); 
\draw[line width=.1em,red!80!black!70,line cap=round] (1,0) circle(.55em); 
\draw[line width=.1em,blue!80!black!70,line cap=round] (2,1) circle(.55em); 
\draw[line width=.1em,red!80!black!70,line cap=round] (4,1) circle(.55em); 
\node[font=\scriptsize] at (3,-1.2) {$[47] [14]$}; 
\end{scope} 
\begin{scope}[xshift=24em] 
\draw[line width=0,fill=black!15,black!15] (-1,0) circle(.35em); 
\draw[line width=0,fill=black!15,black!15] (4,1) circle(.35em); 
\draw[line width=.4pt] (-2,1) circle(.25ex); 
\draw[line width=.4pt,fill=black] (-1,0) circle(.25ex); 
\draw[line width=.4pt] (-1,0) circle(.5ex); 
\draw[line width=.4pt,fill=black]  (0,1) circle(.25ex); 
\draw[line width=.4pt,fill=black]  (1,0) circle(.25ex); 
\draw[line width=.4pt]  (2,1) circle(.25ex); 
\draw[line width=.4pt]  (3,0) circle(.25ex); 
\draw[line width=.4pt]  (3,0) circle(.5ex); 
\draw[line width=.4pt]  (4,1) circle(.25ex); 
\draw[line width=.4pt]  (5,0) circle(.25ex); 
\draw[line width=.4pt]  (6,1) circle(.25ex); 
\draw[line width=.4pt]  (7,0) circle(.25ex); 
\draw[line width=.1em,red!80!black!70, line cap=round] (-2.5,1.55) -- (-2,1.55) arc[start angle=90, end angle=-45, radius=.55em] -- ++(-.9,-.9); 
\draw[line width=.1em,blue!80!black!70,line cap=round] (7.5,1.55) -- (2,1.55) arc[start angle=90, end angle=225, radius=.55em] -- ++(1,-1) arc[start angle=225, end angle=270, radius=.55em] -- ++(4.5,0); 
\draw[line width=.1em,red!80!black!70,line cap=round] (1,0) circle(.55em); 
\draw[line width=.1em,blue!80!black!70,line cap=round] (-1,0) circle(.55em); 
\node[font=\scriptsize] at (3,-1.2) {$[14] [58]$}; 
\end{scope} 
\begin{scope}[xshift=24em,yshift=-4em] 
\draw[line width=0,fill=black!15,black!15] (0,1) circle(.35em); 
\draw[line width=0,fill=black!15,black!15] (5,0) circle(.35em); 
\draw[line width=.4pt] (-2,1) circle(.25ex); 
\draw[line width=.4pt,fill=black] (-1,0) circle(.25ex); 
\draw[line width=.4pt,fill=black]  (0,1) circle(.25ex); 
\draw[line width=.4pt]  (0,1) circle(.5ex); 
\draw[line width=.4pt,fill=black]  (1,0) circle(.25ex); 
\draw[line width=.4pt]  (2,1) circle(.25ex); 
\draw[line width=.4pt]  (3,0) circle(.25ex); 
\draw[line width=.4pt]  (4,1) circle(.25ex); 
\draw[line width=.4pt]  (4,1) circle(.5ex); 
\draw[line width=.4pt]  (5,0) circle(.25ex); 
\draw[line width=.4pt]  (6,1) circle(.25ex); 
\draw[line width=.4pt]  (7,0) circle(.25ex); 
\draw[line width=.1em,red!80!black!70, line cap=round] (-2.5,1.55) -- (-2,1.55) arc[start angle=90, end angle=45, radius=.55em] -- ++(1,-1) arc[start angle=45, end angle=-90, radius=.55em] -- ++(-1.5,0); 
\draw[line width=.1em,blue!80!black!70,line cap=round] (7.5,1.55) -- (4,1.55) arc[start angle=90, end angle=135, radius=.55em] -- ++(-1,-1) arc[start angle=135, end angle=270, radius=.55em] -- ++(4.5,0); 
\draw[line width=.1em,red!80!black!70,line cap=round] (2,1) circle(.55em); 
\draw[line width=.1em,blue!80!black!70,line cap=round] (0,1) circle(.55em); 
\node[font=\scriptsize] at (3,-1.2) {$[16] [25]$}; 
\end{scope} 
\begin{scope}[xshift=24em,yshift=-8em] 
\draw[line width=0,fill=black!15,black!15] (1,0) circle(.35em); 
\draw[line width=0,fill=black!15,black!15] (6,1) circle(.35em); 
\draw[line width=.4pt] (-2,1) circle(.25ex); 
\draw[line width=.4pt,fill=black] (-1,0) circle(.25ex); 
\draw[line width=.4pt,fill=black]  (0,1) circle(.25ex); 
\draw[line width=.4pt,fill=black]  (1,0) circle(.25ex); 
\draw[line width=.4pt]  (1,0) circle(.5ex); 
\draw[line width=.4pt]  (2,1) circle(.25ex); 
\draw[line width=.4pt]  (3,0) circle(.25ex); 
\draw[line width=.4pt]  (4,1) circle(.25ex); 
\draw[line width=.4pt]  (5,0) circle(.25ex); 
\draw[line width=.4pt]  (5,0) circle(.5ex); 
\draw[line width=.4pt]  (6,1) circle(.25ex); 
\draw[line width=.4pt]  (7,0) circle(.25ex); 
\draw[line width=.1em,red!80!black!70, line cap=round] (-2.5,1.55) -- (0,1.55) arc[start angle=90, end angle=-45, radius=.55em] -- ++(-1,-1) arc[start angle=-45, end angle=-90, radius=.55em] -- ++(-1.5,0); 
\draw[line width=.1em,blue!80!black!70,line cap=round] (7.5,1.55) -- (4,1.55) arc[start angle=90, end angle=225, radius=.55em] -- ++(1,-1) arc[start angle=225, end angle=270, radius=.55em] -- ++(2.5,0); 
\draw[line width=.1em,blue!80!black!70,line cap=round] (1,0) circle(.55em); 
\draw[line width=.1em,red!80!black!70,line cap=round] (3,0) circle(.55em); 
\node[font=\scriptsize] at (3,-1.2) {$[36] [27]$}; 
\end{scope} 
\begin{scope}[xshift=24em,yshift=-12em] 
\draw[line width=0,fill=black!15,black!15] (2,1) circle(.35em); 
\draw[line width=0,fill=black!15,black!15] (7,0) circle(.35em); 
\draw[line width=.4pt] (-2,1) circle(.25ex); 
\draw[line width=.4pt,fill=black] (-1,0) circle(.25ex); 
\draw[line width=.4pt,fill=black]  (0,1) circle(.25ex); 
\draw[line width=.4pt,fill=black]  (1,0) circle(.25ex); 
\draw[line width=.4pt]  (2,1) circle(.25ex); 
\draw[line width=.4pt]  (2,1) circle(.5ex); 
\draw[line width=.4pt]  (3,0) circle(.25ex); 
\draw[line width=.4pt]  (4,1) circle(.25ex); 
\draw[line width=.4pt]  (5,0) circle(.25ex); 
\draw[line width=.4pt]  (6,1) circle(.25ex); 
\draw[line width=.4pt]  (6,1) circle(.5ex); 
\draw[line width=.4pt]  (7,0) circle(.25ex); 
\draw[line width=.1em,red!80!black!70, line cap=round] (-2.5,1.55) -- (0,1.55) arc[start angle=90, end angle=45, radius=.55em] -- ++(1,-1) arc[start angle=45, end angle=-90, radius=.55em] -- ++(-3.5,0); 
\draw[line width=.1em,blue!80!black!70,line cap=round] (7.5,1.55) -- (6,1.55) arc[start angle=90, end angle=135, radius=.55em] -- ++(-1,-1) arc[start angle=135, end angle=270, radius=.55em] -- ++(2.5,0); 
\draw[line width=.1em,red!80!black!70,line cap=round] (4,1) circle(.55em); 
\draw[line width=.1em,blue!80!black!70,line cap=round] (2,1) circle(.55em); 
\node[font=\scriptsize] at (3,-1.2) {$[38] [47]$}; 
\end{scope} 
\end{tikzpicture} 
\end{center} 
\caption{The twelve 2-intermediate $t$-structures of $\dq{A_2}$, their hearts (shaded), projective objects (circled vertices) and their contribution to the scattering amplitude} 
\label{hearts2X} 
\end{figure}
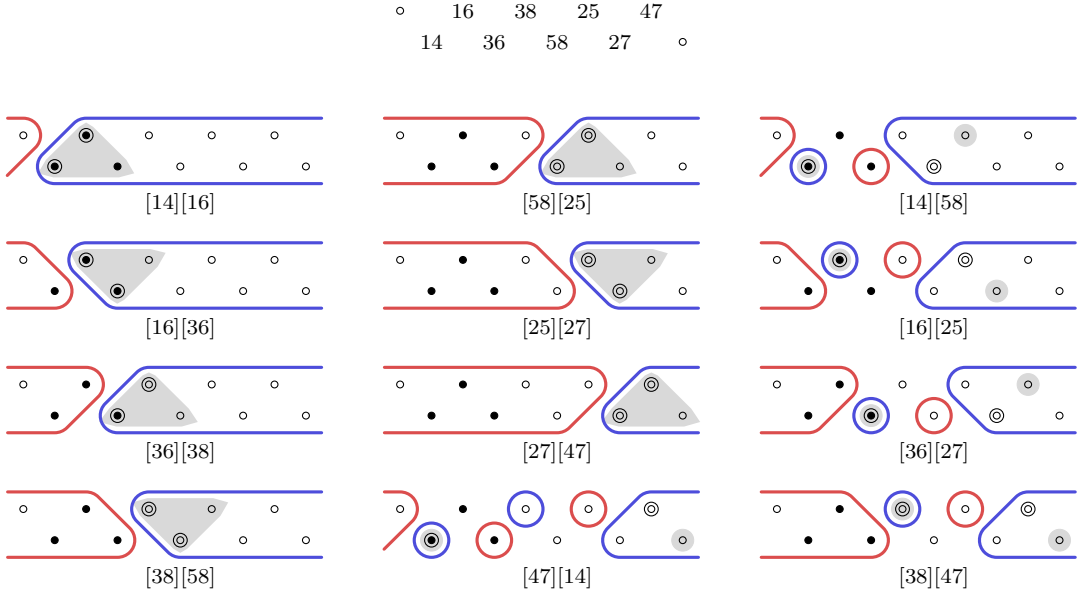 
\begin{example}[Quintic theory] 
\label{ex:quintic}
The first non-trivial case in a quintic theory with potential 
$V(\phi)=\lambda_5\phi^5$
is $N=11$ with $n=3$ and $\Lambda=(3,3,3)$. The mesh for this case is
\begin{equation}  
\label{n3m3} 
\begin{tikzpicture}
\matrix (m) [matrix of math nodes, minimum width=1.4em, row 
sep={3.6em,between origins}, text height=1.4ex, column sep={2.6em,between 
origins}, text depth=.15ex, font=\scriptsize,ampersand replacement=\&] 
{ 
\& (1,8) \&\& (4,11) \&\& (3,7) 
\&\& (6,10) \&\& 
(2,9) \&\& (1,5) \&\&\\ 
(1,5) \&\& (4,8) \&\& (7,11) \&\& (3,10)  
\&\& (2,6) \&\& (5,9) \&\& (1,8) \&\&\\ 
}; 
\foreach \i/\j in {1/2,3/4,5/6,7/8,9/10,11/12} 
\draw[->] (m-2-\i)--(m-1-\j); 
%
\foreach \i/\j in {2/3,4/5,6/7,8/9,10/11,12/13} 
\draw[->] (m-1-\i)--(m-2-\j); 
%
\end{tikzpicture} 
\end{equation} 
matching the $3$-cluster category  $\C^3_{A_2}$ in \eq{A2:C3}.
The scattering amplitude is given as $22$ terms obtained in terms 
of the projectives marked in \figref{hearts23}
using the correspondence between planar variables in \eq{n3m3} and the
representations in \eq{A2:C3} as
\begin{multline}
S_{11}(\phi^5)=
\frac{1}{X_{1,5}X_{1,8}}+\frac{1}{X_{1,8}X_{4,8}}+\frac{1}{X_{4,8}X_{4,11}}
+\frac{1}{X_{4,11}X_{7,11}}+\frac{1}{X_{7,11}X_{3,7}}
+\frac{1}{X_{3,7}X_{3,10}}\\
+\frac{1}{X_{3,10}X_{6,10}}+\frac{1}{X_{6,10}X_{2,6}}
+\frac{1}{X_{2,6}X_{2,9}}+\frac{1}{X_{2,9}X_{5,9}}+\frac{1}{X_{1,5}X_{7,11}}
+\frac{1}{X_{1,8}X_{3,7}}\\
+\frac{1}{X_{4,8}X_{3,10}}+\frac{1}{X_{4,11}X_{6,10}}
+\frac{1}{X_{7,10}X_{2,6}}+\frac{1}{X_{3,7}X_{2,9}}
+\frac{1}{X_{3,10}X_{5,9}}+\frac{1}{X_{1,5}X_{6,10}}\\
+\frac{1}{X_{1,8}X_{2,6}}+\frac{1}{X_{4,8}X_{2,9}}+
\frac{1}{X_{4,11}X_{5,9}}+\frac{1}{X_{1,5}X_{5,9}}.
\end{multline} 
\end{example}
\subsection{General potentials} 
As mentioned earlier, the description of 
the category arising in the computation of  
scattering amplitudes 
for theories with general polynomial  potentials \eq{vphi} with more than 
one terms is more involved. In terms of planar
polygons the angulations are now effected by smaller polygons of different
number of edges at once, specifically, with $n_i$  number of of $m_i$-gons.
We refer to it as a mixed angulation. 
This makes the enumeration of angulations cumbersome. While still less
straightforward than the previous cases with equal $m_i$'s, 
we present a method to write the
scattering amplitudes using the category theoretic paraphernalia
which can be developed into a computer code. 


The procedure to obtain the mesh diagram described in \eq{mesh:hop}
and exemplified in \eq{ex:mesh} using the partition $\Lambda$ yields the
pseudo-periodic category described in section~\ref{dercat:orb}.
For an $N$-particle scattering with potential \eq{vphi} the pseudo-periodic
category is modelled on the periodic category ${\mathcal P}^N_{A_{n-1}}$,
with $n$ defined in \eq{ni}. 
We continue, however, provisionally
using the category-theoretic parlance in the sequel.

In order to obtain the projectives of hearts of $t$-structures as in the
previous cases, we start with the cluster category of an $A_{N-3}$ quiver,
which corresponds to the triangulations of an $N$-gon as in the $\phi^3$
theory \cite{br}. The diagonals corresponding to the triangulations are given
by the projectives of hearts of intermediate $t$-structures of this cluster
category. We first observe that the set of projectives for a given
intermediate $t$-structure can be of two types. The first type is a set 
connected in the AR quiver in the sense that it directly furnishes a copy of the
$A_{N-3}$ quiver, with the projectives graded by their positions in the
quiver. The other type of sets are  disconnected in this sense. However, the projectives
in this case can be graded by noting that they can be obtained from the
connected ones by successively replacing the projectives, one each time,
with the substitute bestowed with the grade from the predecessor. 

We write the projectives for each intermediate $t$-structure ordered by the
grading. To each end of an ordered set we then append a node corresponding
to the null entries $X_{i,i+1}$ and $X_{1N}$. We then use all the
permutations of the partition
$\Lambda$ to choose projectives from each of these ordered sets. These give
rise to the projectives of hearts of intermediate $t$-structure of
the pseudo-periodic category.
\begin{example}[$(3,4)$ theory]
Triangulations and mesh diagram for this has been discussed in
Example~\ref{ex:34}. 
For $N=6$, the mesh diagram \eq{mesh:34-6} is a pseudo-periodic category
modelled on its periodic counterpart
$\P^6_{A_2}$ given  in \eq{A2:P6}. The entries $X_{1,3}$, $X_{1,4}$, $X_{1,5}$
$X_{2,4}$, $X_{2,6}$, $X_{3,5}$, $X_{3,6}$ are repeated at intermediate
steps.
We need to list the projectives of hearts
of the intermediate $t$-structures. To this end, as described above, we start
from the cluster category of the $A_{N-3}=A_3$ quiver. It corresponds to
the triangulations of a hexagon. 
The projectives in the present case are obtained by deleting a subset from
each set of this bigger quiver. 
The cluster category of the $A_3$ quiver \eq{A3:C1} written in terms of
$X_{i,j}$ is 
\begin{equation}  
\label{eq:34-6} 
\begin{tikzpicture}[baseline=-4pt] 
\matrix (m) [matrix of math nodes, row sep={2em,between origins}, text 
height=1.4ex,  
column sep={1.62em,between origins}, text depth=.15ex, font=\scriptsize, 
ampersand replacement=\&]  
{ 
\& \& 1,5 \& \& 2,6 \& \& 1,3 \& \& \\
\& 1,4 \& \& 2,5 \& \& 3,6 \& \& 1,4 \&\\
1,3 \& \& 2,4 \& \& 3,5 \& \& 4,6 \& \& 1,5\\
}; 
\end{tikzpicture}
\end{equation} 
The intermediate $t$-structures, their hearts and projectives are given in
Fig~\ref{hearts3}.
Let us consider the set of projectives in Fig~\ref{hearts3}. The first
one furnishes the set $[13]-[14]-[15]$. This is a copy of the $A_3$ quiver,
thus a connected one. We append the null entries to obtain
$\circ-[13]-[14]-[15]-\circ$. For this case $\Lambda=(1,1,2)$. We choose
subsets using the three permutations of $\Lambda$ as
\begin{equation}
\begin{tikzpicture}[baseline=-4pt] 
\matrix (m) [matrix of math nodes, row sep={3em,between origins}, text 
height=1.4ex,  
column sep={1.62em,between origins}, text depth=.15ex, font=\scriptsize, 
ampersand replacement=\&]  
{ 
\circ\&\& 1,3\&\& 1,4 \&\& 1,5\&\&\circ\\
};
\draw[->,blue,bend left=20] (m-1-1) to  (m-1-3);
\draw[->,blue,bend left=20] (m-1-3) to  (m-1-5);
\draw[->,blue,bend left=20] (m-1-5) to  (m-1-9);
\draw[->,red,bend left=40] (m-1-1) to  (m-1-3);
\draw[->,red,bend left=40] (m-1-3) to  (m-1-7);
\draw[->,red,bend left=40] (m-1-7) to  (m-1-9);
\draw[->,green!70!black,bend right=40] (m-1-1) to  (m-1-5);
\draw[->,green!70!black,bend right=40] (m-1-5) to  (m-1-7);
\draw[->,green!70!black,bend right=40] (m-1-7) to  (m-1-9);
\end{tikzpicture}
\end{equation} 
Thus, we have three sets of projectives $\{[1,3],[1,4]\}$, 
$\{[1,3],[1,5]\}$ and $\{[1,4],[1,5]\}$ arising from this. They contribute 
$\tfrac{1}{X_{1,3}X_{1,4}}$, $\tfrac{1}{X_{1,3}X_{1,5}}$ and 
$\tfrac{1}{X_{1,4}X_{1,5}}$, respectively, to the scattering amplitude. 
The first eight cases can be similarly analysed. Let us consider the ninth
one in Fig.~\ref{hearts3}. The projectives $[1,3]-[1,4]-[4,6]$ may be obtained
from the first one by replacing $[1,5]$ by $[4,6]$. We thus have the grading
$\circ-[1,3]-[1,4]-[4,6]-\circ$, where we have also appended the null
entries. From it using the permutations of $\Lambda$ to select projectives
we obtain 
\begin{equation}
\begin{tikzpicture}[baseline=-4pt] 
\matrix (m) [matrix of math nodes, row sep={3em,between origins}, text 
height=1.4ex,  
column sep={1.62em,between origins}, text depth=.15ex, font=\scriptsize, 
ampersand replacement=\&]  
{ 
\circ\&\& 1,3\&\& 1,4 \&\& 4,6\&\&\circ\\
};
\draw[->,blue,bend left=20] (m-1-1) to  (m-1-3);
\draw[->,blue,bend left=20] (m-1-3) to  (m-1-5);
\draw[->,blue,bend left=20] (m-1-5) to  (m-1-9);
\draw[->,red,bend left=40] (m-1-1) to  (m-1-3);
\draw[->,red,bend left=40] (m-1-3) to  (m-1-7);
\draw[->,red,bend left=40] (m-1-7) to  (m-1-9);
\draw[->,green!70!black,bend right=40] (m-1-1) to  (m-1-5);
\draw[->,green!70!black,bend right=40] (m-1-5) to  (m-1-7);
\draw[->,green!70!black,bend right=40] (m-1-7) to  (m-1-9);
\end{tikzpicture}
\end{equation} 
This furnishes two further sets, namely, $[1,3],[4,6]$ and $[1,4],[4,6]$
of projectives and hence two more terms to the scattering amplitude. The
repeated set $[1,3],[1,4]$ is not to be counted again. 
As a third example, let us consider the thirteenth diagram in
Fig.~\ref{hearts3}. The set of projectives $\circ-[1,3]-[3,5]-[1,5]-\circ$ is
obtained again from the first one, by replacing $[1,4]$ by $[3,5]$.
The choice of projectives for the pseudo-periodic quiver is 
\begin{equation}
\begin{tikzpicture}[baseline=-4pt] 
\matrix (m) [matrix of math nodes, row sep={3em,between origins}, text 
height=1.4ex,  
column sep={1.62em,between origins}, text depth=.15ex, font=\scriptsize, 
ampersand replacement=\&]  
{ 
\circ\&\& 1,3\&\& 3,5 \&\& 1,5\&\&\circ\\
};
\draw[->,blue,bend left=20] (m-1-1) to  (m-1-3);
\draw[->,blue,bend left=20] (m-1-3) to  (m-1-5);
\draw[->,blue,bend left=20] (m-1-5) to  (m-1-9);
\draw[->,red,bend left=40] (m-1-1) to  (m-1-3);
\draw[->,red,bend left=40] (m-1-3) to  (m-1-7);
\draw[->,red,bend left=40] (m-1-7) to  (m-1-9);
\draw[->,green!70!black,bend right=40] (m-1-1) to  (m-1-5);
\draw[->,green!70!black,bend right=40] (m-1-5) to  (m-1-7);
\draw[->,green!70!black,bend right=40] (m-1-7) to  (m-1-9);
\end{tikzpicture}
\end{equation} 
This yields yet two more sets of projectives, namely, $[1,3],[3,5]$ and
$[3,5],[1,5]$.
Proceeding in this
manner we obtain all the sets of projectives of hearts of intermediate
$t$-structures of the pseudo-periodic orbit category. 
The resulting scattering amplitude is \cite{ajk}
\begin{multline}
S_6(\phi^3+\phi^4) = \frac{1}{X_{1,3}X_{1,4}}+
\frac{1}{X_{1,3}X_{1,5}}+ \frac{1}{X_{1,3}X_{3,5}}+ \frac{1}{X_{1,3}X_{3,6}}+
\frac{1}{X_{1,3}X_{4,6}}+ \frac{1}{X_{1,4}X_{1,5}}\\
+\frac{1}{X_{1,4}X_{2,4}}+ \frac{1}{X_{1,4}X_{4,6}}+
\frac{1}{X_{1,5}X_{2,4}}+ \frac{1}{X_{1,5}X_{2,5}}+ \frac{1}{X_{1,5}X_{3,5}}+
\frac{1}{X_{2,4}X_{2,5}}\\
+\frac{1}{X_{2,4}X_{2,6}} + \frac{1}{X_{2,4}X_{4,6}}+
\frac{1}{X_{2,5}X_{2,6}}  + \frac{1}{X_{2,5}X_{3,5}} +
\frac{1}{X_{2,6}X_{3,5}} + \frac{1}{X_{2,6}X_{3,6}}\\
+ \frac{1}{X_{2,6}X_{4,6}} + \frac{1}{X_{3,5}X_{3,6}} +
\frac{1}{X_{3,6}X_{4,6}}.
\end{multline}

Two cases arise for $N=7$.
The mesh diagrams \eq{mesh:34-71} and \eq{mesh:34-72} 
are the pseudo-periodic categories 
corresponding to the periodic categories 
$\P^7_{A_2}$ given in \eq{A2:P7} and 
$\P^7_{A_3}$, given in \eq{A3:P7}, respectively.
The mesh diagram for $N=7$ with $\Lambda=(1,2,2)$ is  
\begin{equation}  
\label{mesh:34-71} 
\begin{tikzpicture}[x=1.4em] 
\node at (0,1) (a1) {$\scriptstyle 1,5$}; 
\node at (2,1) (a2) {$\scriptstyle 2,7$}; 
\node at  (4,1) (a3) {$\scriptstyle 1,4$}; 
\node at  (6,1) (a4) {$\scriptstyle 3,6$}; 
\node at (8,1)  (a5) {$\scriptstyle 5,7$}; 
\node at (10,1)  (a6) {$\scriptstyle 2,6$}; 
\node at (12,1)  (a7) {$\scriptstyle 1,4$}; 
\node at (14,1)  (a8) {$\scriptstyle 3,5$}; 
\node at (16,1)  (a9) {$\scriptstyle 4,7$}; 
\node at (18,1)  (a10) {$\scriptstyle 2,6$}; 
\node at (20,1)  (a11) {$\scriptstyle 1,3$}; 
\node at  (-1,0) (b1) {$\scriptstyle 1,3$}; 
\node at (1,0) (b2) {$\scriptstyle 2,5$}; 
\node at (3,0) (b3) {$\scriptstyle 4,7$}; 
\node at  (5,0)  (b4) {$\scriptstyle 1,6$}; 
\node at (7,0) (b5) {$\scriptstyle 3,7$}; 
\node at  (9,0) (b6) {$\scriptstyle 2,5$}; 
\node at (11,0) (b7) {$\scriptstyle 4,6$}; 
\node at (13,0) (b8) {$\scriptstyle 1,5$}; 
\node at  (15,0)  (b9) {$\scriptstyle 3,7$}; 
\node at (17,0) (b10) {$\scriptstyle 2,4$}; 
\node at (19,0) (b11) {$\scriptstyle 3,6$}; 
\node at (21,0) (b12) {$\scriptstyle 1,5$}; 
\draw[->] (b1) -- (a1); 
\draw[->] (b2) -- (a2); 
\draw[->] (b3) -- (a3); 
\draw[->] (b4) -- (a4); 
\draw[->] (b5) -- (a5); 
\draw[->] (b6) -- (a6); 
\draw[->] (b7) -- (a7); 
\draw[->] (b8) -- (a8); 
\draw[->] (b9) -- (a9); 
\draw[->] (b10) -- (a10); 
\draw[->] (b11) -- (a11); 
\draw[->] (a1) -- (b2) ; 
\draw[->] (a2) -- (b3); 
\draw[->] (a3) -- (b4); 
\draw[->] (a4) -- (b5); 
\draw[->] (a5) -- (b6); 
\draw[->] (a6) -- (b7); 
\draw[->] (a7) -- (b8); 
\draw[->] (a8) -- (b9); 
\draw[->] (a9) -- (b10); 
\draw[->] (a10) -- (b11); 
\draw[->] (a11) -- (b12); 
\end{tikzpicture} 
\end{equation} 
The scattering amplitude is obtained by starting with the cluster category of
an $A_4$ quiver. It has terms with two $X$'s in the denominator, for example,
$\tfrac{1}{X_{1,3}X_{1,5}}$.
\begin{equation}
\end{equation}
The other one for $N=7$ for $\Lambda=(1,1,1,2)$ is  
\begin{equation}  
\label{mesh:34-72} 
\begin{tikzpicture}[x=1.2em] 
\node at (1,2) (c1) {$\scriptstyle 1,5$}; 
\node at (3,2) (c2) {$\scriptstyle 2,7$}; 
\node at  (5,2) (c3) {$\scriptstyle 1,3$}; 
\node at  (7,2) (c4) {$\scriptstyle 2,4$}; 
\node at (9,2)  (c5) {$\scriptstyle 3,6$}; 
\node at (11,2)  (c6) {{$\scriptstyle 5,7$}}; 
\node at (13,2)  (c7) {\eqmakebox[1,8]{$\scriptstyle 1,6$}}; 
\node at (15,2)  (c8) {\eqmakebox[(,8]{$\scriptstyle 2,7$}}; 
\node at (17,2)  (c9) {\eqmakebox[1,8]{$\scriptstyle 1,4$}}; 
\node at (19,2)  (c10) {\eqmakebox[1,8]{$\scriptstyle 3,5$}}; 
\node at (21,2)  (c11) {\eqmakebox[1,8]{$\scriptstyle 4,6$}}; 
\node at (23,2)  (c12) {\eqmakebox[1,8]{$\scriptstyle 5,7$}}; 
\node at (25,2)  (c13) {\eqmakebox[1,8]{$\scriptstyle 2,6$}}; 
\node at (27,2)  (c14) {\eqmakebox[1,8]{$\scriptstyle 1,3$}}; 
\node at (0,1)  (b1) {$\scriptstyle 1,4$}; 
\node at (2,1)  (b2) {$\scriptstyle 2,5$}; 
\node at (4,1)  (b3) {$\scriptstyle 3,7$}; 
\node at (6,1)  (b4) {$\scriptstyle 1,4$}; 
\node at (8,1)  (b5) {$\scriptstyle 2,6$}; 
\node at (10,1) (b6) {$\scriptstyle 3,7$}; 
\node at (12,1) (b7) {$\scriptstyle 1,5$}; 
\node at (14,1) (b8) {$\scriptstyle 2,6$}; 
\node at (16,1) (b9) {$\scriptstyle 4,7$}; 
\node at (18,1) (b10) {$\scriptstyle 1,5$}; 
\node at (20,1) (b11) {$\scriptstyle 3,6$}; 
\node at (22,1) (b12) {$\scriptstyle 4,7$}; 
\node at (24,1) (b13) {$\scriptstyle 2,5$}; 
\node at (26,1) (b14) {$\scriptstyle 3,6$}; 
\node at (28,1) (b15) {$\scriptstyle 1,4$}; 
\node at  (-1,0) (a1) {$\scriptstyle 1,3$}; 
\node at (1,0) (a2) {$\scriptstyle 2,4$}; 
\node at (3,0) (a3) {$\scriptstyle 3,5$}; 
\node at  (5,0) (a4) {$\scriptstyle 4,7$}; 
\node at (7,0) (a5) {$\scriptstyle 1,6$}; 
\node at (9,0) (a6) {$\scriptstyle 2,7$}; 
\node at (11,0) (a7) {$\scriptstyle 1,3$}; 
\node at (13,0) (a8) {$\scriptstyle 2,5$}; 
\node at (15,0) (a9) {$\scriptstyle 4,6$}; 
\node at (17,0) (a10) {$\scriptstyle 5,7$}; 
\node at (19,0) (a11) {$\scriptstyle 1,6$}; 
\node at (21,0) (a12) {$\scriptstyle 3,7$}; 
\node at (23,0) (a13) {$\scriptstyle 2,4$}; 
\node at (25,0) (a14) {$\scriptstyle 3,5$}; 
\node at (27,0) (a15) {$\scriptstyle 4,6$}; 
\node at (29,0) (a16) {$\scriptstyle 1,5$}; 
\draw[->] (a1) -- (b1); 
\draw[->] (a2) -- (b2); 
\draw[->] (a3) -- (b3); 
\draw[->] (a4) -- (b4); 
\draw[->] (a5) -- (b5); 
\draw[->] (a6) -- (b6); 
\draw[->] (a7) -- (b7); 
\draw[->] (a8) -- (b8); 
\draw[->] (a9) -- (b9); 
\draw[->] (a10) -- (b10); 
\draw[->] (a11) -- (b11); 
\draw[->] (a12) -- (b12); 
\draw[->] (a13) -- (b13); 
\draw[->] (a14) -- (b14); 
\draw[->] (a15) -- (b15); 
\draw[->] (b1) -- (c1); 
\draw[->] (b2) -- (c2); 
\draw[->] (b3) -- (c3); 
\draw[->] (b4) -- (c4); 
\draw[->] (b5) -- (c5); 
\draw[->] (b6) -- (c6); 
\draw[->] (b7) -- (c7); 
\draw[->] (b8) -- (c8); 
\draw[->] (b9) -- (c9); 
\draw[->] (b10) -- (c10); 
\draw[->] (b11) -- (c11); 
\draw[->] (b12) -- (c12); 
\draw[->] (b13) -- (c13); 
\draw[->] (b14) -- (c14); 
\draw[->] (b1) -- (a2); 
\draw[->] (b2) -- (a3); 
\draw[->] (b3) -- (a4); 
\draw[->] (b4) -- (a5); 
\draw[->] (b5) -- (a6); 
\draw[->] (b6) -- (a7); 
\draw[->] (b7) -- (a8); 
\draw[->] (b8) -- (a9); 
\draw[->] (b9) -- (a10); 
\draw[->] (b10) -- (a11); 
\draw[->] (b11) -- (a12); 
\draw[->] (b12) -- (a13); 
\draw[->] (b13) -- (a14); 
\draw[->] (b14) -- (a15); 
\draw[->] (b15) -- (a16); 
\draw[->] (c1) -- (b2); 
\draw[->] (c2) -- (b3); 
\draw[->] (c3) -- (b4); 
\draw[->] (c4) -- (b5); 
\draw[->] (c5) -- (b6); 
\draw[->] (c6) -- (b7); 
\draw[->] (c7) -- (b8); 
\draw[->] (c8) -- (b9); 
\draw[->] (c9) -- (b10); 
\draw[->] (c10) -- (b11); 
\draw[->] (c11) -- (b12); 
\draw[->] (c12) -- (b13); 
\draw[->] (c13) -- (b14); 
\draw[->] (c14) -- (b15); 
\end{tikzpicture} 
\end{equation} 
The scattering amplitude with terms having three $X$'s in the denominator as
$\tfrac{1}{X_{1,3}X_{1,4}X_{1,5}}$
is obtained once again by starting with the cluster category of
an $A_4$ quiver, but this time projectives are chosen with permutations
of $\Lambda=(1,1,1,2)$.
\end{example}
\begin{example}[$(4,5)$ theory]
The scattering amplitude for the theory with potential
$V(\phi)=\lambda_4\phi^4+\lambda_5\phi^5$ is obtained similarly. For the
simplest non-trivial case
with $N=9$ and $\Lambda=(2,2,3)$, the mesh is the pseudo-periodic category
corresponding to \eq{A2:P7}, with $X$'s different from \eq{mesh:34-71}.
\begin{equation}
\label{mesh:45-9}
\begin{tikzpicture}[baseline=-4pt] 
\matrix (m) [matrix of math nodes, row sep={3em,between origins}, text 
height=1.4ex, text centered, 
column sep={1em,between origins}, text depth=.15ex, font=\scriptsize, 
ampersand replacement=\&]  
{ 
\& 1,6\&\& 3,9\&\& 2,5\&\& 4,8\&\& 1,7\&\& 3,9\&\&
2,6\&\&5,8\&\&1,7\&\&4,9\&\&3,6\&\&5,8\&\&2,7\&\&1,4 \&\\
1,4\&\&3,6\&\& 5,9\&\&2,8\&\& 1,4\&\& 3,7\&\& 6,9\&\&
2,8\&\&1,5\&\&4,7\&\&6,9\&\&3,8\&\&2,5\&\&4,7\&\&1,6
\&\&\\
};
\foreach \i/\j in
{1/2,3/4,5/6,7/8,9/10,11/12,13/14,15/16,17/18,19/20,21/22,23/24,25/26,27/28} 
{ \draw[->] (m-2-\i) to (m-1-\j); }
\foreach \i/\j in
{2/3,4/5,6/7,8/9,10/11,12/13,14/15,16/17,18/19,20/21,22/23,24/25,26/27,28/29}  
{ \draw[->] (m-1-\i) to (m-2-\j); }
\end{tikzpicture}
\end{equation}
The scattering amplitude 
\begin{multline}
\label{S:45-9}
S_9(\phi^4+\phi^5) = 
\frac{1}{X_{1,4}X_{1,6}}+\frac{1}{X_{2,5}X_{2,7}}+\frac{1}{X_{3,6}X_{3,8}}+\frac{1}{X_{4,7}X_{4,9}}+\frac{1}{X_{5,8}X_{1,5}}+\frac{1}{X_{6,9}X_{2,6}}\\
+\frac{1}{X_{1,7}X_{3,7}}+\frac{1}{X_{2,8}X_{4,8}}+\frac{1}{X_{3,9}X_{5,9}}+\frac{1}{X_{1,4}X_{5,9}}+\frac{1}{X_{2,5}X_{1,6}}+\frac{1}{X_{3,6}X_{2,7}}\\
+\frac{1}{X_{4,7}X_{3,8}}+\frac{1}{X_{5,8}X_{4,9}}+\frac{1}{X_{6,9}X_{1,5}}+\frac{1}{X_{1,7}X_{2,6}}+\frac{1}{X_{2,8}X_{3,7}}+\frac{1}{X_{3,9}X_{4,8}}\\
+\frac{1}{X_{1,4}X_{4,8}}+\frac{1}{X_{2,5}X_{5,9}}+\frac{1}{X_{3,6}X_{1,6}}+\frac{1}{X_{4,7}X_{2,7}}+\frac{1}{X_{5,8}X_{3,8}}+\frac{1}{X_{6,9}X_{4,9}}\\
+\frac{1}{X_{1,7}X_{1,5}}+\frac{1}{X_{2,8}X_{2,6}}+\frac{1}{X_{3,9}X_{3,7}}+\frac{1}{X_{3,6}X_{3,9}}+\frac{1}{X_{4,7}X_{1,4}}+\frac{1}{X_{5,8}X_{2,5}}\\
+\frac{1}{X_{6,9}X_{3,6}}+\frac{1}{X_{1,7}X_{4,7}}+\frac{1}{X_{2,8}X_{5,8}}+\frac{1}{X_{3,9}X_{6,9}}+\frac{1}{X_{1,4}X_{1,7}}+\frac{1}{X_{2,5}X_{2,8}}\\
+\frac{1}{X_{3,6}X_{2,8}}+\frac{1}{X_{4,7}X_{3,9}}+\frac{1}{X_{5,8}X_{1,4}}+\frac{1}{X_{6,9}X_{2,5}}+\frac{1}{X_{1,7}X_{3,6}}+\frac{1}{X_{2,8}X_{4,7}}\\
+\frac{1}{X_{3,9}X_{5,8}}+\frac{1}{X_{1,4}X_{6,9}}+\frac{1}{X_{2,5}X_{1,7}}+\frac{1}{X_{3,6}X_{1,7}}+\frac{1}{X_{4,7}X_{2,8}}+\frac{1}{X_{5,8}X_{3,9}}\\
+\frac{1}{X_{6,9}X_{1,4}}+\frac{1}{X_{1,7}X_{2,5}}+\frac{1}{X_{2,8}X_{3,6}}+\frac{1}{X_{3,9}X_{4,7}}+\frac{1}{X_{1,4}X_{5,8}}+\frac{1}{X_{2,5}X_{6,9}}
\end{multline}
is obtained by starting with the projectives of the
hearts of intermediate $t$-structures of an $A_6$ quiver and selecting
subsets using $\Lambda$.

For $N=12$ and Feynman diagrams with $n=4$ vertices, two quartic and two
quintic, the mesh is the pseudo-periodic category \eq{m2233} corresponding
to \eq{A3:P12}. The scattering amplitude can be obtained starting with 
the cluster category of an $A_9$ quiver.
\end{example}
As the last example, let us consider scattering with a potential with three
terms. 
\begin{example}[$(3,4,5)$ theory] 
For $V(\phi)=\lambda_3\phi^3+\lambda_4\phi^4+\lambda_5\phi^5$ the mesh for
$N=8$ and three vertices, one each
of cubic, quartic and quintic ones, the mesh is
the pseudo-periodic category  
\begin{equation} 
\label{A21} 
\begin{tikzpicture}[baseline=-4pt] 
\matrix (m) [matrix of math nodes, minimum width=1.4em, row 
sep={3.6em,between origins}, text height=1.4ex, column sep={1.1em,between 
origins}, text depth=.15ex, ampersand replacement=\&,font=\scriptsize] 
{ 
\& 1,5 \&\&   2,8 \&\&   1,4 
\&\&   3,7 \&\&   6,8  \&\&   2,7 
\&\&   1,5  \&\&   4,6  \&\&   5,8  
\&\&   3,7  \&\&   2,4 \&\&   3,6  
\&\&   1,5   
\\ 
1,3 \&\&   2,5 \&\&   4,8  
\&\&   1,7 \&\&   3,8  \&\&   2,6 
\&\&   5,7  \&\&   1,6  \&\&   4,8  
\&\&   3,5  \&\&   4,7 \&\&   2,6  
\&\&   1,3 
\\ 
}; 
\draw[->] (m-2-1)--(m-1-2)  ;
\draw[->] (m-1-4)--(m-2-5) ;
\draw[->] (m-2-7)--(m-1-8) ;
\draw[->] (m-1-10)--(m-2-11); 
\draw[->] (m-2-13)--(m-1-14) ;
\draw[->] (m-1-16)--(m-2-17) ;
\draw[->] (m-2-19)--(m-1-20) ;
\draw[->] (m-1-22)--(m-2-23) ;
\draw[->] (m-2-25)--(m-1-26) ;
\draw[->] (m-2-3)--(m-1-4) ;
\draw[->] (m-1-6)--(m-2-7); 
\draw[->] (m-2-9)--(m-1-10); 
\draw[->] (m-1-12)--(m-2-13); 
\draw[->] (m-2-15)--(m-1-16); 
\draw[->] (m-1-18)--(m-2-19); 
\draw[->] (m-2-21)--(m-1-22); 
\draw[->] (m-1-24)--(m-2-25); 
\draw[->] (m-1-2)--(m-2-3); 
\draw[->] (m-2-5)--(m-1-6); 
\draw[->] (m-1-8)--(m-2-9); 
\draw[->] (m-2-11)--(m-1-12); 
\draw[->] (m-1-14)--(m-2-15); 
\draw[->] (m-2-17)--(m-1-18); 
\draw[->] (m-1-20)--(m-2-21); 
\draw[->] (m-2-23)--(m-1-24); 
\end{tikzpicture} 
\end{equation} 
The scattering amplitude
\begin{multline}
\label{eq:123-8}
S_8(\phi^3+\phi^4+\phi^5) = \frac{1}{X_{1,3}X_{1,5}} +\frac{1}{X_{2,4}X_{2,6}}+\frac{1}{X_{3,5}X_{3,7}}+\frac{1}{X_{4,6}X_{4,8}}+\frac{1}{X_{5,7}X_{1,5}}+\frac{1}{X_{6,8}X_{2,6}}\\
+\frac{1}{X_{1,7}X_{3,7}} +\frac{1}{X_{2,8}X_{4,8}}+\frac{1}{X_{1,3}X_{4,8}}+\frac{1}{X_{2,4}X_{1,5}}+\frac{1}{X_{3,5}X_{2,6}}+\frac{1}{X_{4,6}X_{3,7}}\\
+\frac{1}{X_{5,7}X_{4,8}} +\frac{1}{X_{6,8}X_{1,5}}+\frac{1}{X_{1,7}X_{2,6}}+\frac{1}{X_{2,8}X_{3,7}}+\frac{1}{X_{1,3}X_{3,7}}+\frac{1}{X_{2,4}X_{4,8}}\\
+\frac{1}{X_{3,5}X_{1,5}} +\frac{1}{X_{4,6}X_{2,6}}+\frac{1}{X_{5,7}X_{3,7}}+\frac{1}{X_{6,8}X_{4,8}}+\frac{1}{X_{1,7}X_{1,5}}+\frac{1}{X_{2,8}X_{2,6}}\\
+\frac{1}{X_{2,5}X_{2,8}} +\frac{1}{X_{3,6}X_{1,3}}+\frac{1}{X_{4,7}X_{2,4}}+\frac{1}{X_{5,8}X_{3,5}}+\frac{1}{X_{1,6}X_{4,6}}+\frac{1}{X_{2,7}X_{5,7}}\\
+\frac{1}{X_{3,8}X_{6,8}} +\frac{1}{X_{1,4}X_{1,7}}+\frac{1}{X_{2,5}X_{1,7}}+\frac{1}{X_{3,6}X_{2,8}}+\frac{1}{X_{4,7}X_{1,3}}+\frac{1}{X_{5,8}X_{2,4}}\\
+\frac{1}{X_{1,6}X_{3,5}} +\frac{1}{X_{2,7}X_{4,6}}+\frac{1}{X_{3,8}X_{5,7}}+\frac{1}{X_{1,4}X_{6,8}}+\frac{1}{X_{2,5}X_{6,8}}+\frac{1}{X_{3,6}X_{1,7}}\\
+\frac{1}{X_{4,7}X_{2,8}} +\frac{1}{X_{5,8}X_{1,3}}+\frac{1}{X_{1,6}X_{2,4}}+\frac{1}{X_{2,7}X_{3,5}}+\frac{1}{X_{3,8}X_{4,6}}+\frac{1}{X_{1,4}X_{5,7}}\\
+\frac{1}{X_{2,5}X_{5,7}} +\frac{1}{X_{3,6}X_{6,8}}+\frac{1}{X_{4,7}X_{1,7}}+\frac{1}{X_{5,8}X_{2,8}}+\frac{1}{X_{1,6}X_{1,3}}+\frac{1}{X_{2,7}X_{2,4}}\\
+\frac{1}{X_{3,8}X_{3,5}} +\frac{1}{X_{1,4}X_{4,6}}+\frac{1}{X_{4,8}X_{1,4}}+\frac{1}{X_{1,5}X_{2,5}}+\frac{1}{X_{2,6}X_{3,6}}+\frac{1}{X_{3,7}X_{4,7}}\\
+\frac{1}{X_{4,8}X_{5,8}} +\frac{1}{X_{1,5}X_{1,6}}+\frac{1}{X_{2,6}X_{2,7}}+\frac{1}{X_{3,7}X_{3,8}}+\frac{1}{X_{4,8}X_{3,8}}+\frac{1}{X_{1,5}X_{1,4}}\\
+\frac{1}{X_{2,6}X_{2,5}} +\frac{1}{X_{3,7}X_{3,6}}+\frac{1}{X_{4,8}X_{4,7}}+\frac{1}{X_{1,5}X_{5,8}}+\frac{1}{X_{2,6}X_{1,6}}+\frac{1}{X_{3,7}X_{2,7}}\\
\end{multline} 
is obtained from the cluster category of an $A_5$ quiver with the projectives
chosen using $\Lambda=(1,2,3)$.
\end{example}
\section{Summary}
To conclude, in this article we obtain a method to write down the scattering
amplitudes of scalar field theories with any polynomial potential \eq{vphi}
in a categorical framework, generalizing the case of cubic theory \cite{br}. 
One important advantage of the approach is in
bookkeeping and hence the algorithm presented can be developed into a code. 
The description is in terms of certain triangulated categories of
representations of quivers of type $A$. 
Planar Feynman diagrams of scalar theories with potential \eq{vphi} are dual
to subdivisions of poygons. The latter is known to be associated to (higher) 
cluster categories of quivers of type $A$. 
We show that the terms of the scattering 
amplitude of $N$ particles in a theory with a single-term potential 
$V(\phi)=\lambda\phi^{m+2}$ at an order $\lambda^n$ 
are given by the projectives of the hearts of the
$m$-intermediate $t$-structures of $m$-cluster category of an $A_{n-1}$
quiver. The $m$-cluster category arises as the mesh diagram obtained in terms
of the kinematic planar variables $X$.
For a generic potential \eq{vphi} the terms in the scattering
amplitude are given by a subset of the projectives of hearts of intermediate
$t$-structure of the cluster category of an $A_{N-3}$ quiver. The mesh
diagram then is the pseudo-periodic category modelled on the $N$-periodic 
category with extra intermediate identification of the planar variables. It
may be possible to view this as a quotient of the triangulated category of
the AR quiver by  a group of homomorphisms with more than one generators
\eq{Feq}.

\begin{landscape}
\section*{Appendix}
The category $\P^{12}_{A_3}$ is  
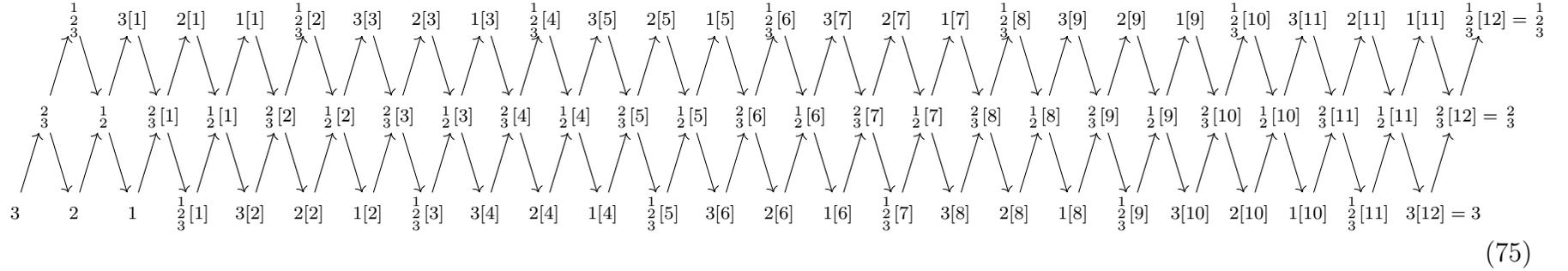
\begin{equation}  
\label{A3:P12} 
\kern-1cm
\begin{tikzpicture}[baseline=-4pt] 
\matrix (m) [matrix of math nodes, minimum width=1.4em, row 
sep={3.6em,between origins}, text height=1.4ex, column sep={1.1em,between 
origins}, text depth=.15ex, ampersand replacement=\&,font=\scriptsize] 
{ 
\&\&   \nodethree{1}{2}{3} \&\&   3[1] \&\&   2[1] 
\&\&   1[1] \&\&   \nodethree{1}{2}{3}[2] \&\&  3[3] 
\&\&   2[3] \&\&   1[3] \&\&   \nodethree{1}{2}{3}[4]
\&\&   3[5] \&\&   2[5] \&\&   1[5] 
\&\&   \nodethree{1}{2}{3}[6] \&\&  3[7] \&\&  2[7]  
\&\&   1[7] \&\& \nodethree{1}{2}{3}[8] \&\&  3[9]
\&\&   2[9] \&\&   1[9] \&\&   \nodethree{1}{2}{3}[10] 
\&\&   3[11] \&\&   2[11] \&\&   1[11] 
\&\&   \nodethree{1}{2}{3}[12]\mathrlap{{}=\nodethree{1}{2}{3}}\\ 
%
\& \nodetwo{2}{3} \&\&  \nodetwo{1}{2} 
\&\&  \nodetwo{2}{3}[1] \&\&  \nodetwo{1}{2}[1] 
\&\&  \nodetwo{2}{3}[2] \&\&  \nodetwo{1}{2}[2] 
\&\&  \nodetwo{2}{3}[3] \&\&  \nodetwo{1}{2}[3] 
\&\&  \nodetwo{2}{3}[4] \&\&  \nodetwo{1}{2}[4] 
\&\&  \nodetwo{2}{3}[5] \&\&  \nodetwo{1}{2}[5] 
\&\&  \nodetwo{2}{3}[6] \&\&  \nodetwo{1}{2}[6] 
\&\&  \nodetwo{2}{3}[7] \&\&  \nodetwo{1}{2}[7] 
\&\&  \nodetwo{2}{3}[8] \&\&  \nodetwo{1}{2}[8] 
\&\&  \nodetwo{2}{3}[9] \&\&  \nodetwo{1}{2}[9] 
\&\&  \nodetwo{2}{3}[10] \&\& \nodetwo{1}{2}[10] 
\&\&  \nodetwo{2}{3}[11]\&\&  \nodetwo{1}{2}[11] 
\&\&  \nodetwo{2}{3}[12]\mathrlap{{}=\nodetwo{2}{3}}\\ 
%
\nodeone{3} \&\& \nodeone{2} \&\& \nodeone{1} 
\&\&  \nodethree{1}{2}{3}[1] \&\&  \nodeone{3}[2] \&\&  2[2]
\&\&  1[2] \&\& \nodethree{1}{2}{3}[3] \&\&  3[4] 
\&\&  2[4] \&\&  1[4] \&\&  \nodethree{1}{2}{3}[5] 
\&\&  3[6] \&\&  2[6] \&\&  1[6] 
\&\&  \nodethree{1}{2}{3}[7]\&\&  3[8] \&\&  2[8] 
\&\&  1[8] \&\& \nodethree{1}{2}{3}[9] \&\&  3[10]
\&\&  2[10] \&\&  1[10] \&\&  \nodethree{1}{2}{3}[11] 
\&\&  3[12]\mathrlap{{}=\nodeone{3}} \\ 
}; 
\draw[->] (m-3-1)--(m-2-2); 
\draw[->] (m-3-3)--(m-2-4); 
\draw[->] (m-3-5)--(m-2-6); 
\draw[->] (m-3-7)--(m-2-8); 
\draw[->] (m-3-9)--(m-2-10); 
\draw[->] (m-3-11)--(m-2-12); 
\draw[->] (m-3-13)--(m-2-14); 
\draw[->] (m-3-15)--(m-2-16); 
\draw[->] (m-3-17)--(m-2-18); 
\draw[->] (m-3-19)--(m-2-20); 
\draw[->] (m-3-21)--(m-2-22); 
\draw[->] (m-3-23)--(m-2-24); 
\draw[->] (m-3-25)--(m-2-26); 
\draw[->] (m-3-27)--(m-2-28); 
\draw[->] (m-3-29)--(m-2-30); 
\draw[->] (m-3-31)--(m-2-32); 
\draw[->] (m-3-33)--(m-2-34); 
\draw[->] (m-3-35)--(m-2-36); 
\draw[->] (m-3-37)--(m-2-38); 
\draw[->] (m-3-39)--(m-2-40); 
\draw[->] (m-3-41)--(m-2-42); 
\draw[->] (m-3-43)--(m-2-44); 
\draw[->] (m-3-45)--(m-2-46); 
\draw[->] (m-3-47)--(m-2-48); 
\draw[->] (m-3-49)--(m-2-50); 
%
\draw[->] (m-2-2)--(m-1-3); 
\draw[->] (m-2-4)--(m-1-5); 
\draw[->] (m-2-6)--(m-1-7); 
\draw[->] (m-2-8)--(m-1-9); 
\draw[->] (m-2-10)--(m-1-11); 
\draw[->] (m-2-12)--(m-1-13); 
\draw[->] (m-2-14)--(m-1-15); 
\draw[->] (m-2-16)--(m-1-17); 
\draw[->] (m-2-18)--(m-1-19); 
\draw[->] (m-2-20)--(m-1-21); 
\draw[->] (m-2-22)--(m-1-23); 
\draw[->] (m-2-24)--(m-1-25); 
\draw[->] (m-2-26)--(m-1-27); 
\draw[->] (m-2-28)--(m-1-29); 
\draw[->] (m-2-30)--(m-1-31); 
\draw[->] (m-2-32)--(m-1-33); 
\draw[->] (m-2-34)--(m-1-35); 
\draw[->] (m-2-36)--(m-1-37); 
\draw[->] (m-2-38)--(m-1-39); 
\draw[->] (m-2-40)--(m-1-41); 
\draw[->] (m-2-42)--(m-1-43); 
\draw[->] (m-2-44)--(m-1-45); 
\draw[->] (m-2-46)--(m-1-47); 
\draw[->] (m-2-48)--(m-1-49); 
\draw[->] (m-2-50)--(m-1-51); 
%
%
%
\draw[->] (m-2-2)--(m-3-3); 
\draw[->] (m-2-4)--(m-3-5); 
\draw[->] (m-2-6)--(m-3-7); 
\draw[->] (m-2-8)--(m-3-9); 
\draw[->] (m-2-10)--(m-3-11); 
\draw[->] (m-2-12)--(m-3-13); 
\draw[->] (m-2-14)--(m-3-15); 
\draw[->] (m-2-16)--(m-3-17); 
\draw[->] (m-2-18)--(m-3-19); 
\draw[->] (m-2-20)--(m-3-21); 
\draw[->] (m-2-22)--(m-3-23); 
\draw[->] (m-2-24)--(m-3-25); 
\draw[->] (m-2-26)--(m-3-27); 
\draw[->] (m-2-28)--(m-3-29); 
\draw[->] (m-2-30)--(m-3-31); 
\draw[->] (m-2-32)--(m-3-33); 
\draw[->] (m-2-34)--(m-3-35); 
\draw[->] (m-2-36)--(m-3-37); 
\draw[->] (m-2-38)--(m-3-39); 
\draw[->] (m-2-40)--(m-3-41); 
\draw[->] (m-2-42)--(m-3-43); 
\draw[->] (m-2-44)--(m-3-45); 
\draw[->] (m-2-46)--(m-3-47); 
\draw[->] (m-2-48)--(m-3-49); 
%
%
\draw[->] (m-1-3)--(m-2-4); 
\draw[->] (m-1-5)--(m-2-6); 
\draw[->] (m-1-7)--(m-2-8); 
\draw[->] (m-1-9)--(m-2-10); 
\draw[->] (m-1-11)--(m-2-12); 
\draw[->] (m-1-13)--(m-2-14); 
\draw[->] (m-1-15)--(m-2-16); 
\draw[->] (m-1-17)--(m-2-18); 
\draw[->] (m-1-19)--(m-2-20); 
\draw[->] (m-1-21)--(m-2-22); 
\draw[->] (m-1-23)--(m-2-24); 
\draw[->] (m-1-25)--(m-2-26); 
\draw[->] (m-1-27)--(m-2-28); 
\draw[->] (m-1-29)--(m-2-30); 
\draw[->] (m-1-31)--(m-2-32); 
\draw[->] (m-1-33)--(m-2-34); 
\draw[->] (m-1-35)--(m-2-36); 
\draw[->] (m-1-37)--(m-2-38); 
\draw[->] (m-1-39)--(m-2-40); 
\draw[->] (m-1-41)--(m-2-42); 
\draw[->] (m-1-43)--(m-2-44); 
\draw[->] (m-1-45)--(m-2-46); 
\draw[->] (m-1-47)--(m-2-48); 
\draw[->] (m-1-49)--(m-2-50); 
\end{tikzpicture} 
\end{equation} 
The mesh diagram for the $\phi^4+\phi^5$ theory for $N=12$ with $n=4$ and
$\Lambda=(2,2,3,3)$  is
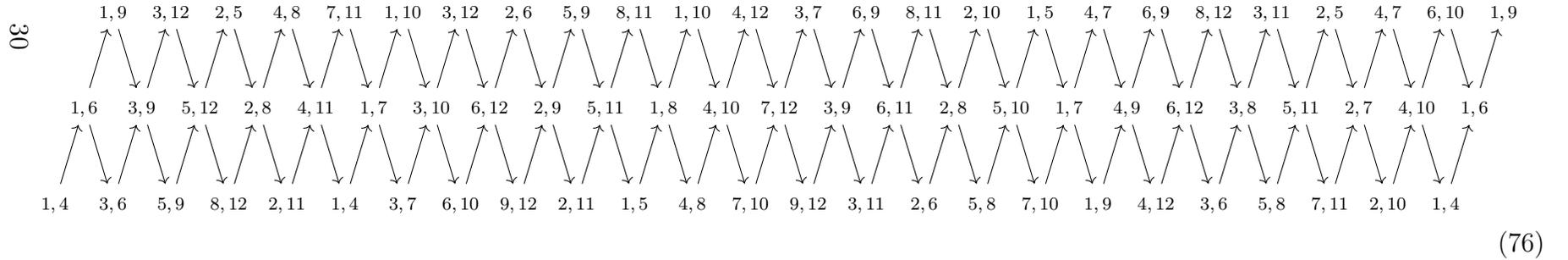
\begin{equation}  
\label{m2233} 
\kern-1cm
\begin{tikzpicture}[baseline=-4pt] 
\matrix (m) [matrix of math nodes, minimum width=1.4em, row 
sep={3.6em,between origins}, text height=1.4ex, column sep={1.1em,between 
origins}, text depth=.15ex, ampersand replacement=\&,font=\scriptsize] 
{ 
\&\&   1,9 \&\&   3,12 \&\&   2,5 
\&\&   4,8 \&\&   7,11 \&\&   1,10 
\&\&   3,12 \&\&   2,6 \&\&   5,9 
\&\&   8,11 \&\&   1,10 \&\&   4,12 
\&\&   3,7 \&\&   6,9 \&\&   8,11 
\&\&   2,10 \&\&   1,5 \&\&   4,7 
\&\&   6,9 \&\&   8,12 \&\&   3,11 
\&\&   2,5 \&\&   4,7 \&\&   6,10 
\&\&   1,9\\ 
%
\&   1,6 \&\&   3,9 \&\&   5,12 
\&\&   2,8 \&\&   4,11 \&\&   1,7 
\&\&   3,10 \&\&   6,12 \&\&   2,9 
\&\&   5,11 \&\&   1,8 \&\&   4,10 
\&\&   7,12 \&\&   3,9 \&\&   6,11 
\&\&   2,8 \&\&   5,10 \&\&   1,7 
\&\&   4,9 \&\&   6,12 \&\&   3,8 
\&\&   5,11 \&\&   2,7 \&\&   4,10 
\&\&   1,6\\ 
%
  1,4 \&\&  3,6 \&\&  5,9 
\&\&  8,12 \&\&  2,11 \&\&  1,4 
\&\&  3,7 \&\&  6,10 \&\&  9,12 
\&\&  2,11 \&\&  1,5 \&\&  4,8 
\&\&  7,10 \&\&  9,12 \&\&  3,11 
\&\&  2,6 \&\&  5,8 \&\&  7,10 
\&\&  1,9 \&\&  4,12 \&\&  3,6 
\&\&  5,8 \&\&  7,11 \&\&  2,10 
\&\&  1,4\\ 
}; 
\draw[->] (m-3-1)--(m-2-2);
\draw[->] (m-3-3)--(m-2-4); 
\draw[->] (m-3-5)--(m-2-6); 
\draw[->] (m-3-7)--(m-2-8); 
\draw[->] (m-3-9)--(m-2-10); 
\draw[->] (m-3-11)--(m-2-12); 
\draw[->] (m-3-13)--(m-2-14); 
\draw[->] (m-3-15)--(m-2-16); 
\draw[->] (m-3-17)--(m-2-18); 
\draw[->] (m-3-19)--(m-2-20); 
\draw[->] (m-3-21)--(m-2-22); 
\draw[->] (m-3-23)--(m-2-24); 
\draw[->] (m-3-25)--(m-2-26); 
\draw[->] (m-3-27)--(m-2-28); 
\draw[->] (m-3-29)--(m-2-30); 
\draw[->] (m-3-31)--(m-2-32); 
\draw[->] (m-3-33)--(m-2-34); 
\draw[->] (m-3-35)--(m-2-36); 
\draw[->] (m-3-37)--(m-2-38); 
\draw[->] (m-3-39)--(m-2-40); 
\draw[->] (m-3-41)--(m-2-42); 
\draw[->] (m-3-43)--(m-2-44); 
\draw[->] (m-3-45)--(m-2-46); 
\draw[->] (m-3-47)--(m-2-48); 
\draw[->] (m-3-49)--(m-2-50); 
%
\draw[->] (m-2-2)--(m-1-3);  
\draw[->] (m-2-4)--(m-1-5); 
\draw[->] (m-2-6)--(m-1-7); 
\draw[->] (m-2-8)--(m-1-9); 
\draw[->] (m-2-10)--(m-1-11); 
\draw[->] (m-2-12)--(m-1-13); 
\draw[->] (m-2-14)--(m-1-15); 
\draw[->] (m-2-16)--(m-1-17); 
\draw[->] (m-2-18)--(m-1-19); 
\draw[->] (m-2-20)--(m-1-21); 
\draw[->] (m-2-22)--(m-1-23); 
\draw[->] (m-2-24)--(m-1-25); 
\draw[->] (m-2-26)--(m-1-27); 
\draw[->] (m-2-28)--(m-1-29); 
\draw[->] (m-2-30)--(m-1-31); 
\draw[->] (m-2-32)--(m-1-33); 
\draw[->] (m-2-34)--(m-1-35); 
\draw[->] (m-2-36)--(m-1-37); 
\draw[->] (m-2-38)--(m-1-39); 
\draw[->] (m-2-40)--(m-1-41); 
\draw[->] (m-2-42)--(m-1-43); 
\draw[->] (m-2-44)--(m-1-45); 
\draw[->] (m-2-46)--(m-1-47); 
\draw[->] (m-2-48)--(m-1-49); 
\draw[->] (m-2-50)--(m-1-51); 
%
%
\draw[->] (m-2-2)--(m-3-3); 
\draw[->] (m-2-4)--(m-3-5); 
\draw[->] (m-2-6)--(m-3-7); 
\draw[->] (m-2-8)--(m-3-9); 
\draw[->] (m-2-10)--(m-3-11); 
\draw[->] (m-2-12)--(m-3-13); 
\draw[->] (m-2-14)--(m-3-15); 
\draw[->] (m-2-16)--(m-3-17); 
\draw[->] (m-2-18)--(m-3-19); 
\draw[->] (m-2-20)--(m-3-21); 
\draw[->] (m-2-22)--(m-3-23); 
\draw[->] (m-2-24)--(m-3-25); 
\draw[->] (m-2-26)--(m-3-27); 
\draw[->] (m-2-28)--(m-3-29); 
\draw[->] (m-2-30)--(m-3-31); 
\draw[->] (m-2-32)--(m-3-33); 
\draw[->] (m-2-34)--(m-3-35); 
\draw[->] (m-2-36)--(m-3-37); 
\draw[->] (m-2-38)--(m-3-39); 
\draw[->] (m-2-40)--(m-3-41); 
\draw[->] (m-2-42)--(m-3-43); 
\draw[->] (m-2-44)--(m-3-45); 
\draw[->] (m-2-46)--(m-3-47); 
\draw[->] (m-2-48)--(m-3-49); 
%
%
\draw[->] (m-1-3)--(m-2-4); 
\draw[->] (m-1-5)--(m-2-6); 
\draw[->] (m-1-7)--(m-2-8); 
\draw[->] (m-1-9)--(m-2-10); 
\draw[->] (m-1-11)--(m-2-12); 
\draw[->] (m-1-13)--(m-2-14); 
\draw[->] (m-1-15)--(m-2-16); 
\draw[->] (m-1-17)--(m-2-18); 
\draw[->] (m-1-19)--(m-2-20); 
\draw[->] (m-1-21)--(m-2-22); 
\draw[->] (m-1-23)--(m-2-24); 
\draw[->] (m-1-25)--(m-2-26); 
\draw[->] (m-1-27)--(m-2-28); 
\draw[->] (m-1-29)--(m-2-30); 
\draw[->] (m-1-31)--(m-2-32); 
\draw[->] (m-1-33)--(m-2-34); 
\draw[->] (m-1-35)--(m-2-36); 
\draw[->] (m-1-37)--(m-2-38); 
\draw[->] (m-1-39)--(m-2-40); 
\draw[->] (m-1-41)--(m-2-42); 
\draw[->] (m-1-43)--(m-2-44); 
\draw[->] (m-1-45)--(m-2-46); 
\draw[->] (m-1-47)--(m-2-48); 
\draw[->] (m-1-49)--(m-2-50); 
\end{tikzpicture} 
\end{equation} 
\end{landscape}

\paragraph*{Acknowledgements}

This collaboration started when the first and last named authors were visiting researchers at the Hausdorff Research Institute for Mathematics in Bonn as part of the Junior Trimester Program ``New Trends in Representation Theory'' and we would like to thank the institute for the hospitality and the excellent working conditions. S.~Barmeier was also supported by the Research Training Group GK1821 ``Cohomological Methods in Geometry'' at the University of Freiburg funded by the Deutsche Forschungsgemeinschaft (DFG, German Research Foundation). H.~Treffinger was partially funded by the DFG under Germany's Excellence Strategy Programme -- EXC-2047/1 -- 390685813. H.~Treffinger is also supported by the European Union's Horizon 2020 research and innovation programme under the Marie Sk\l{}odowska-Curie grant agreement No 893654. H.~Treffinger would also like to thank the Isaac Newton Institute for Mathematical Sciences, Cambridge, for support and hospitality during the programme ``Cluster Algebras and Representation Theory'' where part of the work on this paper was undertaken. This work was supported by EPSRC grant no EP/K032208/1 and the Simons Foundation.


\begin{thebibliography}{99} 
\bibitem{abhy} 
N. Arkani-Hamed, Y. Bai, S. He, G. Yan, 
``Scattering Forms and the Positive Geometry of Kinematics, Color and the 
Worldsheet", JHEP {\bf 05} (2018) 096. 
\bibitem{dfgk} 
J. Drummond, J. Foster, \"O. G\"urdo\u gan, C. Kalousios, 
``Tropical Grassmannians, Cluster Algebras and 
Scattering Amplitudes", JHEP {\bf 04} (2020) 146. 
\bibitem{pppp} 
A. Padrol, Y. Palu, V. Pilaud, P.-G. Plamondon, 
``Associahedra for finite type cluster algebras and minimal relations between 
$g$-vectors", 
arXiv:1906.06861 [math.RT]. 
\bibitem{bdmty} 
V. Bazier-Matte, G. Douville, K. Mousavand, H. Thomas, 
E. Y\i ld\i r\i m, 
``ABHY associahedra and Newton polytopes of $F$-polynomials for finite type 
cluster algebras", 
arXiv:1808.09986 [math-RT]. 
\bibitem{hbcgpt} 
N. Arkani-Hamed, J. L. Bourjaily, F. Cachazo, A. B. 
Goncharov, A. Postnikov, J. Trnka, 
``Scattering amplitudes and the positive Grassmannian", 
arXiv:1212.5605 [hep-th]. 
\bibitem{blr} 
P. Banerjee, A. Laddha, P. Raman, 
``Stokes Polytopes: The positive geometry for $\phi^4$ interactions",
JHEP {\bf 08} (2019) 067. 
\bibitem{raman} 
P. Raman, 
``The positive geometry for $\phi^p$ interactions", 
JHEP {\bf 10} (2019) 271. 
\bibitem{kojima} 
R. Kojima,  
``Weights and recursion relations for $\phi^p$ tree amplitudes from the 
positive geometry", JHEP {\bf 08} (2020) 054. 
\bibitem{ishan} 
I. Srivastava, 
``Constraining the weights of Stokes polytopes using BCFW recursions for 
$\phi^4$", arXiv:2005.12886 [hep-th]. 
\bibitem{abjjlm} 
P. B. Aneesh, P. Banerjee, M. Jagadale, R. R. John,  
A. Laddha, S. Mahato, 
``On positive geometries of quartic interactions II: Stokes polytopes, lower 
forms on associahedra and worldsheet forms", 
JHEP {\bf 04} (2020) 149. 
\bibitem{ajk} 
P. B. Aneesh, M. Jagadale, N. Kalyanapuram, 
``Accordiohedra as positive geometries for generic scalar field theories",  
Phys.\ Rev.\ D {\bf 100} (2019) 106013. 
\bibitem{jl3} 
M. Jagadale and A. Laddha, 
``On the positive geometry of quartic interactions III: 
One loop integrands from polytopes", arXiv:2007.12145 [hep-th]. 
\bibitem{ahs} 
Md. Abhishek, S. Hegde, A. P. Saha, 
``One-loop integrand from generalised scattering 
equations", JHEP {\bf 05} (2021) 012.
\bibitem{causal}  
N. Arkani-Hamed, S. He, G. Salvatori, H. Thomas, 
``Causal diamonds, cluster polytopes and scattering amplitudes", 
arXiv:1912.12948 [hep-th]. 
\bibitem{br} 
S. Barmeier and K. Ray, 
``Learning scattering amplitudes by heart", Phys. Lett. B {\bf 820} 
(2021) 136594.
\bibitem{BM1} 
K. Baur and B. Marsh, 
``A geometric description of the $m$-cluster categories of type $D_n$", 
Int.\ Math.\ Res.\ Not.\ (2007) rnm011.
\bibitem{baur} K. Baur,
``Cluster categories, $m$-cluster categories and diagonals in polygons",
Geometric methods in representation theory II, 261--273, S\'emin.\ Congr.\ 24-II, Soc. Math. France, Paris, 2012.
\bibitem{BM2} 
K. Baur and R. J. Marsh, 
``A geometric description of $m$-cluster categories",
Trans.\ Amer.\ Math.\ Soc.\ {\bf 360} (2008) 5789.
\bibitem{keller} 
B. Keller, 
``Cluster algebras, quiver representations and triangulated categories", 
Triangulated categories, 76--160, Cambridge University Press, 2010. 
\bibitem{schiffler}  
R. Schiffler, ``Quiver representations", Canadian Mathematical Society, 2010. 
\bibitem{bmrrt} 
A. B. Buan, R. J. Marsh, M. Reineke, I. Reiten, G. Todorov,  
``Tilting theory and cluster combinatorics", 
Adv.\ Math.\ {\bf 204} (2006) 572. 
\bibitem{air} 
T. Adachi, O. Iyama, I. Reiten, 
``$\tau$-tilting theory", 
Compos.\ Math.\ {\bf 150} (2014) 415. 
\bibitem{thomas}
R. P. Thomas,
``Derived categories for the working mathematician'',
Winter School on Mirror Symmetry, Vector Bundles and Lagrangian Submanifolds (Cambridge, MA, 1999), 349--361,
American Mathematical Society, Providence, RI, 2001. 
\bibitem{subir} 
S. Mukhopadhyay and K. Ray, 
``Branes in hearts with perverse sheaves", 
Indian J. Phys. A {\bf 80} (2006) 1109.
\bibitem{by} 
T.~Br\"ustle, D.~Yang, 
``Ordered exchange graphs", Advances in Representation Theory of Algebras, 135--193, 
European Mathematical Society, 2013. 
\bibitem{keller5} 
B. Keller, ``On triangulated orbit categories",
Doc.\ Math.\ {\bf 10} (2005) 551.
\bibitem{saito} 
S. Saito, ``Tilting objects in periodic triangulated categories", 
arXiv:2011.14096 [math.RT]. 
\bibitem{brt}
A. B. Buan, I. Reiten, H. Thomas,
``Three kinds of mutation'',
J.\ Algebra {\bf 339} (2011) 97.
\end{thebibliography}
\end{document}